\documentclass[12pt]{article}
\usepackage{cite}
\usepackage{graphicx}
\usepackage{pifont}
\usepackage{float}
\usepackage{subfig}
\usepackage[all]{xy}
\usepackage{multicol}
\usepackage{amsfonts}
\usepackage{picture}
\usepackage{amssymb}
\usepackage{amsmath}
\usepackage{color}
\usepackage{clay}
\usepackage{hyperref}
\hypersetup{colorlinks=true}
\hypersetup{linkcolor=black}
\hypersetup{citecolor=black}
\hypersetup{urlcolor=black}
\usepackage{setspace}
\usepackage{wrapfig}
\usepackage{rotating}
\usepackage{array}
\usepackage{verbatim}
\usepackage{tikz}
\usepackage{braids}

\newcommand*{\vcenteredhbox}[1]{\begingroup
\setbox0=\hbox{#1}\parbox{\wd0}{\box0}\endgroup}

\newcommand{\cN}{\mathcal{N}}
\newcommand{\cW}{\mathcal{W}}

\newcommand{\cV}{\mathcal{V}}
\newcommand{\cM}{\mathcal{M}}

\newcommand{\cZ}{\mathcal{Z}}

\def\IR{{\mathbb R}}

\def\IC{{\mathbb C}}

\def\IK{{\mathbb K}}

\newcommand{\IZ}{{\mathbb Z}}

\makeatletter
\renewcommand\paragraph{\@startsection{paragraph}{4}{\z@}%
  {-3.25ex\@plus -1ex \@minus -.2ex}%
  {1.5ex \@plus .2ex}%
  {\normalfont\normalsize\bfseries}}
\makeatother
\numberwithin{equation}{section}
\begin{document}    

\date{November, 2012}

\institution{Fellows}{\centerline{${}^{1}$Society of Fellows, Harvard University, Cambridge, MA, USA}}
\institution{HarvardU}{\centerline{${}^{2}$Jefferson Physical Laboratory, Harvard University, Cambridge, MA, USA}}

\title{Tangles,  Generalized Reidemeister Moves, and Three-Dimensional Mirror Symmetry}

\authors{Clay C\'{o}rdova,\worksat{\Fellows}\footnote{e-mail: {\tt cordova@physics.harvard.edu}} Sam Espahbodi,\worksat{\HarvardU}\footnote{e-mail:{\tt espahbodi@physics.harvard.edu}} Babak Haghighat,\worksat{\HarvardU}\footnote{e-mail:{\tt babak@physics.harvard.edu}} Ashwin Rastogi,\worksat{\HarvardU}\footnote{e-mail:{\tt rastogi@physics.harvard.edu}} and Cumrun Vafa \worksat{\HarvardU}\footnote{e-mail:{\tt vafa@physics.harvard.edu}} }

\abstract{Three-dimensional $\mathcal{N}=2$ superconformal field theories are constructed by compactifying M5-branes on three-manifolds.  In the infrared the branes recombine, and the physics is captured by a single M5-brane on a branched cover of the original ultraviolet geometry.   The branch locus is a \emph{tangle}, a one-dimensional knotted submanifold of the ultraviolet geometry.  A choice of branch sheet for this cover yields a Lagrangian for the theory, and varying the branch sheet provides dual descriptions.  Massless matter arises from vanishing size M2-branes and appears as singularities of the tangle where branch lines collide.  Massive deformations of the field theory correspond to resolutions of singularities resulting in distinct smooth manifolds connected by geometric transitions.  A generalization of Reidemeister moves for singular tangles captures mirror symmetries of the underlying theory yielding a geometric framework where dualities are manifest.  }

\maketitle

\setcounter{tocdepth}{2}
\begin{spacing}{.9}
\tableofcontents
\end{spacing}
\section{Introduction}
\label{intro}
The $(2,0)$ superconformal field theories in six dimensions, in particular the theory of $N$ parallel M5-branes, are among the most important quantum systems, and yet they remain poorly understood.  Their importance
stems not only from the fact that they represent the highest possible dimension in which superconformal field theories can exist \cite{Nahm}, but also from the observation that their compactifications to lower dimensions yield a rich class of quantum field theories whose dynamics are encoded by geometry.  For example, four-dimensional $\mathcal{N}=2$ theories arise upon compactification on a Riemann surface  \cite{KLMVW,DW,Witten1997,Gaiotto}, and provide a geometric explanation for Seiberg-Witten theory \cite{SW1, SW2}.

It is natural to expect that more general compactifications will provide more information about these mysterious six-dimensional theories.  One way to do this is to increase the dimension of the compactification geometry.   Thus, the next cases of interest would be compactifications with dimensions $d\geq 3$ resulting at low-energies in effective quantum field theories in dimensions $6-d$.   The aim of this paper is to focus on the situation where $d=3$ with ${\cal N}=2$ supersymmetry. Examples of this type have been recently considered in \cite{DGG1, DGG2, CCV} for the situation where 2 M5-branes wrap some ultraviolet geometry.  In such constructions, as advocated in \cite{CCV}, the infrared dynamics of the system is described by a single recombined brane, similar to the situation studied in \cite{TBranes}, that can be viewed as a double cover of the original compactifiaction manifold.  This infrared geometry is captured by describing the branching strands for the cover which in general are knotted.  When the branching strands collide the cover becomes singular and on that locus an M2-brane of vanishing size can end on the M5-branes leading to massless charged matter fields.  The goal of this paper is to clarify and extend the rules discussed in \cite{CCV} and find the correspondence between the knotted branch locus encoding the geometry of the double cover, and the underlying ${\cal N}=2$ quantum field theory.

\begin{figure}[here!]
  \centering
  \subfloat[A Tangle]{\label{fig:tangex}\includegraphics[width=0.45\textwidth]{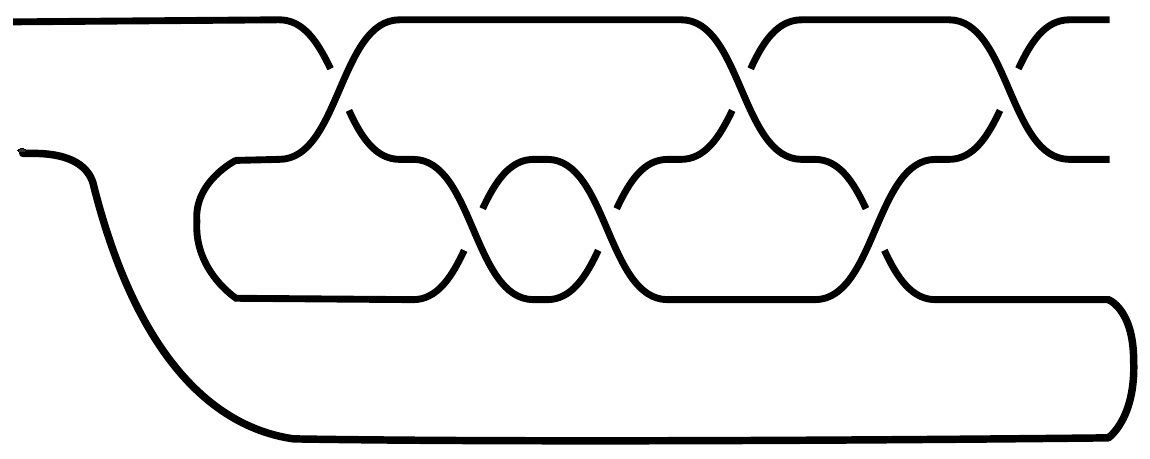}}     
  \hspace{.25in}    
  \subfloat[A Singular Tangle]{\label{fig:stangex}\includegraphics[width=0.45\textwidth]{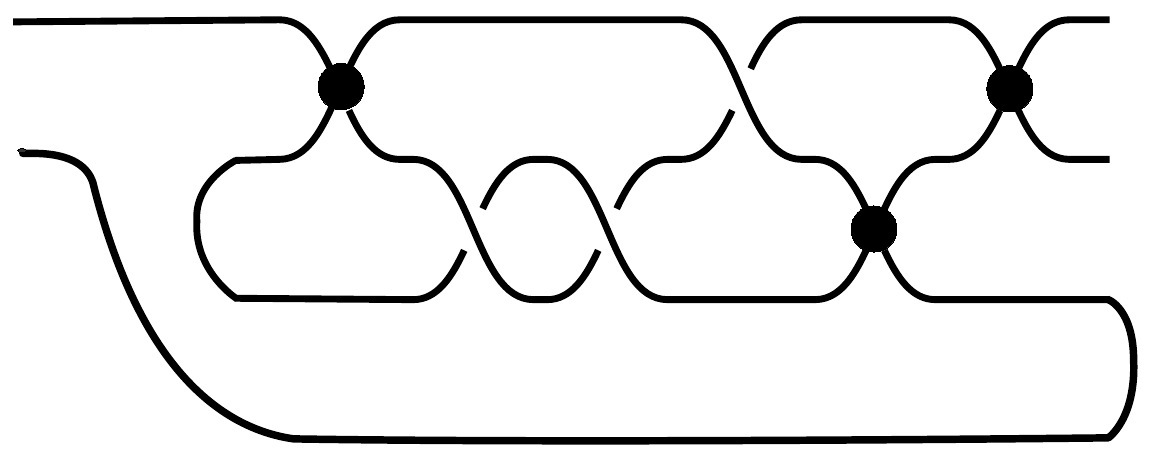}}
  \caption{Examples of tangles and their singularities.  In (a) a tangle in the $\mathbb{R}^{3}$.  In (b) a singular tangle where the strands have merged at various points.  The singularities are modeled in quantum field theory by charged massless matter. }
  \label{fig:exs}
\end{figure}

With this background we can phrase more precisely what we wish to do: we would like to uncover the relationship between three-dimensional $\mathcal{N}=2$ supersymmetric conformal field theories and a class of mathematical objects called \emph{singular tangles}.  In words, a tangle is a generalization of a knot to allow for open ends, and a singular tangle is the situation where the pieces of string are permitted to merge and loose their individual identity. Examples are illustrated in Figure \ref{fig:exs}.

The class of three-manifolds $M$ where the infrared M5-brane resides are defined as double covers of  $\mathbb{R}^{3}$ branched along a singular tangle.  The reduction of the theory of a single M5-brane along $M$ will result in the three-dimensional quantum field theories under investigation.  The simplest class of examples are associated to non-singular tangles.  In this situation $M$ is a smooth manifold and a single M5-brane on $M$ constructs a free Abelian $\mathcal{N}=2$ Chern-Simons theory in the macroscopic dimensions.  Light matter, appearing in chiral multiplets in three dimensions, arises in the theory from  M2-brane discs which end along $M$.   When such matter becomes massless, the associated cycle shrinks and $M$ develops a singularity.  The collapsing of this cycle can be described by the geometry of a singular tangle.

A conceptual slogan for the program described above is that we are investigating a three-dimensional analog of Seiberg-Witten theory.  In the ultraviolet, one may envision an unknown non-Abelian three-dimensional field theory arising from the interacting theory of two M5-branes on $\mathbb{R}^{3}$ with suitable boundary conditions at infinity.   Moving onto the moduli space of this theory is accomplished geometrically by allowing the pair of M5-branes to fuse together into a single three-manifold $M$.  The long-distance Abelian physics can then be directly extracted from the geometry of $M$.  The situation we have described should be compared with the case of four-dimensional $\mathcal{N}=2$ theories whose infrared moduli space physics can be extracted from a Seiberg-Witten curve.  In that case, charged matter fields are described by BPS states and can be constructed in M-theory from M2-branes.  The case of an interacting conformal field theory can arise when the M2-brane particles become massless and the Seiberg-Witten curve develops a singularity, directly analogous to the three-dimensional setup outlined above.

An important feature of the constructions carried out in this paper, familiar from many constructions of field theories by branes, is that non-trivial quantum properties of field theories are mapped to simpler geometric properties of the compactification manifold.  In the case of $\mathcal{N}=2$ Abelian Chern-Simons matter theories the quantum features which are apparent in geometry are the following.
\begin{itemize}
\item $Sp(2F,\mathbb{Z})$ Theory Multiplets:  

The set of three dimensional theories with $\mathcal{N}=2$ supersymmetry and $U(1)^{F}$ flavor symmetry is naturally acted on by the group $Sp(2F,\mathbb{Z})$  \cite{WittenSL2,KapusStrass}.  This group does not act by dualities.  It provides us with a simple procedure for building complicated theories out of simpler ones by a sequence of shifts in Chern-Simons levels and gauging operations.
\item Anomalies: 

In three dimensions, charged chiral multiplets have non-trivial parity anomalies.  This means that upon integrating out a massive chiral field the effective Chern-Simons levels are shifted by half-integral amounts \cite{Redlich}.
\item Dualities:

Three dimensional $\mathcal{N}=2$ conformal field theories enjoy mirror symmetry dualities.  Thus, distinct $\mathcal{N}=2$ Abelian Chern-Simons matter theories may flow in the infrared to the same conformal field theory.  In the case of three-dimensional Abelain Chern-Simons matter theories there are essentially three building block mirror symmetries which we may compose to engineer more complicated dualities.
\begin{itemize}
\item Equivalences amongst pure CS theories.  These theories are free and characterized by a matrix of  integral levels $K$.  It may happen that two distinct classical theories given by matrices $K_{1}$ and $K_{2}$ nevertheless give rise to equivalent correlation functions and hence are quantum mechanically equivalent.
\item Gauged $U(1)$ at level $1/2$ with a charge one chiral multiplet is mirror to the theory of a free chiral multiplet \cite{KapusStrass}.
\item Super-QED with one flavor of electron is mirror to a theory of three chiral multiplets, no gauge symmetry, and a cubic superpotential \cite{IS,XYZ}.
\end{itemize}
\end{itemize}

One way non-trivial dualities appear stems from the fact that the M5-brane theory reduced on $M$ does not have a preferred classical Lagrangian.  To obtain a Lagrangian description of the dynamics requires additional choices.  In our context such a choice is a Seifert surface, which is a Riemann surface with boundary the given tangle.  For any given tangle there exist infinitely many distinct choices of Seifert surfaces each of which corresponds to a distinct equivalent Lagrangian description of the physics.  This fact is closely analogous to the choice of triangulation appearing in the approach of \cite{DGG1} for studying the same theories, as well as the choice of pants decomposition required to provide a Lagrangian description of M5-branes on Riemann surfaces \cite{Gaiotto}.

Throughout the paper, our discussion of duality will be guided by a particular invariant of the infrared conformal field theory, the squashed  three-sphere partition function
\begin{equation}
\mathcal{Z}_{b}(x_{1},\cdots, x_{F}).
\end{equation}
This is a complex-valued function of a squashing parameter $b$ (which we frequently suppress in notation) as well as $F$ chemical potentials $x_{i}$.  It is an invariant of a field theory with prescribed couplings to $U(1)^{F}$ background flavor fields.  This partition function gives us a strong test for two theories to be mirror and as such it is useful to build into the formalism techniques for computing $\mathcal{Z}$.

One method of explicit computation is provided by supersymmetric localization formulas.  At the classical level, an Abelian Chern-Simons matter theory coupled to background flavor fields is determined by the following data:
\begin{itemize}
\item Integers $G$ and $F$ specifying that the theory in question has a $U(1)^{G}$ gauge group and a $U(1)^{F}$ flavor group,
\item An $(G+F)\times(G+F)$ matrix of Chern-Simons levels,
\item A set of chiral multiplets $\Phi_{\mathfrak{a}}$, with $(G+F)$ dimensional charge vectors $q_{\mathfrak{a}}$,
\item A superpotential $W(\Phi_{\mathfrak{a}})$; a holomorphic function of chiral fields.
\end{itemize}
Given such data, the three-sphere partition function for the infrared conformal field theory can be presented as a finite dimensional integral\footnote{In the following formula, certain details about R-charge assignments are suppressed.  These will be dealt with more fully in section \ref{particles}.} \cite{Jafferis, Kapustin}
\begin{equation} 
\mathcal{Z}(x_{i})=\int d^{G}y \exp\left(-\pi i (y \  \ x)K\left(\begin{array}{c}y \\ x\end{array}\right)\right)\prod_{\mathfrak{a}}E(q_{\mathfrak{a}}\cdot (y \ \ x)). \label{generalpart}
\end{equation}
In the above, $E(x)$ denotes a certain transcendental function, the so-called non-compact quantum dilogarithm, which will be discussed in detail in section \ref{particles}.  The superpotential $W$ enters the discussion only in so far as it restricts the flavor symmetries of the theory.  The real integration variables $y$ appearing in the formula can be interpreted as parameterizing fluctuations of the real scalars in the $\mathcal{N}=2$ vector multiplets.  

We will be interested in computation of $\mathcal{Z}$ up to multiplication by an overall phase independent of all flavor variables.  Physically this means in particular that throughout this work we will ignore all framing anomalies of Chern-Simons terms.  We will see that the partition function in \eqref{generalpart} can be usefully viewed as a wavefunction in a certain finite dimensional quantum mechanics and develop this interpretation throughout.

One important test of the ideas that we develop can be found in their application to a class of three-manifolds $M$ of the form $\Sigma_{t} \times \mathbb{R}_{t}$, where the Reimann surface $\Sigma$ varies in complex structure along the line parameterized by $t$.  These examples are closely connected to four-dimensional quantum field theories.    At a fixed value of $t$, the situation is that of an M5-brane on $\Sigma$ which can be interpreted as a Seiberg-Witten curve for a four-dimensional $\mathcal{N}=2$ field theory.  As $t$ varies this field theory moves in its parameter space and hence describes a kind of domain wall in four dimensions.  When equipped with suitable boundary conditions, this geometry can engineer a three-dimensional $\mathcal{N}=2$ theory.  

In the context of such examples, one may utilize the machinery of Seiberg-Witten theory and BPS state counting to determine the resulting three-dimensional physics. When the variation of $\Sigma$ takes a particularly natural form, known as $\emph{R-flow}$, the spectrum of three-dimensional chiral multiplets is  in one-to-one correspondence with the BPS states of the underlying four-dimensional model in a particular chamber.  As the moduli of the four-dimensional theory are varied, one may cross walls of marginal stability and hence find distinct spectra of chiral multiplets in three-dimensions. Remarkably, the resulting three-dimensional theories are mirror symmetric.  In this way, the geometry provides a striking confluence between two fundamental quantum phenomena: wall crossing of BPS states, and mirror symmetry.

The organization of this paper is as follows.  In section \ref{nopart} we explain how free Abelian Chern-Simons theories arise from tangles, and how their partition functions are encoded in a simple quantum mechanical setup.  In section \ref{particles} we show how the data of massless chiral fields is encoded in terms of singular tangles where branch loci collide.  Each such singularity can be geometrically resolved in one of three ways, matching the expected deformations of the field theory.  Upon fixing a Seifert surface, a surface with boundary on the tangle, we are able to extract a Lagrangian description of the theory associated to the singular tangle including superpotential couplings.   In section \ref{dualities} we generalize to arbitrary singular tangles, and explore physical redundancy in the geometry.  As a consequence of mirror symmetries, distinct singular tangles can give rise to the same superconformal theory.  These equivalences on field theories can be described geometrically by introducing a set of generalized Reidemeister moves acting on singular tangles.  On deforming away from the critical point by activating relevant deformations of the field theory, we find that the generalized Reidemeister moves resolve to the ordinary  Reidemeister moves familiar from elementary knot theory.  The appearance of Reidemeister moves clarifies the relationship between quantum dilogarithm functions and braids first observed by \cite{Faddeev:2000if}. In section \ref{rflow} we describe how three-dimensional mirror symmetries can be understood from the perspective of four-dimensional $\mathcal{N}=2$ parent theories via R-flow.  Finally, in section \ref{sec:app} we describe three-dimensional $U(1)$ SQED with arbitrary $N_f$.

\section{Abelian Chern-Simons Theory and Tangles}
\label{nopart}

In this section we explore the simplest class of examples: Abelian $\mathcal{N}=2$ Chern-Simons theories without matter fields.  Such theories are free and hence of course conformal.  We find that such models are usefully constructed via reduction of the M5-brane on a non-singular manifold which is conveniently viewed as a double cover of $\mathbb{R}^{3}$ branched over a tangle, and describe the necessary geometric technology for elucidating their structure.  In addition we describe a finite dimensional quantum mechanical framework for evaluating their partition functions.  Throughout we will study the theories with $U(1)^{F}$ flavor symmetries and couple them to $F$ non-dynamical vector multiplets.  The set of such theories is acted upon by $Sp(2F,\mathbb{Z})$ and we describe this action from various points of view.

\subsection{Chern-Simons Actions, $Sp(2F,\mathbb{Z})$, and Quantum Mechanics}
Consider a classical $\mathcal{N}=2$ Abelian Chern-Simons theory.  Let $G$ denote the number of $U(1)$ gauge groups, and $F$ the number of $U(1)$ flavor groups.\footnote{Here and in the following \emph{flavor} symmetries refer to non-$R$ symmetries except when explicitly indicated otherwise. }  The Lagrangian of the theory coupled to $F$ background vector multiplets is specified by a $(G+F)\times (G+F)$, symmetric matrix of levels
\begin{equation}
K=\left(\begin{array}{cc}k_{G} & k_{M}\\ k^{T}_{M} & k_{F}\end{array}\right), \hspace{.5in}k_{G}=k_{G}^{T}, k_{F}=k_{F}^{T}. \label{csmex}
\end{equation} 
Here, $k_{G}$ denotes the ordinary Chern-Simons levels of the $U(1)^{G}$ gauge group, $k_{M}$ indicates the $G\times F$  matrix of mixed gauge-flavor levels, and $k_{F}$ the $F\times F$ matrix of flavor levels.  The action for the theory is
\begin{equation}
\sum_{\alpha \beta}\frac{K_{\alpha \beta}}{4\pi}\int d^{3}x \ A_{\alpha}\wedge dA_{\beta} + \cdots, \hspace{.5in}\alpha, \beta =1, 2, \cdots G+F. \label{csactdef}
\end{equation}
Where in the above the terms $``\cdots "$ indicate the supersymetrization of the Chern-Simons Lagrangian.  $K_{\alpha \beta}$ is integrally quantized with minimal unit one. The first $G$ vector multiplets are dynamical variables in the path integral while the last $F$ are non-dynamical background fields.\footnote{The normalization of the Chern-Simons levels appearing in \eqref{csactdef} indicates that these are \emph{spin} Chern-Simons theories \cite{MooreCS} whose definition depends on a choice of spin structure on spacetime.  Since all the models we consider are supersymmetric and hence contain dynamical fermions, this is no restriction.}

It is worthwhile to note that one might naively think that the matrix $K$ does not completely specify an $\mathcal{N}=2$ Chern-Simons theory.  Indeed, since such theories are conformal they contain a distinguished flavor symmetry, $U(1)_{R}$, whose associated conserved current appears in the same supersymmetry multiplet as the energy-momentum tensor.  One might therefore contemplate Chern-Simons couplings involving background $U(1)_{R}$ gauge fields.  However, such terms while supersymmetric violate conformal invariance.  Thus, as our interest here is superconformal field theories, we are justified in ignoring these couplings.

Already in this simple context of Abelian Chern-Simons theory, we can see the action of $Sp(2F,\mathbb{Z})$ specified as operations on the level matrix $K$ defined in equation \eqref{csmex}.  For later convenience, it is useful to use a slightly unconventional form of the symplectic matrix $J$
\begin{equation}
J=\left(\begin{array}{ccccccc}
0 & 1 & 0 & 0 & \cdots & 0 & 0 \\
-1 & 0 & 1 & 0 & \cdots & 0 & 0 \\
0 & -1 & 0 & 1 & \cdots & 0 & 0 \\
\vdots  & \vdots & \ddots &\ddots  &\ddots &\vdots & \vdots \\
0 & 0 & 0 & 0 & \cdots & 0 & 1 \\
0 & 0 & 0 & 0 & \cdots &-1 &  0 \\
 \end{array}\right).
 \label{jdef}
\end{equation}
In this basis, the integral symplectic group is conveniently generated by $2F$ generators $\sigma_{n}$ with $n=1, 2, \cdots, 2F$  whose matrix elements are given as
\begin{equation}
(\sigma_{n})_{i,j}=\delta_{i,j}+\delta_{i, n}\delta_{n+1, j}-\delta_{i,n}\delta_{n-1,j}, \hspace{.5in} i,j=1,2,\cdots, 2F \label{sigmaspdef}
\end{equation}
To define an action of the symplectic group $Sp(2F,\mathbb{Z})$ on this class of theories, it therefore suffices to specify the action of the generators $\sigma_{n}$.  

The action of the generators with odd labels $\sigma_{2n-1}$ preserves the number of gauge groups and shifts the levels of the $n$-th background field
\begin{equation} 
\sigma_{2n-1}: \hspace{.5in} (k_{G})_{i,j}\rightarrow (k_{G})_{i,j}, \hspace{.5in}(k_{M})_{i,j}\rightarrow (k_{M})_{i,j}, \hspace{.5in} (k_{F})_{i,j}\rightarrow (k_{F})_{i,j}+\delta_{i,n}\delta_{n,j}. \label{oddbraid}
\end{equation}
The action of the even generators, $\sigma_{2n},$ is more complicated and performs a change of basis in the flavor symmetries while at the same time increasing the number of gauge groups by one.  Explicitly, $\sigma_{2n}$ can be factored as $\sigma_{2n}=g_{n}\circ c_{U}$ where $c_{U}$ is a change of basis operation 
\begin{equation}
c_{U}: \hspace{.5in}k_{G}\rightarrow k_{G}, \hspace{.5in} k_{M}\rightarrow k_{M}U, \hspace{.5in} k_{F} \rightarrow U^{T}k_{F}U  \label{evenbraid}
\end{equation}
where in the above, the $F\times F$ matrix $U$ is given by
\begin{equation}
(U)_{i,j}=\delta_{i,j}-\delta_{i-1,j}. \label{udef}
\end{equation}
And the gauging operation $g_{n}$ is given by
\begin{eqnarray}
k_{G}&\rightarrow &\left(\begin{array}{c|c}
k_{G} & (k_{M})_{i,n} \label{gaugingbraider}\\
\hline
(k_{M}^{T})_{n,i} & (k_{F})_{n,n}-1
\end{array}\right), \\
 k_{M}&\rightarrow& \left(\begin{array}{cccccccc}
(k_{M})_{i,1} & (k_{M})_{i,2} & \cdots & (k_{M})_{i,n-1} & 0 & (k_{M})_{i,n+1}& \cdots & (k_{M})_{i,F} \\
\hline
(k_{F})_{n,1} & (k_{F})_{n,2} & \cdots & (k_{F})_{n,n-1} &1 & (k_{F})_{n,n+1}& \cdots & (k_{F})_{n,F} 
\end{array}
\right) , \nonumber\\
k_{F}& \rightarrow & \left(\begin{array}{cccccccc}
(k_{F})_{1,1} & (k_{F})_{1,2} & \cdots & (k_{F})_{1,n-1} & 0 & (k_{F})_{1,n+1}& \cdots & (k_{F})_{1,F} \\
(k_{F})_{2,1} & (k_{F})_{2,2} & \cdots & (k_{F})_{2,n-1} & 0 & (k_{F})_{2,n+1}& \cdots & (k_{F})_{2,F} \\
\vdots & \vdots &\vdots &\vdots &\vdots &\vdots &\vdots &\vdots \\
(k_{F})_{n-1,1} & (k_{F})_{n-1,2} & \cdots & (k_{F})_{n-1,n-1} & 0 & (k_{F})_{n-1,n+1}& \cdots & (k_{F})_{n-1,F} \\
0 & 0 & \cdots & 0 & -1 & 0 & \cdots & 0 \\
(k_{F})_{n+1,1} & (k_{F})_{n+1,2} & \cdots & (k_{F})_{n+1,n-1} & 0 & (k_{F})_{n+1,n+1}& \cdots & (k_{F})_{n+1,F} \\
\vdots & \vdots &\vdots &\vdots &\vdots &\vdots &\vdots &\vdots \\
(k_{F})_{F,1} & (k_{F})_{F,2} & \cdots & (k_{F})_{F,n-1} & 0 & (k_{F})_{F,n+1}& \cdots & (k_{F})_{F,F}
\end{array}\right). \nonumber
\end{eqnarray}

Straightforward calculation using Gaussian path integrals may be used to verify that these operations satisfy the defining relations of $Sp(2F,\mathbb{Z})$.  Notice that, while these relations are simple to prove, they nevertheless involve \emph{quantum} field theory in an essential way.  If $w$ is any word in the generators $\sigma_{i}$ which is equal to the identity element by a relation in the symplectic group, then the action of $w$ on a given matrix of levels $K$ produces a new matrix $w(K)$ which in general is not equal, as a matrix, to $K$.   Nevertheless, the path integral performed with the matrices $K$ and $w(K)$ produce identical correlation functions.  Thus, the relations in $Sp(2F,\mathbb{Z})$ provide us with elementary, provable examples of duality in three-dimensional conformal field theory.

Let us now turn our attention to the partition function $\mathcal{Z}$ for this class of models.  Since Abelian Chern-Simons theory is free, an application of the localization formula \eqref{generalpart} reduces the computation to a simple Gaussian integral which is a function of an $F$-dimensional vector $x$ of chemical potentials for the $U(1)^{F}$ flavor symmetry
\begin{equation}
\mathcal{Z}(x)=\int d^{G}y  \ \exp\left[-\pi i\left(y \  \  x\right) K\left(\begin{array}{c}y \\ x\end{array}\right)\right]. \label{zintdef}
\end{equation}
The integral is trivially done to obtain\footnote{As remarked in the introduction, we are only interested in $\mathcal{Z}$ up to overall phases independent of all flavor variables.  Thus in the following formulas we neglect such phases.}
\begin{equation}
\mathcal{Z}(x)=\frac{1}{\sqrt{|\det(k_{G})|}}\exp\left[-\pi ix^{T}\tau x\right], \hspace{.5in} \tau \equiv k_{F}-k_{M}^{T}k_{G}^{-1}k_{M}. \label{Gaussianres}
\end{equation}
From the resulting formula we see that the partition function is labeled by two invariants
\begin{equation}
|\det(k_{G})|\in\mathbb{N}, \hspace{.5in} \tau \in \mathfrak{gl}(F,\mathbb{Q}\cup \{\infty\}), \hspace{.2in} \tau^{T}=\tau. \label{invts}
\end{equation}
The possibility that the matrix $\tau$ may have infinite entries is included to allow for non-invertible $k_{G}$.   In that case, the associated vector in the kernel of $k_{G}$ describes a massless $U(1)$ vector multiplet and the flavor variable coupling to this multiplet is interpreted as a Fayet-Illiopoulos parameter.  At the origin of this flavor variable the vector multiplet in question has a non-compact cylindrical Coulomb branch.  This flat direction is not lifted when computing the path integral on $S^{3}$ because the $R$-charge assignments do not induce conformal mass terms.  This implies that the partition function $\mathcal{Z}$ has a diveregence.  Meanwhile, away from from the origin the non-zero FI parameter breaks supersymmetry and $\mathcal{Z}$ vanishes.  In total then, the partition function is proportional to a delta function in the flavor variable, and the narrow width limit of the Gaussian, when entries of $\tau$ are infinite, with infinite coefficient, $\det(k_{G})\rightarrow 0$,  should be interpreted as such a delta function.

The partition function formula \eqref{Gaussianres} provides another context to illustrate the symplectic group $Sp(2F,\mathbb{Z})$ on conformal field theories, in this case, via its action on the invariants \eqref{invts}.  A general symplectic matrix can be usefully written in terms of $F\times F$ blocks as
\begin{equation}
R\left(\begin{array}{cc} A & B \\
C & D
\end{array}\right)R^{T}.
\end{equation}
Where $R$ is certain invertible matrix which transforms the standard symplectic form to our choice \eqref{jdef} whose precise form is not important.  Then, the action of symplectic transformations on $\tau$ is simply the standard action of the sympletic group on the Siegel half-space
\begin{equation}
\tau \rightarrow (A\tau+ B)(C\tau + D)^{-1}.
\end{equation}
Meanwhile, $\det(k_{G})$ transforms as a modular form
\begin{equation}
\det(k_{G})\rightarrow \det(C\tau +D )\det(k_{G}).
\end{equation}
Thus the symplectic action on field theories reduces, at the level of partition functions, to the more familiar symplectic action on Gaussian integrals.

Before moving on to additional methods for studying these theories, let us revisit the issue of Chern-Simons couplings involving a background $U(1)_{R}$ gauge field.  As remarked above such couplings are forbidden by superconformal invariance.  Nevertheless, to elucidate the physical content of $\mathcal{Z}(x)$ as well as the partition functions on interacting field theories appearing later in this paper it is useful to examine exactly how such spurious terms would enter the result.  

The squashed three-sphere partition functions under examination are Euclidean path integrals on the manifold
\begin{equation}
b|z_{1}|^{2}+\frac{1}{b}|z_{2}|^{2}=0, \hspace{.5in} (z_{1}, z_{2})\in \mathbb{C}^{2}, \hspace{.2in} b\in \mathbb{R}_{+}.
\end{equation}
This geometry is labelled by a parameter $b$, a positive real number, however the symmetry under $b\rightarrow 1/b$ allows us to restrict our attention to the parameter 
\begin{equation}
c_{b}\equiv \frac{i}{2}\left(b+\frac{1}{b}\right). \label{cbdef}
\end{equation}
In this geometry preservation of supersymmetry requires one to turn on background values for scalars in the supergravity multiplet.  While these fields are normally real, like the real mass variables $x_{i}$ coupling to the ordinary flavors, in this background they are imaginary and proportional to $c_{b}$.  As a result $R-R$ Chern-Simons levels, and $R$-flavor Chern-Simons levels appear as Gaussian prefactors in the partition function of the form
\begin{equation}
\exp\left(i\pi k_{RR}(c_{b})^{2}+2\pi i k_{RF}(c_{b})x\right). \label{badcsterms}
\end{equation}
From the above, we note that the $R-R$ Chern-Simons levels appear as multiplicative constants independent of the flavor variables $x$.  Since we are interested in computation of partition functions up to overall multiplication by phases such terms are not relevant for this work.  On the other hand, the $R-F$ Chern-Simons terms appear as linear terms in $x$ in the exponent.  One can easily see why such terms violate superconformal invariance.  The round three-sphere partition function for the conformal field theory in the absence of background fields is given by evaluating $\mathcal{Z}(x)$ at vanishing $x$ and $c_{b}=i$.  The first derivative with respect to $x$ evaluated at the round three-sphere and vanishing $x$ therefore computes the one-point function of the associated current
\begin{equation}
\partial_{x} \mathcal{Z}(x)|_{x=0,c_{b}=i}\sim k_{RF}\sim \langle j_{F}\rangle.
\end{equation}
As the three-sphere is conformal to flat space, conformal invariance means that this one point function vanishes implying that $k_{RF}$ must also vanish.

Quite generally throughout this paper we encounter examples of partition functions of interacting CFTs where the naive value of $k_{RF}$, as extracted from the first derivative of $\mathcal{Z}(x)$ evaluated at the conformal point, does not vanish.  Superconformal invariance can always be restored in such examples by explicitly including ultraviolet counterterm values for $k_{RF}$ to cancel the spurious contributions \cite{contactterms}.  Thus, from now on we write expressions for partition functions with non-vanishing first derivatives, always keeping in mind that the true physical partition function of the conformal theory is only obtained by including suitable counterterms.

\subsubsection{Quantum Mechanics and Partition Functions}
\label{sec:QM}
The partition function calculations and $Sp(2F,\mathbb{Z})$ action described in the previous section can be phrased in useful way in elementary quantum mechanics.  We consider the Hilbert space of complex valued functions of $F$ real variables and aim to interpret $\mathcal{Z}(x)$ as a wavefunction.\footnote{As is typical in quantum mechanical settings we have need of wavefunctions which are not square integrable.  Indeed all the partition functions associated to pure Chern-Simons theories are non-normalizable.}

First, introduce position and momentum operators acting on wavefunctions and consistent with the symplectic matrix $J$ introduced in \eqref{jdef}
\begin{equation}
\hat{x_{i}}\rightarrow x_{i}, \hspace{.5in} \hat{p_{j}}\rightarrow -\frac{i}{2\pi}\frac{\partial}{\partial x_{j}}+\frac{i}{2\pi}\frac{\partial}{\partial x_{j-1}}, \hspace{.5in}[\hat{x}_{i}, \hat{p}_{j}]=\frac{i}{2\pi}(\delta_{i,j}-\delta_{i ,j+1}).
\end{equation}
We use Dirac bra-ket notation for states, and let $|y\rangle$ denote a normalized simultaneous eigenstate of the position operators
\begin{equation}
\hat{x}_{i}|y \rangle= y_{i}|y \rangle, \hspace{.5in} \langle x|y\rangle =\delta(x-y), \hspace{.5in} \mathbf{1}=\int dy  \ |y\rangle \langle y|.
\end{equation}
For convenience we also note that the wavefunction of a momentum eigenstate takes the form
\begin{equation}
\langle y|p \rangle=\exp\left[2\pi i\left(y_{1}p_{1}+y_{2}(p_{1}+p_{2})+\cdots + y_{F}(p_{1}+p_{2}+\cdots+p_{F})\right)\right].
\end{equation}

On this Hilbert space there is a natural unitary representation of $Sp(2F,\mathbb{Z})$.  This representation is defined using the generators \eqref{sigmaspdef} as follows\footnote{Technically speaking, the operators above must be multiplied by a certain overall (operator independent) phase.  However, since we are ignoring phases in our partition functions, we will also ignore overall phases in quantum mechanics matrix elements.}
\begin{equation}
\sigma_{2j-1}\mapsto \exp\left(-i\pi \hat{x}_{j}^{2}\right), \hspace{.5in}\sigma_{2j} \mapsto \exp\left(-i \pi \hat{p}_{j}^{2}\right). \label{qmsp2fdef}
\end{equation}
One important feature of this representation is that its action by conjugation on position and momentum operators produces quantized canonical transformations.  Explicitly, if $M$ is any symplectic transformation we have
\begin{equation}
M\left(\sum_{j=1}^{F}a_{2j-1}\hat{x}_{j}+a_{2j}\hat{p}_{j}\right)M^{-1}=\sum_{j=1}^{F}\sum_{k=1}^{2F}\left(M_{2j-1, k}a_{k}\right)\hat{x}_{j}+\left(M_{2j, k}a_{k}\right)\hat{p}_{j}.
\end{equation}
This fact underlies the significance of this representation in all that follows.

We now wish to show that we may interpret the partition function of a theory $\Psi$ as a wavefunction of an associated state $|\Psi\rangle$
\begin{equation}
\mathcal{Z}_{\Psi}(x)= \langle x |\Psi\rangle. \label{statedef}
\end{equation}
Of course both wavefunctions and partition functions are complex-valued functions of a $F$ real variables $x_{i}$ so we are free to make the identification appearing in \eqref{statedef}.  The non-trivial aspect of this identification is that the $Sp(2F,\mathbb{Z})$ action on quantum field theories, defined by the operations appearing in \eqref{oddbraid}-\eqref{evenbraid} can be achieved at the level of the partition function by the action of the operators of the same name defined by the representation given in \eqref{qmsp2fdef}.  To see that these quantum mechanics operators behave correctly, note that given any arbitrary state $|\Psi\rangle$ we have
\begin{equation}
\langle x |\sigma_{2j-1} |\Psi\rangle= \exp(-i\pi x_{j}^{2})\langle x |\Psi \rangle .
\end{equation}
Thus, if the state $|\Psi\rangle$ corresponds to a quantum field theory with partition function $\langle x |\Psi \rangle$, then the integral definition of the partition function given in equation \eqref{zintdef} implies that $\sigma_{2j-1}$ shifts the background Chern-Simons level for the $j$-th flavor by one unit as expected. We can similarly see that the quantum mechanical $\sigma_{2j}$ operator acts as required.  We have
\begin{equation}
\langle x |\sigma_{2j} |\Psi\rangle= \int dy \prod_{k\neq j}\delta(x_{k}-y_{k})e^{i\pi(y_{j}-x_{j})^{2}}\Psi(Uy).
\end{equation}
This is exactly the action expected for the $S$ operation at the level of partition functions.  It performs a change of basis on the flavors, given by the $U$ matrix, and introduces a single new gauge group with specified Chern-Simons levels.

\paragraph{$SL(2,\mathbb{Z})$ Examples}

As a sample application of the above ideas, we present here a simple set of calculations based on $SL(2,\mathbb{Z}),$ relevant for the case of a single flavor symmetry.  Our symplectic transformations acting on quantum field theories are generated by the familiar operators $S$ and $T$  subject to the relations\footnote{As usual we ignore phases in $\mathcal{Z}$ and hence the central element $S^{2}$ can be set to the identity.}
\begin{equation}
S^{2}=(ST)^{3}=1.
\end{equation}
$T$ acts on theories by increasing the Chern-Simons level of the flavor
\begin{equation}
T: k_{G}\rightarrow k_{G}, \hspace{.5in} k_{M}\rightarrow k_{M, \hspace{.5in} }k_{F}\rightarrow k_{F}+1,
\end{equation}
while the $S$ generator acts to gauge the flavor symmetry and introduces a new flavor which is dual to the original symmetry
\begin{equation}
S: k_{G} \rightarrow \left(\begin{array}{cc}k_{G} & k_{M} \\ k_{M}^{T} & k_{F} \end{array}\right), \hspace{.5in}k_{M}^{T}\rightarrow \left(\begin{array}{c c c c c}0 & 0 & \cdots & 0& 1\end{array} \right) , \hspace{.5in}k_{F}\rightarrow 0.
\end{equation}
The relevant quantum mechanics is now single variable for the single $U(1)$ flavor symmetry with standard commutation relations
\begin{equation}
[\hat{x}, \hat{p}]=\frac{i}{2\pi}.
\end{equation}
And the representation of symplectic transformations is given by
\begin{equation}
T\rightarrow \exp\left(-i\pi \hat{x}^{2}\right), \hspace{.5in} STS^{-1}\rightarrow \exp \left(-i\pi \hat{p}^{2}\right).
\end{equation}

A simple class of theories is defined starting from the trivial theory $\Omega$.  This theory has no gauge groups and vanishing flavor Chern-Simons levels.  Its partition function is unity
\begin{equation}
\mathcal{Z}_{\Omega}(x)=\langle x |\Omega\rangle =1.
\end{equation}
More interesting theories can be generated by starting with the trivial theory $\Omega$ and acting with $S$ and $T$.  For a general $SL(2,\mathbb{Z})$ element $\mathcal{O}$ we have the following result for the partition function\footnote{Formula \eqref{cspartclass} is correct when $u$ is non-vanishing.  In the special case where $u=0$ the result is just $\mathcal{Z}_{\mathcal{O}}(x)=\delta(x)$. }
\begin{equation}
\mathcal{O}=\left(\begin{array}{cc} r & t \\ s & u\end{array}\right)\in SL(2,\mathbb{Z})\Longrightarrow \mathcal{Z}_{\mathcal{O}}(x)=\langle x |\mathcal{O}|\Omega\rangle =\frac{1}{\sqrt{|u|}} \exp\left(-i\pi x^{2}t/u\right). \label{cspartclass}
\end{equation}
The answer thus takes the general form $\eqref{Gaussianres}$ with associated invariants
\begin{equation}
|\det(k_{G})|=u,\hspace{.5in} \tau =t/u.
\end{equation}
Notice that, consistent with our general discussion, a particular element $\mathcal{O}$ defines a particular \emph{quantum} Abelian Chern-Simons theory, not a classical Lagrangian presentation of such a theory.  To obtain such a Lagrangian presentation, one must pick a word in the generators $S$ and $T$ which is equal to the given element $M$.  Different words in the generators which are equal to the same fixed $\mathcal{O}$ provide examples of dual theories.
 
\paragraph{Doubled Flavor Variables, Operator Multiplication, and Gauging}
\label{sec:doubledflavor}
The $SL(2,\mathbb{Z})$ examples described above can be readily extended to the case of more flavor symmetry.  For any $F$ we consider the $F$-variable quantum mechanics described in section \ref{sec:QM} and introduce a trivial theory $\Omega$ with unit partition function.  Then, if $\mathcal{O}$ is any element of $Sp(2F,\mathbb{Z})$ we can consider a quantum theory generated by acting with $\mathcal{O}$ on the trivial theory.  The resulting partition function can be expressed as the wavefunction obtained by acting on the  vacuum state $|\Omega\rangle$, a normalized momentum eigenstate with eigenvalue zero 
\begin{equation}
\mathcal{Z}_{\mathcal{O}}(x_{1}, \cdots, x_{F})=\langle x_{1}, \cdots, x_{F}|\mathcal{O}|\Omega \rangle. \label{Zmdef}
\end{equation}
As in the case of a single flavor symmetry discussed above, the resulting quantum field theory and partition function depends only on the element $\mathcal{O}$ in $Sp(2F,\mathbb{Z})$, while a particular Lagrangian realization of the theory requires a choice of word in the generators $\sigma_{n}$ which represents $\mathcal{O}$.

This quantum mechanical setup naturally suggests additional quantities to compute.  Rather than considering the wavefunction of $\mathcal{O}$ acting on the trivial state $|\Omega\rangle$, we may instead double the flavor variables and compute the complete matrix element of $\mathcal{O}$
\begin{equation}
\mathcal{Z}_{\mathcal{O}}^{\mathrm{Op}}(x_{1}, \cdots, x_{F},y_{1},\cdots ,y_{F})\equiv\langle x_{1}, \cdots, x_{F}|\mathcal{O}|y_{1}, \cdots y_{F}\rangle. \label{Zopdef}
\end{equation}
Where in the above the superscript `Op' for operator, is used to distinguish from the partition functions introduced in \eqref{Zmdef}.  For $\mathcal{O}\in Sp(2F,Z)$ a symplectic operator, the matrix element $\mathcal{Z}_{\mathcal{O}}^{\mathrm{Op}}$, is the partition function of an Abelian Chern-Simons theory now coupled to $2F$  background flavor fields.

The construction of \eqref{Zopdef} is not limited to the case of symplectic operators.  Indeed, in section \ref{rflow} we will see that an interesting class of non-symplectic operators $\mathcal{O}$ have matrix elements which are identified with partition functions of interacting three-dimensional conformal field theories coupled to $2F$ flavor fields.  In general, such matrix element partition functions have the following features.
\begin{itemize}
\item If $\mathcal{Z}_{\mathcal{O}}^{Op}(x,y)$ is known then the partition function $\mathcal{Z}_{\mathcal{O}}(x)$ is determined,
\begin{equation}
\mathcal{Z}_{\mathcal{O}}(x)=\int dy \ \mathcal{Z}_{\mathcal{O}}^{Op}(x,y).
\end{equation}
In the physical interpretation we have developed, the integration over the $y$ variables is the gauging of the associated flavor variables at vanishing values of the associated FI parameters.
\item More generally, the quantum-mechanical operation of operator multiplication can be interpreted in field theory.  A product of operators can always be decomposed into a convolution by an insertion of a complete set of states
\begin{equation}
\mathcal{Z}_{\mathcal{O}_{1}\mathcal{O}_{2}}^{Op}(x,y)=\int dz \ \mathcal{Z}_{\mathcal{O}_{1}}^{Op}(x,z)\mathcal{Z}_{\mathcal{O}_{2}}^{Op}(z,y). \label{integrationform}
\end{equation}
Again, the integration is physically interpreted as gauging.  We consider the two theories, whose partition functions are given by the matrix elements of $\mathcal{O}_{i}$, we identify flavors as indicated in \eqref{integrationform} and gauge with no FI-term.
\item $\mathcal{Z}_{\mathcal{O}}^{Op}(x,y)$ is a partition function of a theory coupled to $2F$ background flavor fields.  A general theory of this type is acted on by the symplectic group $Sp(4F,\mathbb{Z})$, however a matrix element is acted on only by the subgroup 
\begin{equation}
Sp(2F,\mathbb{Z}) \times Sp(2F,\mathbb{Z}) \subset Sp(4F,\mathbb{Z})
\end{equation}
which does not mix the $x$ and $y$ variables.  The geometrical and physical interpretation of this splitting will be explained in section \ref{rflow}.
\end{itemize}
\subsection{Tangles}
Our goal in this section is to give a geometric counterpart to the field theory and partition function formalism developed in the previous analysis.  A natural way to develop such an interpretation is to engineer the Abelian Chern-Simons theory by compactification of the M5-brane on a three-manifold $M$.  In six dimensions, the worldvolume of the M5-brane supports a two-form field $B$ with self-dual three-form field strength \cite{WittenFiveBrane}.  When reduced on a three-manifold, the modes of $B$ may engineer an Abelian Chern-Simons theory.  We review aspects of this reduction and explain the three-dimensional geometry required to understand the $Sp(2F,\mathbb{Z})$ action.

\subsubsection{Reduction of the Chiral Two-Form}
\label{chiral2form}
Consider the free Abelian M5-brane theory reduced on a three-manifold $M$.  To formulate the theory of a chiral two-form, $M$ must be endowed with an orientation which we freely use throughout our analysis.  The effective theory in the three macroscopic dimensions is controlled by the integral homology group $H_{1}(M,\mathbb{Z})$.  The simplest way to understand this fact is to note that a massive probe particle in the theory arises from an M2-brane which ends on a one-cycle $\gamma$ in $M$.  In particular the homology class of $\gamma \in H_{1}(M,\mathbb{Z})$ labels the charge of the particle.

In the effective theory in three dimensions, massive charged probes are described by Wilson lines.  Let $C$ denote a one-cycle in the non-compact Minkowski space.  A general Wilson line can be written as 
\begin{equation}
\exp\left(iq_{\alpha} \oint_{C} A_{\alpha}\right).
\end{equation}
If the theory in question has $G$ gauge fields and $F$ flavor fields, then the charge vector $q_{\alpha}$ has $G+F$ components and integral entries.  However, in the presence of non-vanishing Chern-Simons levels, the charge vector $q$ is in general torsion valued.  Thus, distinct values of the integral charge vector $q$ may be physically equivalent.  The allowed distinct values of the charge vector are readily determined by examining the two-point function of Wilson loops in Abelian Chern-Simons theory coupled to background vectors.  The results are summarized as follows.  Let $\mathbb{Z}^{G}\subset \mathbb{Z}^{G+F}$ be the subset of charges uncharged under the flavor group $U(1)^{F}$.  We view the level matrix as specifying a map
\begin{equation}
K: \mathbb{Z}^{G}\longrightarrow \mathbb{Z}^{G+F},
\end{equation}
and those charge vectors in the image of this map are physically equivalent to no charge at all.  

Since we have determined that possible Wilson lines encode the homology of $M$ it follows that
\begin{equation}
H_{1}(M,\mathbb{Z})\cong \mathbb{Z}^{G+F}/\Im(K). \label{homident}
\end{equation}
Equation \eqref{homident} encodes the appropriate generalization of Kaluza-Klein reduction to the case of torsion valued charges.  The fact that we study Chern-Simons theories up to possible framing anomalies (equivalently, overall phases in the partition function) means that the entire theory is characterized by the group \eqref{homident}.  
However, the homology of $M$, and hence the underlying physics has no preferred description via a classical Lagrangian.  Indeed as we will illustrate in the remainder of this section, distinct classical theories, with the same group of Wilson line charges, can in fact arise from compactification on the same underlying manifold $M$.  Thus, already in this elementary discussion of reduction of the two-form we see the important fact that compactification of the M5-brane theory produces a specific \emph{quantum field theory} not, as one might naively expect, a specific Lagrangian presentation of a classical theory which we subsequently quantize. It is for this reason that our geometric constructions of field theories are powerful, dualities are manifest.

Finally, before moving on to discuss explicit examples we remark on the geometry associated to flavor symmetries.  These arise when the manifold $M$ is allowed to become non-compact.  Suppose that $M$ develops cylindrical regions near infinity which take the form of $\mathbb{R}\times\mathbb{R}_{+}\times S^{1}$.  Then on the asymptotic $S^{1}$ cycle we may reduce the two-form field to obtain another gauge field 
\begin{equation}
A=\int_{S^{1}}B.
\end{equation}  
However, unlike the compact cycles in the interior of $M$, the cycle $S^{1}$ has no compact Poincar\'{e} dual and hence $A$ is a  non-dynamical background field;  it provides the effective theory in three dimensions with a $U(1)$ flavor symmetry.   Moreover, since the boundary behavior of $A$ must be specified to obtain a well-defined theory in three dimensions, the resulting theory is of the type we have considered in the introduction: a theory with flavor symmetries and a specified coupling to background gauge fields.  As a result the partition function $\mathcal{Z}(x)$ is a well-defined observable of the theory.  The number of flavor variables on which the result depends is the number of homologically independent cylindrical ends of $M$.  For $F$ flavors we require $F+1$ cylindrical ends.

\subsubsection{Double Covers From Tangles}
\begin{figure}[here!]
  \centering
\includegraphics[width=\textwidth]{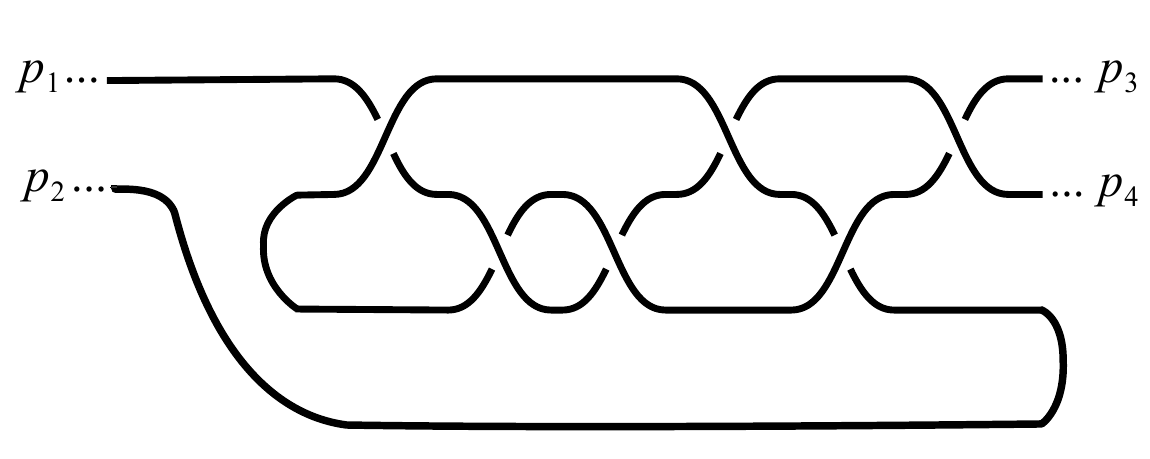}   
  \caption{A tangle.  The four endpoints of $L$ extend forever towards the points at infinity.}
  \label{fig:tangleex}
\end{figure}

The specific class of geometries that we will study are conveniently presented as double covers over the non-compact space $\mathbb{R}^{3}$, branched over a one-dimensional locus $L$
\begin{equation}
\xymatrix{
\mathbb{Z}_{2} \ar[r] & M \ar[d] \\
 & L \subset \mathbb{R}^{3}
}
\end{equation}
Topologically $L$ is simply the union of $F+1$ lines, however its embedding in $\mathbb{R}^{3}$ is constrained.   On the asymptotic two-sphere at the boundary of three-space, we mark $2F+2$ distinct points $p_{1}, \cdots p_{2F+2}$.  The $2F+2$ ends of $L$ at infinity are the points $p_{i}$ .  Meanwhile, in the interior of $\mathbb{R}^{3}$ the components of $L$ may be knotted.  Such an object is known as an $(F+1)$-\emph{tangle}.  An example in the case of $F=1$ is illustrated in Figure \ref{fig:tangleex}.
Given two distinct tangles $L_{1}$ and $L_{2}$, they are considered to be equal topologically when one can be deformed to the other by isotopy in the interior of $\mathbb{R}^{3}$ which keeps the ends at infinity fixed.  

In the following we will also need to be more precise about the behavior near the asymptotes $p_{i}$.  Let $\overline{B}_{r}\subset \mathbb{R}^{3}$ denote the exterior of a closed ball of radius $r$ centered at the origin.  We view $\overline{B}_{r}$ topologically as $S^{2}\times I$ where $I$ is an open interval.  For large $r$ the portion of the tangle $L\cap \overline{B}_{r}$ contained in $\overline{B}_{r}$ consists of $2F+2$ arcs.  We constrain the behavior of these arcs by requiring that the pair $(\overline{B}_{r}, L\cap \overline{B}_{r})$ is homeomorphic to the trivial pair $(S^{2}\times I, \{p_{1}, p_{2}, \cdots, p_{2F+2}\}\times I)$ where the $p_{i}$ are points in $S^{2}$.  This constraint implies that the knotting behavior of the tangle eventually stops as we approach infinity.  In practice it means that any planar projection of the tangle $L$ appears at sufficiently large distances as $2F+2$ disjoint semi-infinite line segment which undergo no crossings.  

For most of the remainder of this section, we will argue that the class of three-manifolds obtained as double covers branched over tangles have exactly the correct properties to engineer the Abelian Chern-Simons theories coupled to a background flavor gauge field which we have discussed in the previous section.  As a first step, observe that such geometries do indeed support $F$ flavor symmetries.  Group the asymptote points into $F+1$ pairs $\{p_{2i-1}, p_{2i}\}$.  The double cover of $\mathbb{R}^{3}$ branched over the two straight arcs emanating from $\{p_{2i-1}, p_{2i}\}$ yields the anticipated cylindrical ends of $M$ required to support flavor symmetry.  The fact that there are $2F+2$ asymptotes ensures that there are exactly $F$ independent flavor symmetries as one linear combination of the asymptotic cycles can be contracted in the interior of the manifold.

In section \ref{surfaces} we explain how to extract a Lagrangian for an Abelian Chern-Simons theory from the geometric data of a tangle.  As we have previously described, the M5-brane on $M$ does not provide a preferred Lagrangian.  Consistent with this fact, we find that a Lagrangian description of the field theory associated to a particular tangle requires additional geometric choices.  In this case the choice is a \emph{Seifert surface}, a surface whose boundary is the given tangle.  For any fixed $L$ there are infinitely many such surfaces each giving rise to a distinct Lagrangian presentation of the same underlying physics.

\begin{figure}[here!]
  \centering
\includegraphics[width=0.5\textwidth]{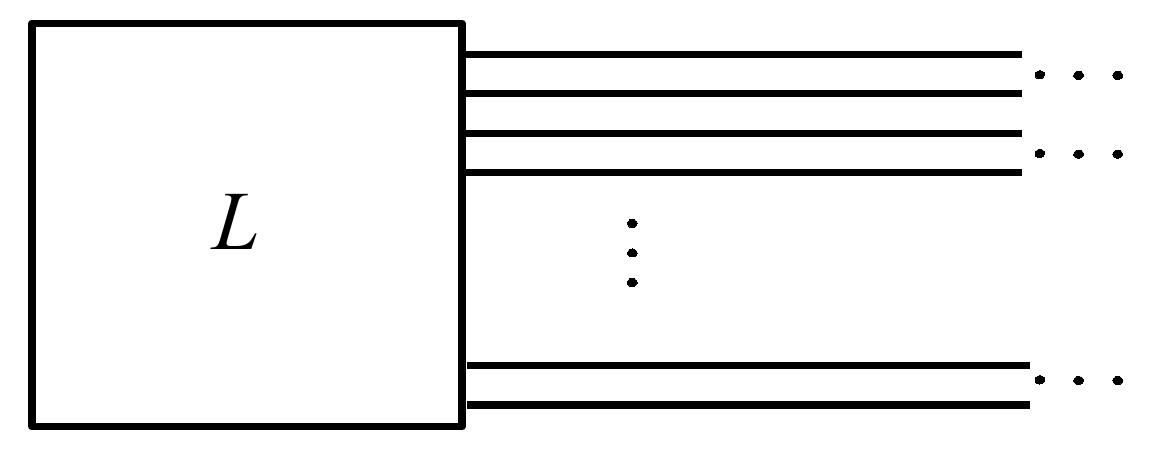}     
    \caption{A generic tangle $L$ in $\mathbb{R}^{3}$.  The ellipsis indicate that the strands continue to infinity with no additional crossings.  In the interior of the box, the strands are in general knotted in an arbitrary way. }
  \label{fig:standardtangle}
\end{figure}

\begin{figure}[here!]
  \centering
  \subfloat[Odd Braid Moves]{\label{fig:oddb}\includegraphics[width=0.49\textwidth]{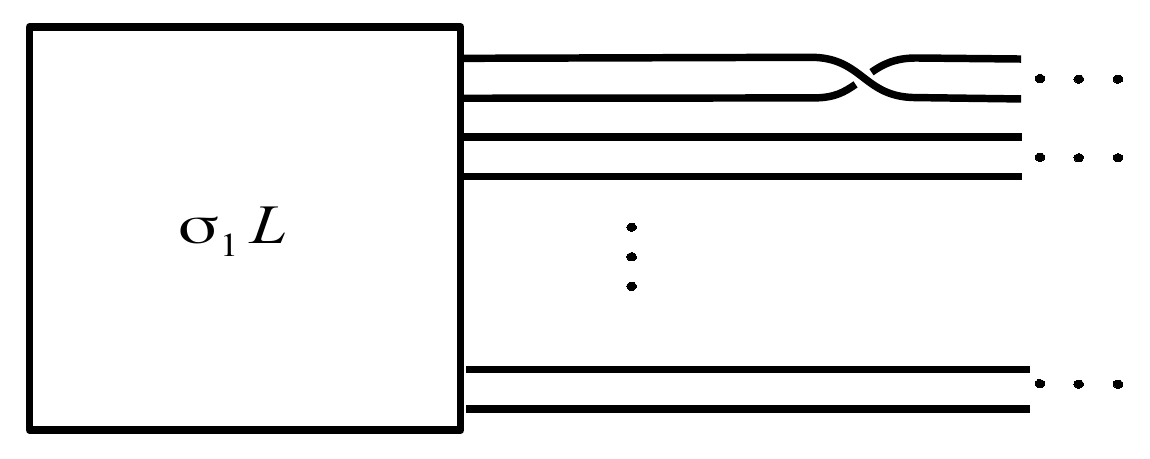}}     
  \hspace{0in}    
  \subfloat[Even Braid Moves]{\label{fig:evenb}\includegraphics[width=0.49\textwidth]{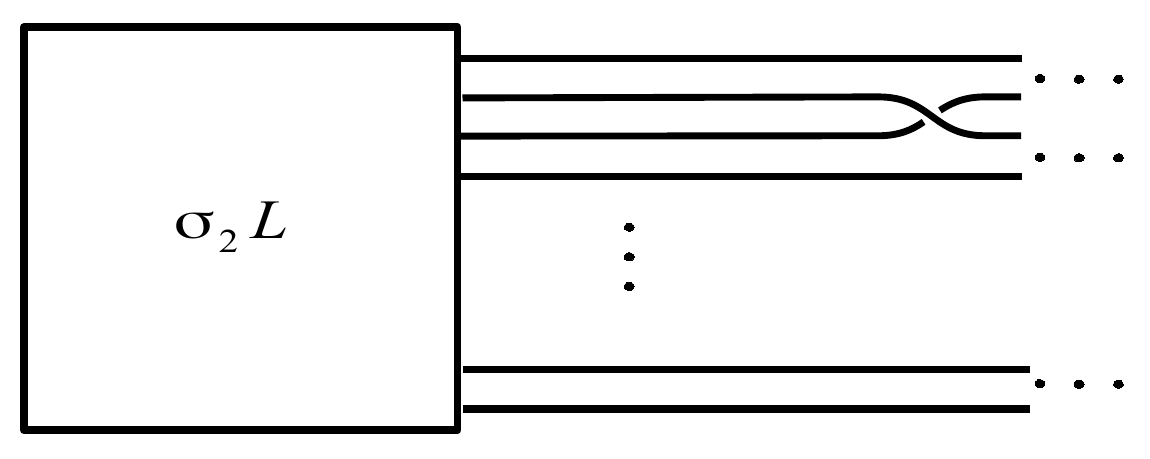}}
  \caption{The symplectic action on the tangles.  In (a) a typical odd generator $\sigma_{2n+1}$.  In (b) an even generator $\sigma_{2n}$. }
  \label{fig:symplecticaction}
\end{figure}

Finally we argue that tangles, and hence the class of three-manifolds described as double covers branched over tangles, enjoy a natural action by $Sp(2F,\mathbb{Z})$.  To illustrate this action, we draw a generic tangle with $F+1$ strands as in Figure \ref{fig:standardtangle}.
Then, the action of the symplectic group is defined by the generators $\sigma_{j}$ where $j=1, \cdots 2F$, which act on the tangles by braid moves in a neighborhood of the asymptotes $p_{i}$.  Several examples are illustrated in Figure \ref{fig:symplecticaction}.  

One way to understand this three-dimensional geometry is to note that the boundary at infinity of $M$ is a double cover of $S^{2}$ branched over $2F+2$ points, and hence is a Riemann surface of genus $F$.  The action defined in Figure \ref{fig:symplecticaction} is a surgery on $M$ which in general changes its topology.  This surgery is induced by mapping class group transformations in a neighborhood of the boundary of $M$.  In particular, as is clear from the illustrations, what we have defined is not, a priori, an action of the symplectic group, but rather an action of the braid group, $B_{2F+1}$ on $2F+1$ strands \cite{Birman}.\footnote{The `last strand' appearing at the bottom of the diagram in Figure \ref{fig:standardtangle} is stationary under all braid moves.  Alternatively one may work with the spherical braid group and impose additional relations.  For simplicity we stick with the more familiar planar braids.}  The braid group and the symplectic group are related by a well-known exact sequence
\begin{equation} 
1\rightarrow \mathcal{T}_{2F+1}\rightarrow B_{2F+1}\rightarrow Sp(2F,\mathbb{Z})\rightarrow 1.
\end{equation}
Where $\mathcal{T}_{2F+1}$ is the Torrelli group.  To make contact with our discussion of field theories, we wish to illustrate that the action of the braid group defined by Figure \ref{fig:symplecticaction} reduces to an action of the symplectic group on the associated field theories.  This implies that any two elements of $B_{2F+1}$ that differ by multiplication by a Torelli element must give rise equivalent actions on the field theories extracted from an arbitrary tangle.  More bluntly, the Torelli group generates dualities.  One of the outcomes of this section is a proof of this fact.

\subsection{Seifert Surfaces}
\label{surfaces} 
\begin{figure}[here!]
  \centering
  \subfloat[Pretzel Knot]{\label{fig:pknot}\includegraphics[width=0.25\textwidth]{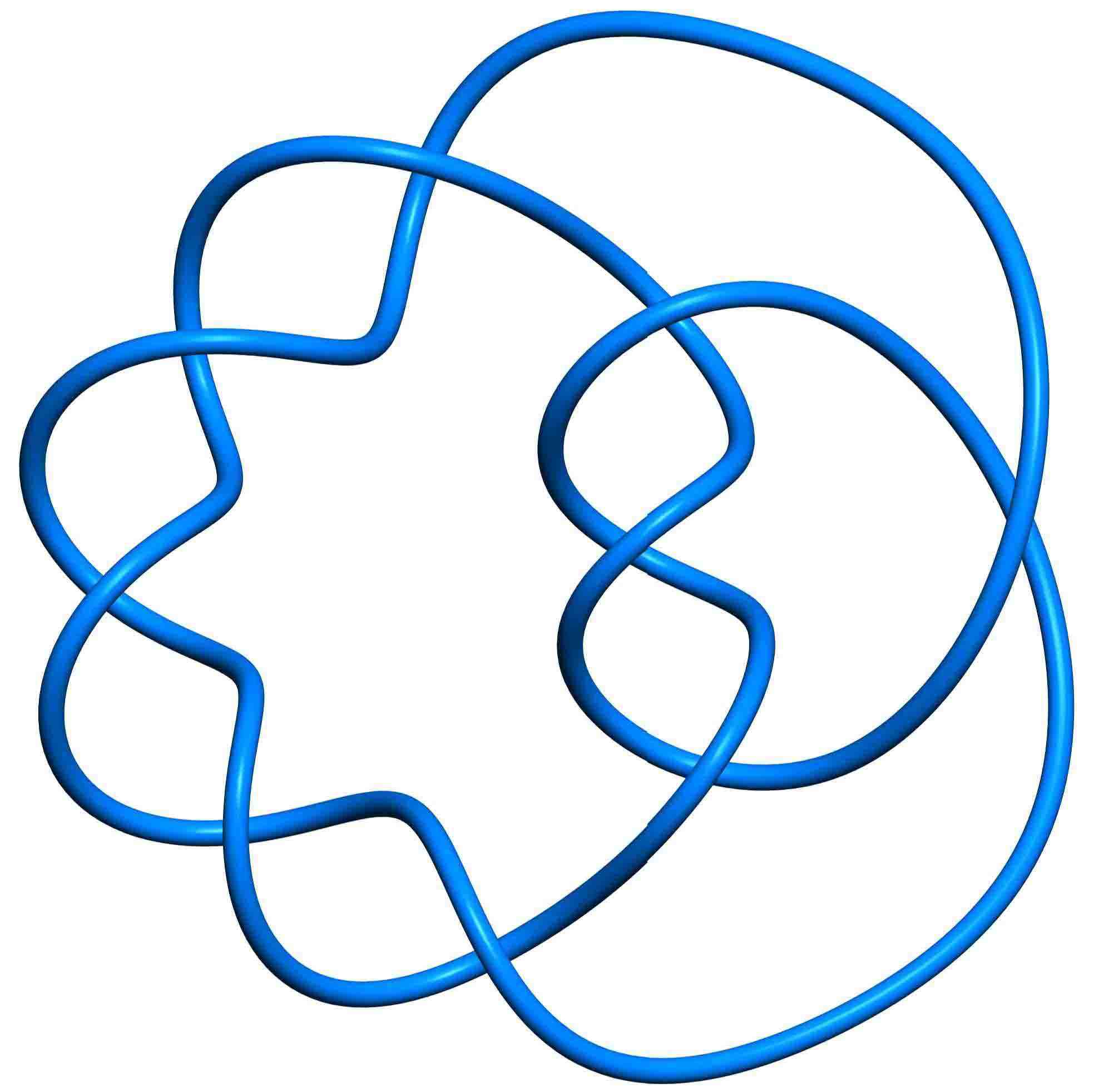}}     
  \hspace{.5in}    
  \subfloat[Seifert Surface]{\label{fig:pnotsurf}\includegraphics[width=0.25\textwidth]{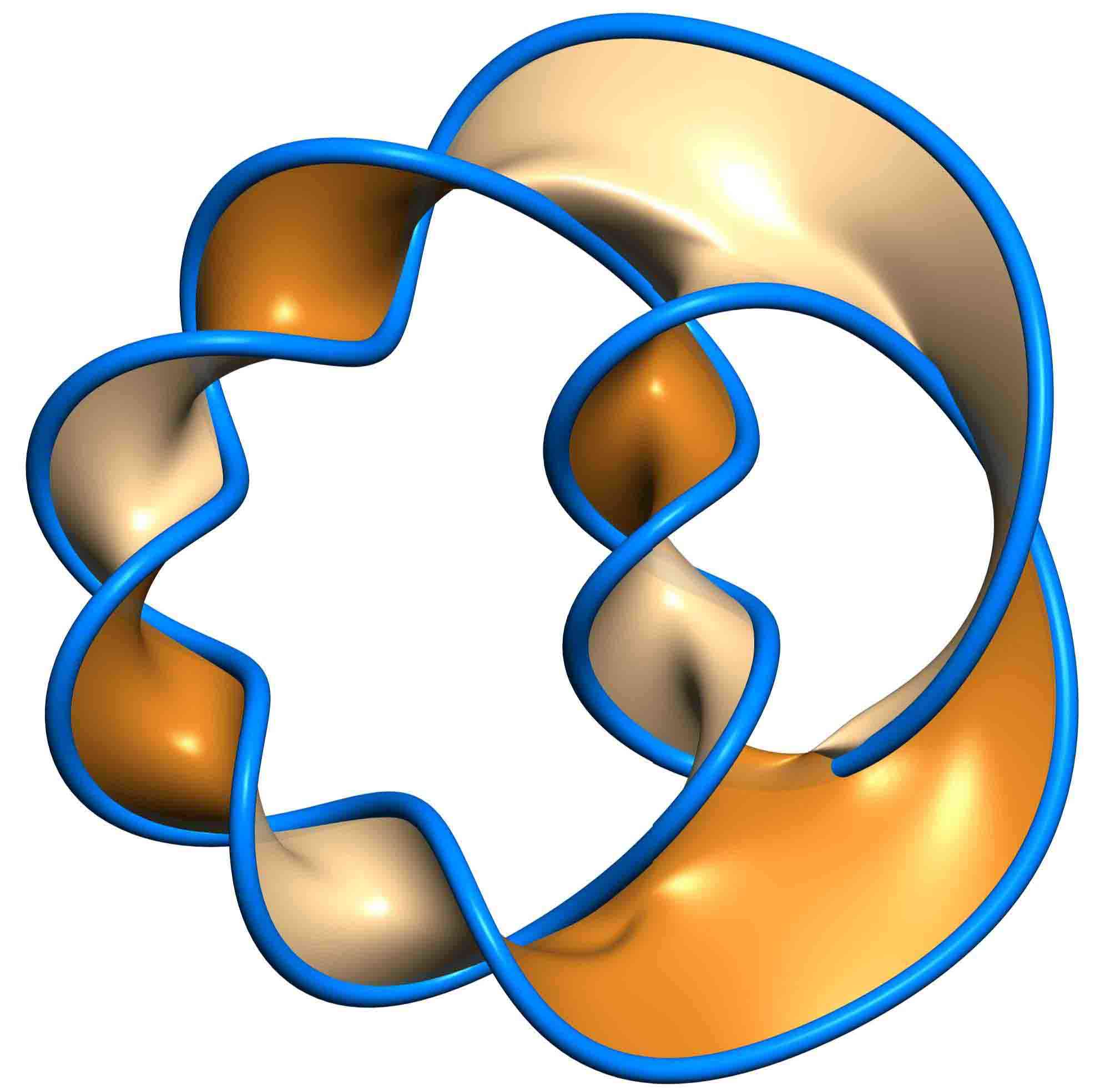}}
  \caption{A sample Seifert surface.  In (a) a pretzel knot in the three-sphere.  In (b) as Seifert surface for the knot.}
  \label{fig:seifertex}
\end{figure}

To understand the physics encoded by a tangle we need control over the homology of the cover manifold $M$.  The appropriate tool for this task is a Seifert surface.  In general given any knot\footnote{In this paper the term \emph{knot} will be used broadly to include both knots and multicomponent links. }, a Seifert surface $\Sigma$ for the knot is a connected Riemann surface with boundary the given knot.  An example is illustrated in Figure \ref{fig:seifertex}.  In the mathematics literature it is common to impose the additional requirement that $\Sigma$ be oriented.  In our context there is no natural orientation for $\Sigma$ and hence we proceed generally allowing possibly non-orientable Seifert surfaces.

For any knot, there exist infinitely many distinct Seifert surfaces and given a knot diagram a number of simple algorithms exists to construct a $\Sigma$ \cite{Lickorish}.  We describe one useful algorithm in section \ref{checkerboards}.  The reason that Seifert surfaces are relevant for our discussion is that if one wishes to construct a double cover branched over a knot then a choice of $\Sigma$ is equivalent to a choice of branch sheet.  As such, features of the homology of the branched cover $M$ can be extracted from a knowledge of a Seifert surface.  However, the resulting three-manifold $M$ depends only on the branch locus $L$ and hence the homology and ultimately the associated physical theory are independent of the choice of $\Sigma$.  In the following we explain how any fixed choice of Seifert surface allows us to extract a set of gauge and flavor groups and a matrix of Chern-Simons levels from the geometry.

To begin for simplicity, we assume that we are dealing with a knot in $S^{3}$, as opposed the non-compact tangles in $\mathbb{R}^{3}$ needed to support flavor symmetry.  The generalizations to the present non-compact situation will then be straightforward.  The detailed statements that we require are as follows.  Any cycle in $H_{1}(M,\mathbb{Z})$ can be thought of as a cycle on the base $S^{3}$ which encircles $\Sigma$.  This can be viewed as a direct parallel with the theory of branched covers of the two-sphere.\footnote{In making this comparison it is crucial that the branch locus is connected.}  Thus, we deduce that there is a surjective map
\begin{equation}
H_{1}(S^{3}-\Sigma, \mathbb{Z})\rightarrow H_{1}(M,\mathbb{Z})\rightarrow0. \label{seqbegin}
\end{equation} 
Meanwhile, there is a linking number pairing between cycles in $H_{1}(S^{3}-\Sigma, \mathbb{Z})$ and cycles in $H_{1}(\Sigma,\mathbb{Z})$.  This linking number pairing is perfect and hence we may extend \eqref{seqbegin} to
\begin{equation}
H_{1}(\Sigma, \mathbb{Z})\cong H_{1}(S^{3}-\Sigma, \mathbb{Z})\rightarrow H_{1}(M,\mathbb{Z})\rightarrow0. \label{seqbegin1}
\end{equation} 
Our task is thus reduced to determining which cycles on the Seifert surface correspond to trivial cycles in the homology of $M$.  

To this end, we define a symmetric bilinear form, the so-called \emph{Trotter} form
\begin{equation}
 K: H_{1}(\Sigma, \mathbb{Z}) \times H_{1}(\Sigma, \mathbb{Z}) \rightarrow \mathbb{Z}.
 \end{equation}
 Our choice of notation is intentional: we will see that the Trotter form defines the Chern-Simons levels.  To extract $K$ we let $\alpha \in H_{1}(\Sigma,\mathbb{Z})$, and set $\widetilde{\alpha}$ to be the cycle in $S^{3}$ obtained from locally pushing $\alpha$ off of $\Sigma$ in both directions.  The cycle $\widetilde{\alpha}$ is a two-to-one cover of $\alpha$.  If $\Sigma$ is orientable then $\widetilde{\alpha}$ consists of two disconnected cycles each on a given side of $\Sigma$ (as determined by the orientation), however in general $\widetilde{\alpha}$ is connected.  The definition of the Trotter form is
 \begin{equation}
 K(\alpha, \beta)= lk_{\#}(\widetilde{\alpha},\beta), \label{trotterdef}
 \end{equation}
 Where $lk_{\#}$ denotes the linking number pairing of cycles in $S^{3}$.  A simple calculation illustrates that $K$ is symmetric.  A slightly less trivial argument shows that the image of $K$ is exactly the set of cycles on $\Sigma$ which are trivial in $M$.  Thus, the completion of the sequence \eqref{seqbegin1} is
 \begin{equation}
 0\rightarrow \mathrm{Im}(K)\rightarrow H_{1}(\Sigma, \mathbb{Z})\cong H_{1}(S^{3}-\Sigma, \mathbb{Z})\rightarrow H_{1}(M,\mathbb{Z})\rightarrow0. \label{seqend}
 \end{equation}
In particular we conclude that $H_{1}(M,\mathbb{Z})\cong H_{1}(\Sigma,\mathbb{Z})/\mathrm{Im}(K)$.

Double covers of $S^{3}$ branched over knots are exactly the geometries we expect to engineer Abelian Chern-Simons theories without flavor symmetries and we may relate theorem \eqref{seqend} to physics as follows:
\begin{itemize}
\item A choice of Seifert surface $\Sigma$ and a set of generators of homology $\alpha_{1}, \cdots, \alpha_{G}$ determines a set of $G$ Abelian gauge fields.
\item The Trotter form pairing on cycles in $H_{1}(\Sigma,\mathbb{Z})$ is equal to the Chern-Simons levels matrix on the associated gauge fields.
\end{itemize}
Distinct choices of Seifert surfaces are physically related by duality transformations.  This fact is easy to verify directly.  For example, distinct choices of $\Sigma$ which differ by gluing in handles or Mobius bands add new gauge cycles and compensating levels to keep the underlying physics unmodified.

\begin{figure}[here!]
  \centering
\includegraphics[width=.5\textwidth]{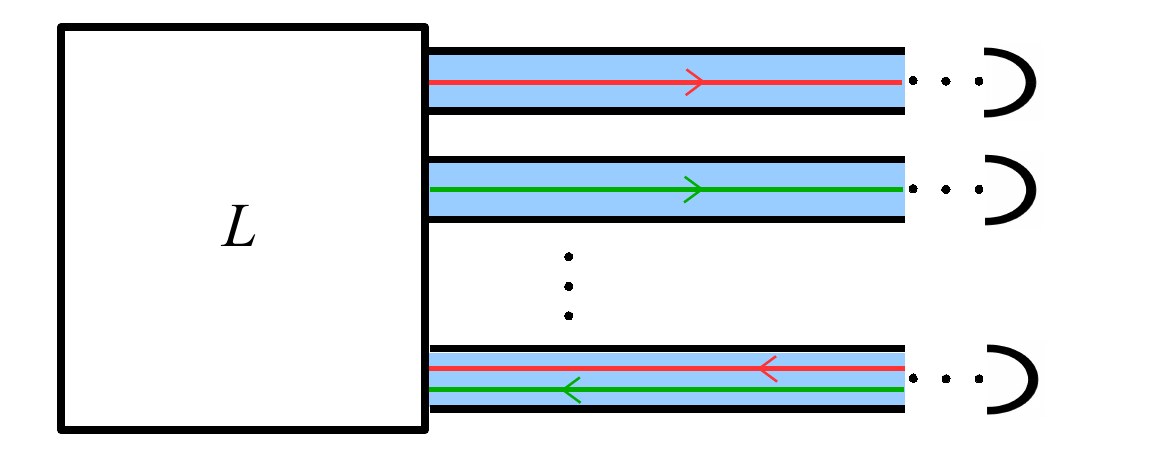}     
    \caption{The asymptotic geometry of a Seifert surface for a generic tangle.  The shaded blue region indicates the interior of $\Sigma$.   The arcs at infinity indicate the compactification of $L$ and $\Sigma$.  The non-compact cycles on $\Sigma$ give rise to flavor symmetries. }
  \label{fig:standardsurf}
\end{figure}

Finally, we generalize our discussion of Seifert surfaces and homology to the case of non-compact geometries required to discuss flavor symmetries.  Let $L$ denote a tangle in $\mathbb{R}^{3}$.  We introduce non-compact Seifert surfaces $\Sigma$ again defined by the condition that they are connected surfaces with boundary $L$.  However, now to compute flavor data we must fix a compactification of both $L$ and $\Sigma$.  We achieve this by identifying the points $p_{i}$ in pairs and glueing in arcs near infinity as illustrated in Figure \ref{fig:standardsurf}.  

Let $\delta$ indicate the union of the arcs at infinity, and $\Sigma_{c}$ the compactified Seifert surface including $\delta$.  The surface $\Sigma_{c}$ should be viewed as embedded inside $S^{3},$ the one-point compactification of $\mathbb{R}^{3}$ and calculations of linking numbers etc. take place inside $S^{3}$.  For simplicity in future diagrams we often leave the compactification data of the Seifert surface implicit by setting the convention that whenever a non-compact Seifert surface consists of strips extending to infinity in $\mathbb{R}^{3}$ the intended compactification is the one where the strips are capped off with arcs as in Figure \ref{fig:standardsurf}.

With these preliminaries about compactifications fixed, we may now state the required generalization of the sequence \eqref{seqend}
 \begin{equation}
 0\rightarrow \mathrm{Im}(K)\rightarrow H_{1}(\Sigma_{c},\delta, \mathbb{Z})\cong H_{1}(\mathbb{R}^{3}-\Sigma, \mathbb{Z})\rightarrow H_{1}(M,\mathbb{Z})\rightarrow0. \label{seqend1}
 \end{equation} 
 Note that in addition to the boundaryless cycles in $\Sigma_{c}$ which give rise to gauge groups, $H_{1}(\Sigma_{c},\delta, \mathbb{Z})$ also contains $F$ cycles with boundary in $\delta$.  In the uncompactified Seifert surface these cycles are non-compact and illustrated in Figure \ref{fig:standardsurf}.  They correspond physically to the $U(1)^{F}$ flavor symmetry. To complete the construction it thus remains to extend the definition of the Trotter form.  For boundaryless cycles in $\Sigma_{c}$ the definition is as before.  Meanwhile to evaluate the Trotter form on cycles with boundary we again push them out locally in both directions from $\Sigma_{c}$ and compute the local linking number from the interior of $\Sigma$.  Alternatively, one may simply think of the pair of points in the boundary of a flavor cycle in $\Sigma_{c}$ as formally identified.  In this way we obtain a closed cycle in $S^{3}$ and we compute its Trotter pairings as before.  In this way we obtain a bilinear form $K$ defined on $H_{1}(\Sigma_{c},\delta,\mathbb{Z}),$ and the image of this form restricted to the boundaryless cycles in $H_{1}(\Sigma_{c},\delta,\mathbb{Z})$ defines the term $\mathrm{Im}(K)$ appearing in \eqref{seqend1}.

To summarize, given any tangle $L$ in $\mathbb{R}^{3}$, we extract a Lagrangian description of the effective Abelian Chern-Simons theory as follows:
\begin{itemize}
\item A choice of Seifert surface $\Sigma$ and a set of generators of the relative homology $H_{1}(\Sigma_{c},\delta,\mathbb{Z})$,  $\alpha_{1}, \cdots, \alpha_{G+F},$ determines a set of Abelian vector fields.  Generators corresponding to boundaryless one-cycles correspond to gauged $U(1)$'s while those corresponding to one-cycles with boundary in $\delta$ are background flavor fields.
\item The Trotter form pairing on cycles in $H_{1}(\Sigma_{c}, \delta, \mathbb{Z})$ is equal to the Chern-Simons levels pairing on the associated vector fields.  We denote by $\mathrm{Im}(K)$ the image of this pairing restricted to the subset of boundaryless cycles in $\Sigma_{c}$, and we have
\begin{equation}
\mathrm{\{Wilson \ Line \ Charges\}} = H_{1}(\Sigma_{c}, \delta, \mathbb{Z})/\mathrm{Im(K)}=H_{1}(M,\mathbb{Z}).
\end{equation}
\end{itemize}

\subsubsection{Checkerboards}
\label{checkerboards}
The previous discussion of Seifert surfaces is complete but abstract.  For computations with explicit examples it is useful to have a fast algorithm for computing the relevant linking numbers and hence extracting a set of Chern-Simons levels from geometry.  One such method, described in this section, is provided by so-called checkerboard Seifert surfaces. 

\begin{figure}[here!]
  \centering
  \subfloat[]{\label{fig:bdp}\includegraphics[width=0.25\textwidth]{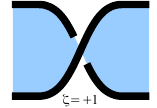}}     
  \hspace{1in}    
  \subfloat[]{\label{fig:bdm}\includegraphics[width=0.25\textwidth]{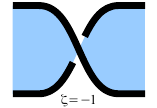}}
  \caption{Checkerboard colorings and their associated signs.  In (a), a positive crossing.  In (b), a negative crossing. }
  \label{fig:boards}
\end{figure}

To begin, fix a planar projection of the tangle $L\subset \mathbb{R}^{3}$.  In such a planar diagram the information about the knotting behavior of $L$ is contained in the crossings in the diagram.  Each crossing locally divides the plane into four quadrants.  We construct a Seifert surface for $L$ by coloring the two of the four quadrants at each crossing in checkerboard fashion and extending consistently to all crossings.  The colored region then defines $\Sigma$.  Note that each crossing $c$ in the diagram is endowed with a sign $\zeta(c)=\pm1$ depending on whether the cross-product of the over-strand with the under-strand through $\Sigma$ at $c$ is in or out of the plane as shown in Figure \ref{fig:boards}.

To compute the Trotter form, we first assume that $\Sigma_{c}$ appears compactly in the plane.\footnote{This assumption cannot in general be relaxed.  Indeed when $\Sigma_{c}$ is non-comact in the plane one must take into account the fact that in the compactification procedure, the plane becomes and embedded $S^{2}$ inside $S^{3}$ and hence may endow $\Sigma_{c}$ with additional topology.}  Then, there is a natural basis of boundaryless cycles in $\Sigma_{c}$ associated to the compact uncolored regions of the plane.  We orient these cycles counterclockwise.  Similarly, in the diagram of $\Sigma$, non-compact white regions may be associated to flavor cycles.  These cycles are again canonically oriented ``counterclockwise,''  i.e. the cross-product of the tangent vector to the cycle with the outward normal pointing into the associated non-compact uncolored region must be out of the plane.\footnote{There is one linear relation among the flavor cycles obtained in this way.  So a given $\Sigma$ will have $F+1$ non-compact uncolored regions and $F$ independent flavor cycles.}  The Trotter pairing on these cycles is determined by the summing over crossings involving a given pair of cycles weighted by the sign of the crossing.  Explicitly, for $\alpha$ and $\beta$ a pair of generators as defined above we have
\begin{equation}
K(\alpha,\beta)=\begin{cases}
+\displaystyle \sum_{\alpha, \beta \in c} \zeta(c) & \mathrm{if}  \ \alpha \neq \beta, \\
-\displaystyle \sum_{\alpha\in c} \zeta(c) & \mathrm{if} \  \alpha =\beta. \label{goeritz}
\end{cases}
\end{equation}
Equation \eqref{goeritz} provides a convenient way to read off Chern-Simons levels for a given tangle and will be utilized heavily (although often implicitly) throughout the remainder of this work.

\subsection{The Torelli Group of Dualities }
\label{torelli}
We are now equipped to investigate the symplectic action on tangles.  In particular, we wish to prove that the action of the braid group $B_{2F+1}$ on tangles, reduces to an action of the symplectic group $Sp(2F,\mathbb{Z})$ when considered as an action on the corresponding physical theories.  

To prove this statement, we proceed in the most direct way possible.  We compute the action of the braid group generators $\sigma_{n}$, illustrated in Figure \ref{fig:symplecticaction}, on the Chern-Simons levels extracted from any Seifert surface associated to the tangle.  We show that this action matches exactly the previously defined action \eqref{oddbraid}-\eqref{evenbraid}.  Since the later action is symplectic this implies that the former is as well.  In particular, this suffices to prove that the Torelli group acts trivially on the underlying quantum field theory.

\begin{figure}[here!]
  \centering
  \subfloat[The action of $\sigma_{1}$ on $L$]{\label{fig:nobs}\includegraphics[width=0.49\textwidth]{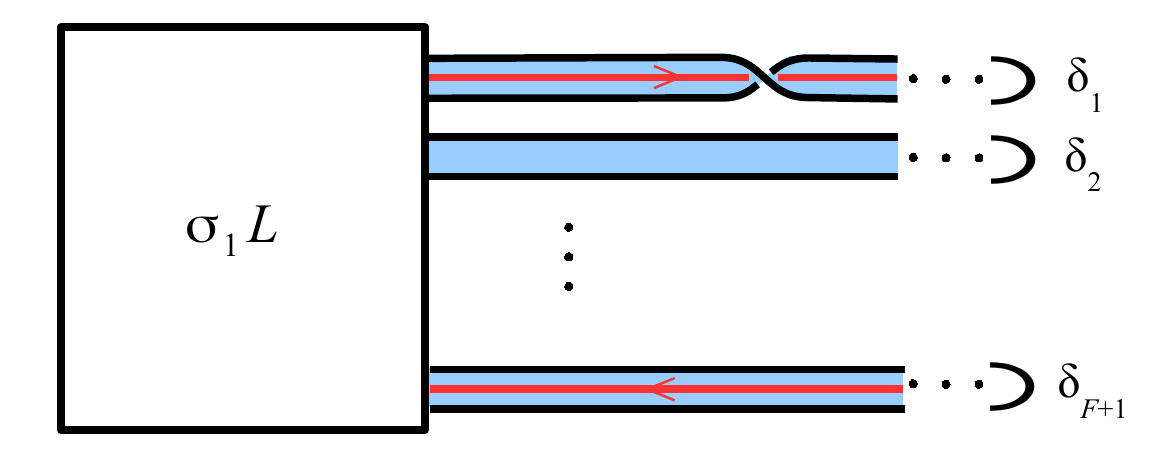}}     
  \hspace{0in}    
  \subfloat[The action of $\sigma_{2}$ on $L$]{\label{fig:nobs1}\includegraphics[width=0.49\textwidth]{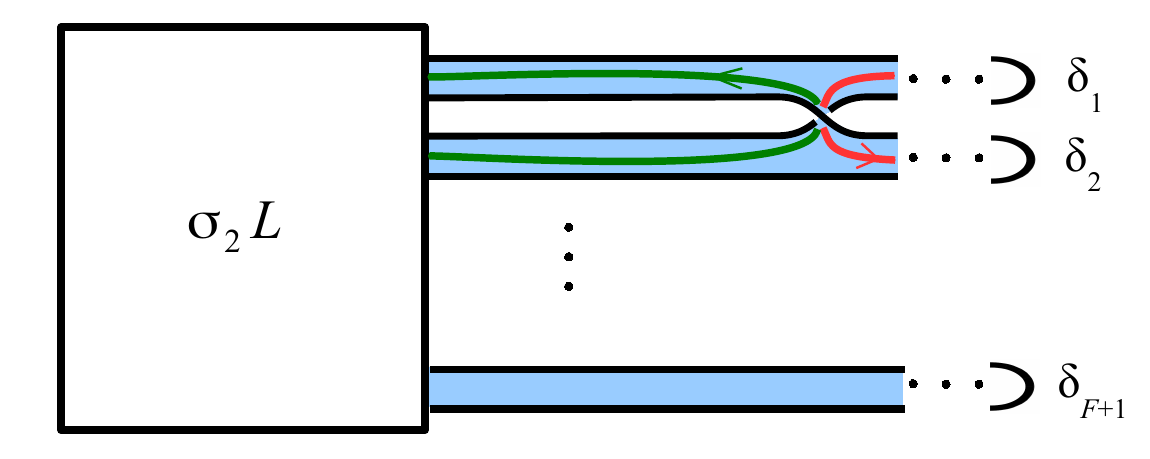}}
  \caption{The action of braid moves on linking numbers.  In (a), all linking number are unmodified except for those of the flavor cycle $\alpha_{1}$ which runs from $\delta_{F+1}$ to $\delta_{1}$,  is illustrated in red, whose self-linking number is increased by one.  In (b), we first change basis of flavor cycles to $\beta_{j}$ which runs from $\delta_{j}$ to $\delta_{j+1}$.  Then we gauge $\beta_{1}$, shown in green, and introduce a new flavor cycle, shown in red, linked with the gauged cycle.}
  \label{fig:torelli1}
\end{figure}

To begin, we fix a Seifert surface with definite compactification data $\delta$.  As we have previously described, $\delta$ is a union of $F+1$ arcs $\delta_{i}$ with $i=1,\cdots, F+1$.  We draw diagrams such that the arcs are ordered down the page, with $\delta_{1}$ appearing at the top, $\delta_{2}$ next and so on.  A basis of flavor cycles in $H_{1}(\Sigma_{c}, \delta,\mathbb{Z})$ is given by $F$ cycles $\alpha_{i}$ each of which begins at $\delta_{F+1}$ and terminates at $\delta_{i}$.  This geometry is shown in Figure \ref{fig:standardsurf}.  With these conventions, the braid moves act as in Figure \ref{fig:torelli1}.

Consider first the odd braid moves $\sigma_{2j-1}$ illustrated in Figure \ref{fig:nobs}.   According to formula \ref{goeritz}, the effect of such a move is to modify the Trotter form by increasing $K(\alpha_{j},\alpha_{j})$ by one while leaving all other entries invariant.  This is exactly the expected action given by \eqref{oddbraid} on Chern-Simons levels for this transformation.

Similarly we may consider the braid moves with even index $\sigma_{2j}$ illustrated in Figure \ref{fig:nobs1}.  To understand this transformation we first change basis on flavor cycles to $\beta_{i}$ which run from $\delta_{i}$ to $\delta_{i+1}$.  The transformation from the basis $\alpha_{i}$ to the basis $\beta_{i}$ is facilitated by the $U$ matrix of equations \eqref{evenbraid}-\eqref{udef}.  Then, the braid move $\sigma_{2j}$ gauges $\beta_{j}$ and introduces a new flavor cycle $\widetilde{\beta}_{j}$.  Finally, we update the Trotter form to account for the new linking numbers apparent in Figure \ref{fig:nobs1}
\begin{equation}
\delta K(\beta_{j},\beta_{j})= -1, \hspace{.5in}\delta K(\widetilde{\beta_{j}},\widetilde{\beta_{j}})= -1, \hspace{.5in} \delta K(\beta_{j},\widetilde{\beta_{j}})= 1.
\end{equation}
This is exactly the gauging operation of equation \eqref{gaugingbraider}.  Thus we have competed the verification of the symplectic action.

As a result of this analysis we conclude that the Torrelli group $\mathcal{T}_{2F+1}$ acts via dualities on Abelian Chern-Simons theories.  Given any tangle one may act on it with a Torrelli element to obtain a new geometry.  Fixing Seifert surfaces, the two geometries in general will have distinct classical Lagrangian descriptions yet their underlying quantum physics is identical.

Moreover, as we see in section \ref{particles} and beyond, the technology of this section generalizes immediately to the more complicated geometries required for constructing interacting field theories.  In particular, the symplectic action we have described arises from braid moves near infinity and hence is enjoyed by any geometry with the same asymptotics.

\subsection{Geometric Origin of Quantum Mechanics}
\label{qmgeo}
To conclude our discussion of Abelian Chern-Simons theories we briefly comment on the origin of the quantum mechanical framework for partition function calculations discussed in section \ref{sec:QM}.  We fix an Abelian Chern-Simons theory $\mathcal{T}(M)$ engineered by reduction of the M5-brane on a three-manifold $M$.  The three-sphere partition function of this theory then has an underlying six-dimensional origin as the M5-brane partition function on the product manifold $M\times S^{3}$,
\begin{equation}
\mathcal{Z}_{S^{3}}^{\mathcal{T}(M)}=\mathcal{Z}^{\mathrm{M5}}_{M\times S^{3}}.
\end{equation}

Thus far, we have viewed $M$ as small and interpreted the long-distance physics as an Abelian Chern-Simons theory coupled to flavors which we subsequently compactify on $S^{3}$.  However, an alternative point of view is to consider $S^{3}$ to be small, and obtain another effective three-dimensional description which is subsequently compactified on $M$.  As $S^{3}$ has vanishing first homology, the resulting three-dimensional description is one with no Wilson line observables and hence from the point of view of this paper which studies partition functions on compact manifolds up to multiplication by overall factors we cannot distinguish the result from the trivial theory.

However, a standing conjecture is that in fact the reduction on $S^{3}$ gives rise to a $U(1)$ Chern-Simons theory at level one.  Assuming the veracity of this statement, we then arrive at a beautiful physical interpretation of the quantum mechanical calculations in section \ref{sec:QM}.  

Recall that $M$ is not a compact manifold, but rather has non-compact cylindrical ends required to support flavor symmetry.  One may equivalently view $M$ as a manifold with boundary at infinity and with specified boundary conditions supplied by the background flavor gauge fields.  On general grounds, the path-integral of $U(1)$ level one Chern-Simons theory on $M$ produces a state in the boundary Hilbert space determined by the quantization of Chern-Simons theory on $\partial M$.  In this case, as a consequence of the conjecture, one is quantizing a space of $U(1)$ flat connections on a Riemann surface with $2F$ independent cycles.  The Hilbert space thus consists of wavefunctions of $F$ real variables $x_{1}, \cdots x_{F}$, which are interpreted as the holonomies of a flat connection around a maximal collection of $F$ non-intersecting homology classes in $\partial M$.  The symplectic action is then the standard action in this Hilbert space induced by the action on the homology of the genus $F$ Riemann surface $\partial M$.

Thus, the quantum mechanical framework which emerged abstractly from supersymmetric localization formulas in section \ref{sec:QM}, takes on a natural physical interpretation when the associated field theories are geometrically engineered.  In particular, the viewpoint of the partition function $\mathcal{Z}_{S^{3}}^{\mathcal{T}(M)}(x)$ as a wavefunction in a Hilbert space is a simple consequence of the six-dimensional origin of the computation and leads to a correspondence of partition functions
\begin{equation}
\mathcal{Z}_{S^{3}}^{\mathcal{T}(M)}(x)=\mathcal{Z}_{M}^{U(1)_{1}}(x).
\end{equation}
This identification is reminiscent to the one studied in \cite{AGT} and was obtained in the case of three-manifolds from different perspectives by \cite{CV09,BDP}.

\section{Particles, Singularities, and Superpotentials}
\label{particles}
In this section we exit the realm of free Abelian Chern-Simons theories and enter the world of interacting quantum systems.  We study conformal field theories described as the terminal point of renormalization group flows from Abelian Chern-Simons matter theories.  Thus, in addition to the vector multiplets describing gauge fields, our field theories will now have charged chiral multiplets.  We will find that, in close analogy with the study of $\mathcal{N}=2$ theories in four-dimensions, such theories can be geometrically encoded by studying the M5-brane on a \emph{singular} manifold.  In the context of three-manifolds branched over tangles the natural class of singularities are those where strands of the tangle collide and lose their individual identity. We refer to such objects as \emph{singular} tangles.  Our main aim in this section is to give a precise description of these objects and explain how they encode non-trivial conformal field theories.  In the process we will also describe how the geometry encodes superpotentials.  A summary of results in the form of a concise set of rules for converting singular tangles to physics appears in section \ref{rules}.
\subsection{Singularities and Special Lagrangians}
\label{joyce}
\begin{figure}[h!]
\center\includegraphics[width=.4\textwidth]{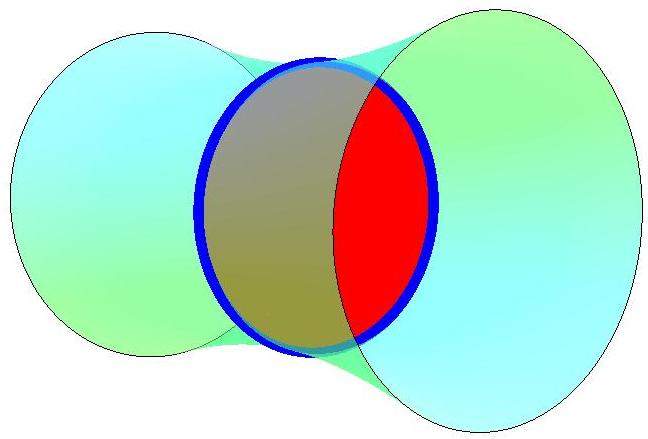}  
\caption{A particle represented by an M2-brane disc ending $M$. The M2-brane is the red disc located in the ambient space $Q,$ while the dark blue circle represents the cycle of the three-manifold on which it ends.}
\label{figure:M2disc}
\end{figure}

We begin with a discussion of the geometric meaning of chiral multiplets and their associated wavefunctions in the three-sphere partition function.   In our M-theory setting the three-manifold $M$ is embedded in an ambient Calabi-Yau $Q$,  and massive particles arise from M2-branes which end along $M$ on a one-cycle.   In the simplest case of a spinless BPS chiral multiplet, supersymmetry implies that $M$ is a special-Lagrangian and the M2-brane is a holomorphic disc as illustrated in Figure \ref{figure:M2disc}\cite{WittenPhase2,AVTOP}.
The mass of the BPS particle is proportional to the area of the disc, and hence in the massless limit the cycle on which the M2-brane ends collapses. 

Thus, when a particle becomes massless the three-manifold $M$ develops a singularity.  A local model for this geometry is a \textit{special Lagrangian cone} on $T^2$ in $\IC^3$. Such a cone is defined to be the subset $L_0$ in $\IC^3$ obeying \cite{Joyce}
\begin{equation} \label{eq:sing}
	L_0 = \left\{ (z_1,z_2,z_3) \in \IC^3 : |z_1|^2 = |z_2|^2 = |z_3|^2, \quad \textrm{Im}(z_1 z_2 z_2) = 0, \quad \textrm{Re}(z_1 z_2 z_3) \geq 0\right\}.
\end{equation}
When the mass of the M2-brane is restored, the singularity is resolved. This can be done in three distinct ways\cite{Joyce}. Let $m > 0$, then the resolutions are
\begin{eqnarray}
	L_m^1 & = & \left\{ (z_1,z_2,z_3) \in \IC^3 : |z_1|^2 - m = |z_2|^2 = |z_3|^2, \quad \textrm{Im}(z_1 z_2 z_2) = 0, \quad \textrm{Re}(z_1 z_2 z_3) \geq 0\right\}, \nonumber \\
	L_m^2 & = & \left\{ (z_1,z_2,z_3) \in \IC^3 : |z_1|^2 = |z_2|^2 - m= |z_3|^2, \quad \textrm{Im}(z_1 z_2 z_2) = 0, \quad \textrm{Re}(z_1 z_2 z_3) \geq 0\right\}, \nonumber \\
	L_m^3 & = & \left\{ (z_1,z_2,z_3) \in \IC^3 : |z_1|^2 = |z_2|^2 = |z_3|^2 - m, \quad \textrm{Im}(z_1 z_2 z_2) = 0, \quad \textrm{Re}(z_1 z_2 z_3) \geq 0\right\}.  \nonumber \\
\end{eqnarray}
The resulting spaces are special Lagrangain three-manifolds in $\IC^3$ \cite{AVTOP} diffeomorphic to $S^1 \times \IR^2$. They differ by the orientation of a closed holomorphic disc in $\IC^3$ with area $\pi m$ which represents the M2-brane. In the case of $L_m^1$ this disc is given by
\begin{equation} \label{eq:holdisc}
	D_m^1 = \left\{(z_1,0,0) : z_1 \in \IC, \quad |z_1|^2 \leq m\right\}.
\end{equation}
The other cases, $D^2_m$ and $D^3_m$, are analogous. We see that the boundary of the disc is an oriented $S^1$ in $L^1_m$ whose homology class generates $H_1(L_m^1, \IZ) \cong \IZ$. In the other cases the boundary is given by an oriented circle around the origin of $z_2$ and $z_3$ respectively. One can thus see that the difference between the three ways the disc appears is determined by the orientation of its central axis in $\IC^3$. 

To make contact with our discussion of tangles we view this local model for the singularity as a double cover over $\mathbb{R}^{3}$.  The special Lagrangians $L_{m}^{a}$ are acted on by the involution
\begin{equation}
	z_i \mapsto \bar{z}_i. \label{inv}
\end{equation}
The quotient space is parametrized by the triple $(x_{1},x_{2},x_{3})\in \mathbb{R}^{3}$ where $x_{i}=\mathrm{Re}(z_{i}).$  Locally the $x_{i}$ provide coordinates on $L_{m}^{a},$ but the global structure of the special Lagrangian is a double cover.  The branch locus is the fixed points of \eqref{inv} and is composed of two strands explicitly given by
\begin{eqnarray} \label{eq:blines}
	L_m^1: x_1 = \sqrt{t^2+m}, ~x_2 = t, ~ x_3 = t, & \textrm{and} & x_1 = -\sqrt{t^2+m}, ~x_2=t, ~x_3 = -t, \nonumber \\
	L_m^2: x_1 = t, ~ x_2 = \sqrt{t^2+m}, ~ x_3=t, & \textrm{and} & x_1 = t, ~ x_2 = - \sqrt{t^2 + m}, ~ x_3 = -t,  \\
	L_m^3: x_1 = t, ~ x_2 = t,~ x_3 = \sqrt{t^2+m},& \textrm{and} &  x_1 = t, ~x_2 = -t, ~x_3 = -\sqrt{t^2+m}. \nonumber
\end{eqnarray}
Where $t\in \mathbb{R}$ provides a coordinate along the strands. 

One way to see that the branched cover is an equivalent description of the original topology is to slice $\mathbb{R}^{3}$ into planes labelled by a time direction.  The coordinate $t$ on the branch lines in \eqref{eq:blines} provides such a foliation and increasing time defines a notion of \emph{flow}.  Each slice is a Riemann surface which is a double-cover of the plane branched over two points and is thus a cylinder. Therefore, including time, we see that topologically the cover is $\IR^2 \times S^1$.  We pursue this perspective on local flows in $M$ and connect them to four-dimensional physics in section \ref{rflow}.

\begin{figure}[here!]
  \centering
  \subfloat[~]{\label{fig:resa}\includegraphics[ width=0.3\textwidth]{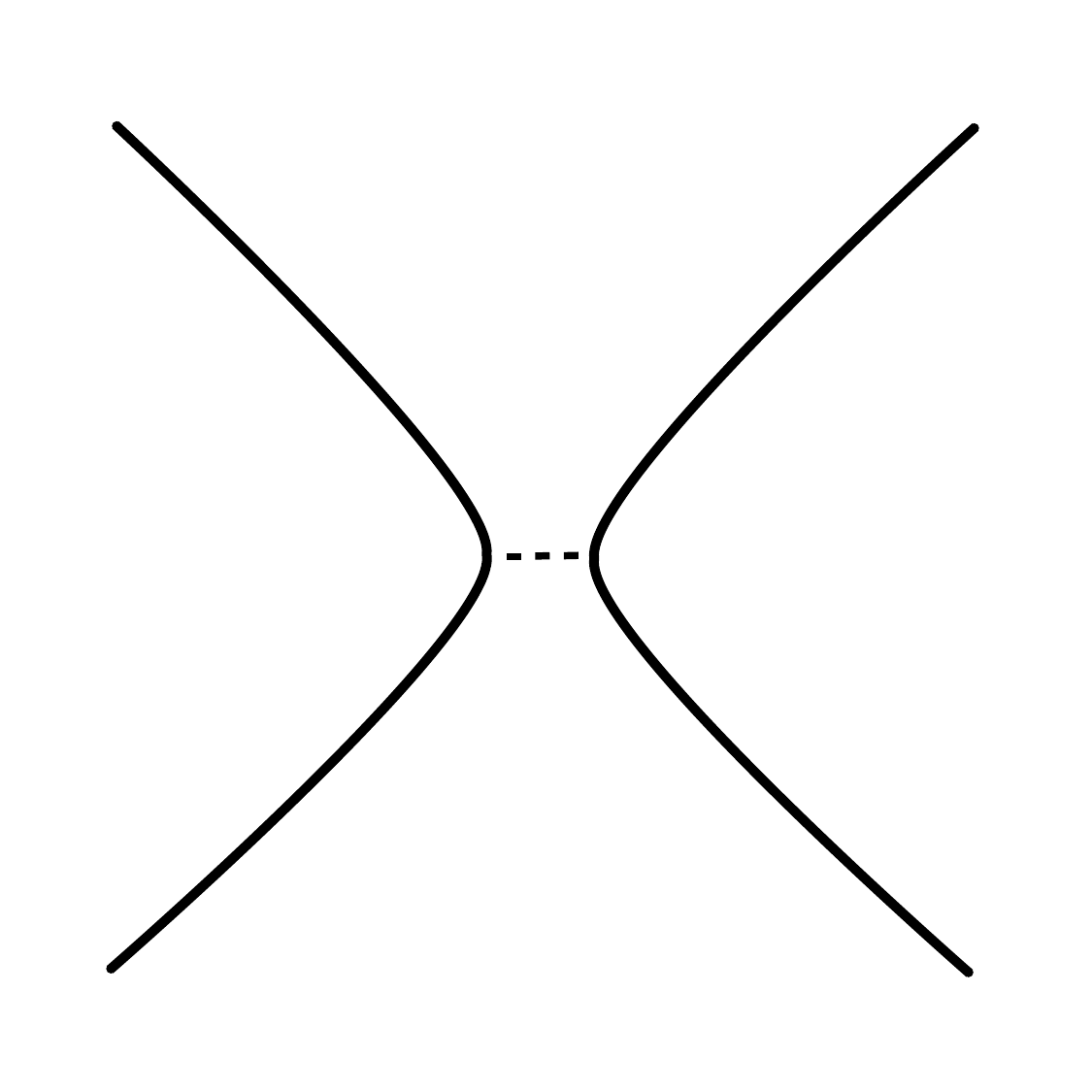}}     
  \hspace{.25in}       
  \subfloat[~]{\label{fig:resb}\includegraphics[ width=0.3\textwidth]{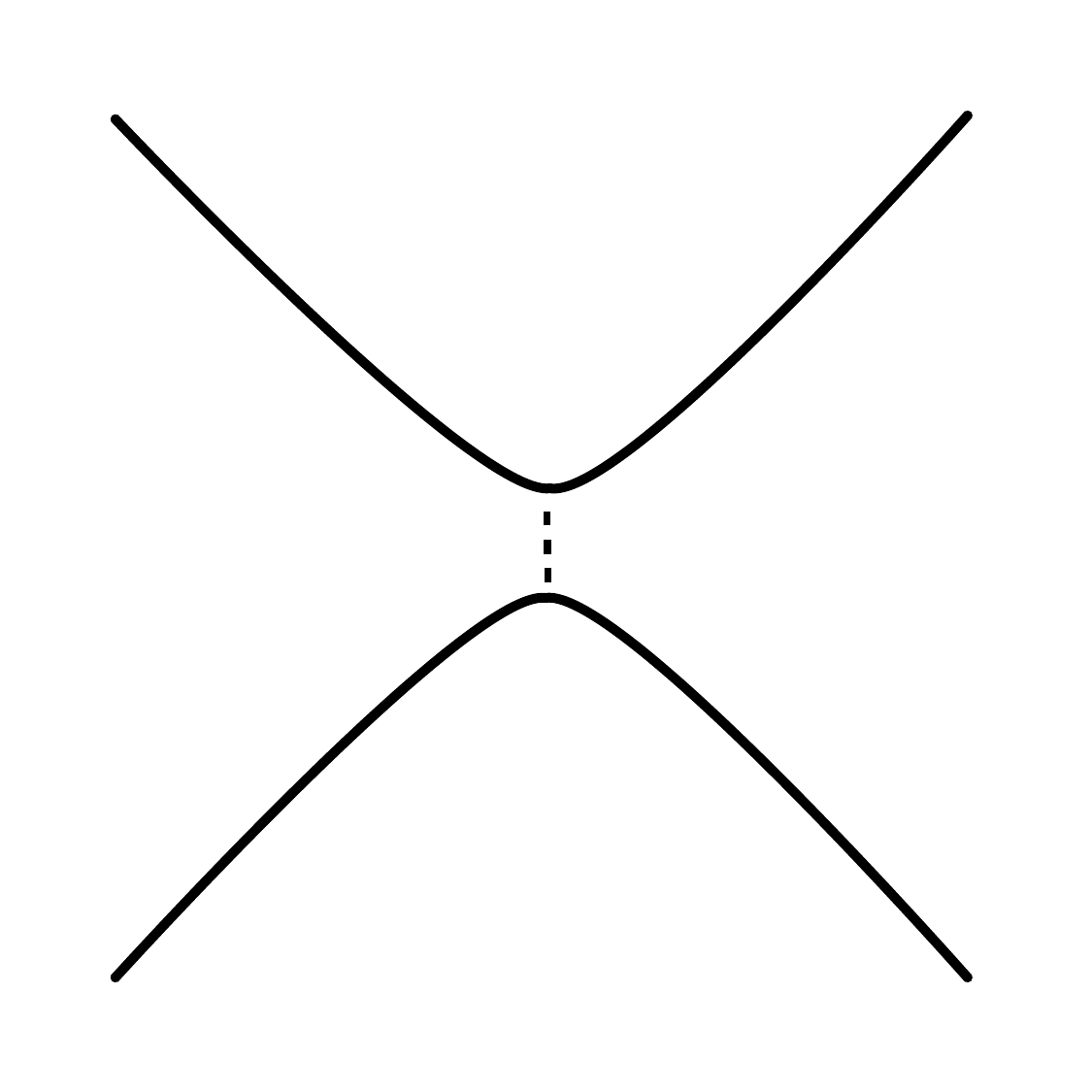}}
  \hspace{.25in}    
  \subfloat[~]{\label{fig:resc}\includegraphics[ width=0.3\textwidth]{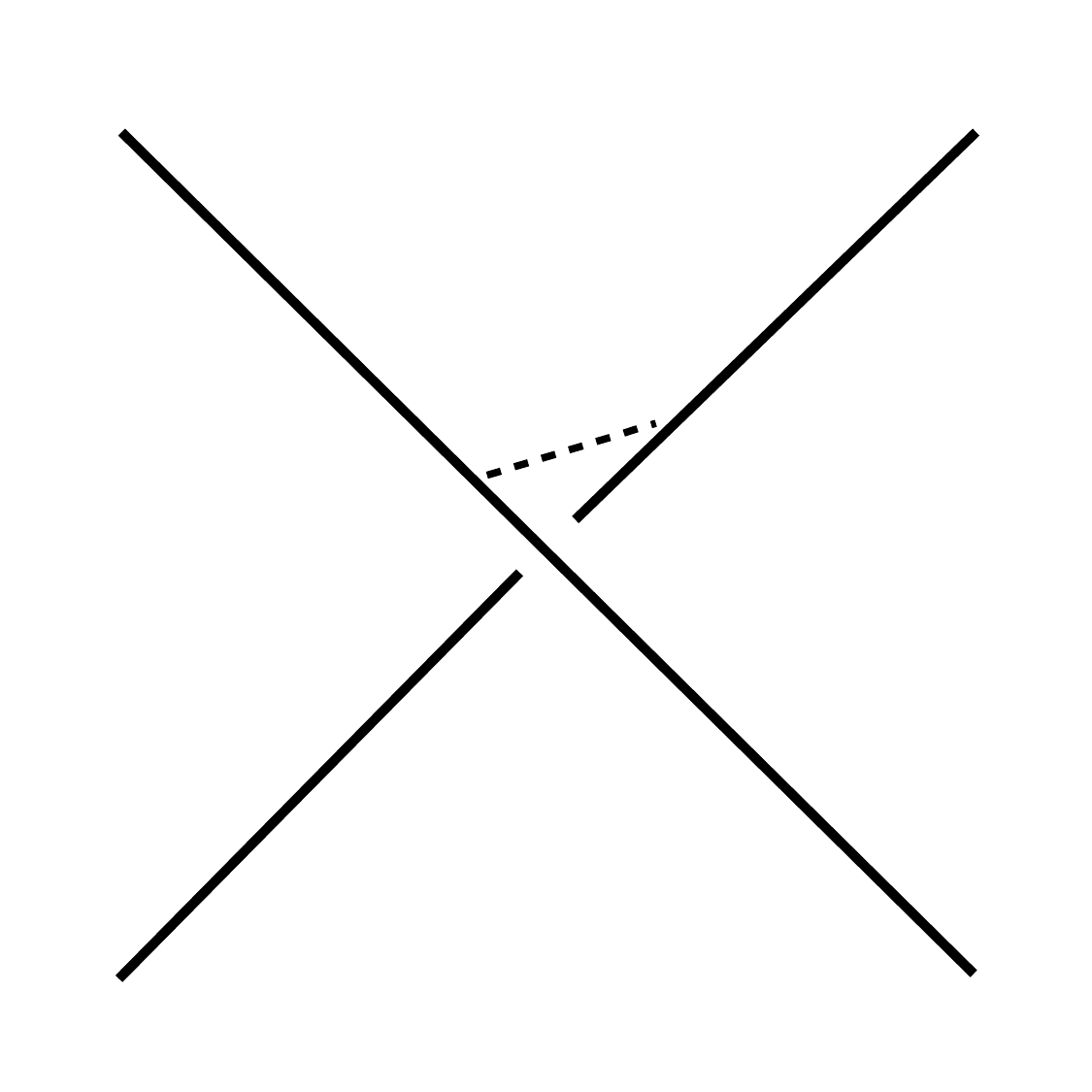}}
  \caption{The three different resolutions together with the M2-brane disc represented by a dashed line. The pictures are drawn in the projection to the $(x_{1},x_{2})$-plane.  In (a) we see the branch locus giving rise to the special Lagrangian $L_m^1$. In (b), the branch locus underlying $L_m^2$.  In (c) the branch locus underlying $L_m^3$.}
  \label{figure:desingularizations}
\end{figure}

Returning to our analysis of the special Lagrangian cone, we note that when viewed as a double cover it is easy to see how the three different resolutions $L_{m}^{a}$ are realized in terms of the configurations of the branch lines \eqref{eq:blines}.  We fix a planar projection of the geometry by declaring $\hat{x}_{3}$ to be the oriented perpendicular direction.   Then, we can depict the geometry as in Figure \ref{figure:desingularizations}.
Note that Figure \ref{fig:resc} only shows the \textit{overcross}. The other choice, where the strand from upper left to lower right goes \textit{under} the second strand, called the \textit{undercross}, does not occur.  This is an artifact of the planar projection which we use to visualize the configuration.  Indeed, exchanging the oriented normal $\hat{x}_{3}$ to $-\hat{x}_{3}$ exchanges the overcross for the undercross.  By contrast, changing the normal direction from $\hat{x}_{3}$ to  $\hat{x}_{1}$ or $\hat{x}_{2}$ permutes the resolutions appearing in \ref{figure:desingularizations} but leaves the triple, as a set, invariant.

In the limit $m \rightarrow 0$ the branch lines collide and we recover the singularity (\ref{eq:sing}). In $\mathbb{R}^{3},$ this appears as four branch half-lines all emanating from the origin. These half-lines approach infinity in four distinct octants and hence specify the vertices of a tetrahedron.  In this way, we see the tetrahedral geometry of \cite{DGG1} emerge from the structure of special Lagrangian singularities.  

\begin{figure}[here!]
  \centering
  \subfloat[~]{\label{fig:overcrosssing}\includegraphics[ width=0.45\textwidth]{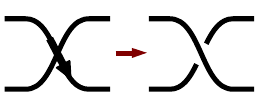}}     
  \hspace{.25in}       
  \subfloat[~]{\label{fig:undercrosssing}\includegraphics[ width=0.45\textwidth]{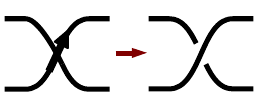}}
  \caption{Two different singularities. In (a) we see how an overcross singularity resolves after applying Figure \ref{fig:resc}. In (b) the corresponding resolution is shown for the undercross singularity. In both cases the two other resolutions of Figure \ref{figure:desingularizations} are also present but not depicted.}
  \label{figure:singularities}
\end{figure}

Having thoroughly analyzed the local model, we may now introduce a precise definition of the concept of a \textit{singular} tangle. It is simply a tangle where we permit pairs of strands to touch at a finite number of points.  The local structure of the cover manifold $M$ at each such point is that of the singular special Lagrangian cone discussed above, and the global identification of strands in the tangle indicates how these local models are glued together.  In specifying the gluing we must keep track of additional pieces of discrete data.
\begin{itemize}
\item We draw singular tangles in planar projections of $\mathbb{R}^{3}$.  Hence each singularity is equipped with an oriented normal vector $\pm \hat{x}_{3}$. Varying the sign of the normal vector changes whether the overcross or undercross appears upon resolution.
\item Fix a sheet labeling 1, and 2, at each singularity.  Then in the gluing we must specify whether the identified sheets are the same or distinct.  Varying between these two choices alters the relative signs of the charges of the particles as determined by the orientation of the M2-branes.
\end{itemize}
Both of the data described above have only a relative meaning:  for a single singularirty they are convention dependent while for multiple singularities they may be compared.  All told then, if we draw singular tangles in a plane, each singularity is one of four possible types.  We encode the four possibilities graphically with a thickened arrow on one of the strands passing through the singularity as in Figure \ref{figure:singularities}.
The thickened strand always resolves out of the page while the direction of the arrow encodes the charge of the massless M2-brane residing at the origin of the singularity. 
 
\subsection{Wavefunctions and Lagrangians}
\label{wavefunctions}
Our next task is to explain in general how to extract a Lagrangian description of the physics defined by a singular tangle.  As in the case of the free Abelian Chern-Simons theories studied in section \ref{nopart}, there is no unique Lagrangian but rather for each choice of Seifert surface we obtain a distinct dual presentation.   In the case of singular tangles, we will see that these changes in Seifert surfaces are related by non-trivial mirror symmetries.

To begin, let us recall the data associated to a chiral multiplet in an Abelian Chern-Simons matter theory.
\begin{itemize}
\item A charge vector $q_{\alpha}\in \mathbb{Z}^{G+F}$ indicating its transformation properties under $U(1)^{G}\times U(1)^{F}$ gauge and flavor rotations.  In all of our examples the vector $q_{\alpha}$ will be primitive meaning that the greatest common divisor of the integers $q_{\alpha}$ is one.  
\item A parity anomaly contribution.  If a chiral multiplet is given a mass $m$, it may be integrated out leaving a residual contribution to the Chern-Simons levels of fields.  The shift in the levels in given by 
\begin{equation} \label{eq:CSshift}
\delta k_{\alpha \beta}=\frac{1}{2}\mathrm{sign}(m)q_{\alpha}q_{\beta}.
\end{equation}
For primitive charge vectors the above shift has at least one non-integral entry.  This implies that the ultraviolet levels are subject to a shifted half-integral quantization law.  We take the associated shift to be part of the definition of the chiral multiplet.
\item An $R$ charge indicating the scaling dimension of the associated chiral operator in the conformal field theory.  This data is fixed by a maximization principle once a superpotential is specified, and hence is not an additional data in the geometry \cite{Jafferis}.  This will be addressed in section \ref{sec:potential}.
\end{itemize}

To encode the partition function of such chiral multiplets we must introduce a new class of wavefunctions depending on these data.  Each is given by a non-compact quantum dilogarithm of the form
\begin{eqnarray}
	E_{+}(z - c_b (1-R)) & \equiv & e^{- i \frac{\pi}{2}z^2} s_b(- z + c_b (1-R)), \label{particlewave}\\
	E_{-}(z - c_b (1-R)) & \equiv & e^{i \frac{\pi}{2}z^2} s_b( z + c_b (1-R)), \nonumber
\end{eqnarray}
where $c_{b}$ is the imaginary constant given in \eqref{cbdef}, and the function $s_b(x)$, defined as
\begin{equation}
	s_b(x) = e^{-i \frac{\pi}{2} x^2} \frac{\displaystyle \prod_{n=0}^{\infty}(1+ e^{(2n+1)\pi i b^2 + 2\pi b x})}{\displaystyle \prod_{n=0}^{\infty}(1+e^{-(2n+1)\pi i b^{-2} + 2\pi b^{-1} x})},
\end{equation}
was obtained through a localization computation on the squashed three-sphere in \cite{HHL} where the numerator and denominator come from vortex partition functions on the two half-spheres \cite{Pasquetti}.  The physical interpretation of this function is read from the variables as follows.
\begin{itemize}
 \item The subscript of $E_{\pm}$ encodes the fractional ultraviolet Chern-Simons level $\pm \frac{1}{2}$ assigned to the particle.
  \item The variable $z$ indicates the linear combination of gauge and flavor fields under which the chiral multiplet is charged.  For $E_{\pm}$ the charge is $z = \pm q \cdot (y~x)$. 
 \item The variable $R$ denotes the $R$-charge.
 \end{itemize}

\begin{figure}[here!]
  \centering
  \subfloat[$E_{+}(x_{\alpha})$]{\label{fig:Epp}\includegraphics[ width=0.25\textwidth]{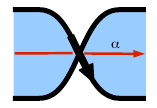}}     
  \hspace{1in}       
  \subfloat[$E_{+}(-x_{\alpha})$]{\label{fig:Epm}\includegraphics[ width=0.25\textwidth]{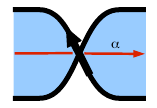}}
  \hspace{1in}
  \subfloat[$E_{-}(x_{\alpha})$]{\label{fig:Emp}\includegraphics[ width=0.25\textwidth]{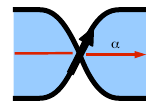}}     
   \hspace{1in}     
     \subfloat[$E_{-}(-x_{\alpha})$]{\label{fig:Emm}\includegraphics[ width=0.25\textwidth]{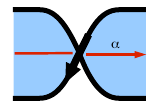}}
  \caption{The dilogarithm assignments for singularities.  The particles are charged under the  $U(1)$ super-field associated to the cycle $\alpha$ indicated in red, and $x_{\alpha}$ is the associated scalar.  The overcross vs. undercross resolution encodes the distinction between $E_{\pm}$ and specifies the fractional part of Chern-Simons levels. The orientation of the arrow on the thickened strand relative to $\alpha$ determines the sign of the particle's charge.}
  \label{figure:Ep}
\end{figure}

Thus, we see that the physical data of a chiral multiplet is completely encoded by the wavefunctions \eqref{particlewave}.  It follows that to assign a definite matter content to a singular tangle, as well as extract the associated contributions to the partition function $\mathcal{Z}$, it suffices to assign a quantum dilogarithm to each singularity.  To proceed, we introduce a singular Seifert surface $\Sigma$ for a singular tangle $L$.  As explained in section \ref{surfaces}, from the homology of $\Sigma$ we extract a basis of gauge and flavor cycles under which particles may be charged.  Let $\alpha$ be such a cycle.  Utilizing the sequence \eqref{seqend1}, we may view $\alpha$ equivalently as a cycle in the cover $M$.  An M2-brane disc $D$ ending on $M$ has a charge determined by its linking numbers
\begin{equation}
q_{\alpha}=lk_{\#}(\alpha, \partial D).
\end{equation}
The extension of this formula to the case of singular $M$ is then depicted in our graphical notation in Figure \ref{figure:Ep}.

These dilogarithm assignments completely determine the matter content of a singular tangle.  However, the assignments require a choice of Seifert surface.  This surface is a choice of branch sheet for the double cover and varying it does not alter the underlying geometry.  As a consequence, our rules are subject to the crucial test: \emph{the underlying quantum physics must be independent of the choice of Seifert surface}.  

\begin{figure}[here!]
  \centering
  \subfloat[$E_{+}(x_{\alpha})$]{\label{fig:Esing}\vcenteredhbox{\includegraphics[ width=0.25\textwidth]{pics/Epp.pdf}}}
    \hspace{.5in}     
  =   
    \hspace{.5in} 
  \subfloat[$E_{-}(x_{\beta})$]{\label{fig:EsingBW}\vcenteredhbox{\includegraphics[ width=0.25\textwidth]{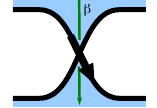}}}
  \caption{A duality results from changing the Seifert surface. In (a) a singularity contributing $E_{+}(x_{\alpha})$ to the partition function. In (b) the Seifert surface is changed and the same singularity contributes $E_{-}(x_{\beta})$.}
  \label{figure:BlackWhite}
\end{figure}

Given the dualities between free Abelian Chern-Simons theories already described in section \ref{nopart}, independence of the choice of Seifert surface is ensured provided we have the equality shown in Figure \ref{figure:BlackWhite}.
There, we see that one and the same singularity may make different contributions to an ultraviolet Lagrangian depending on the choice of Seifert surface. At the level of partition functions, this means that a singularity which contributes as $E_+(x_{\alpha})$ with one choice of branch sheet can contribute with $E_-(x_{\beta})$ with a different choice.  Thus, we see that consistency of our analysis requires a mirror symmetry implying that the same underlying conformal field theory may arise from ultraviolet theories with distinct matter content.

\begin{figure}[here!]
  \centering
  \subfloat[~]{\label{fig:freeChiral}\includegraphics[ width=0.4\textwidth]{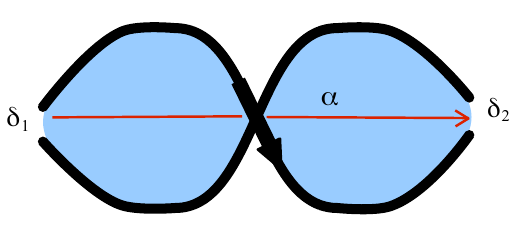}}     
  \hspace{.2in}       
  \subfloat[~]{\label{fig:freeChiralBWflip}\includegraphics[ width=0.4\textwidth]{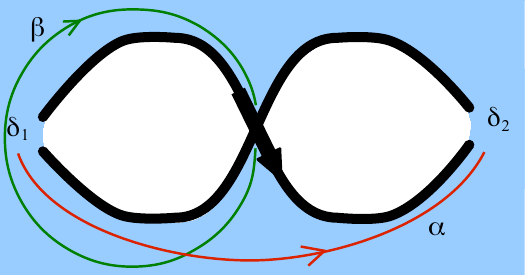}}
  \caption{The duality between a free Chiral multiplet and a $U(1)$ gauge field with a charged chiral field. In  (a) we see the free chiral field couplet to the flavor cycle $\alpha$. In (b) we see the gauge cycle $\beta$ and flavor cycle $\alpha$ of the dual theory.}
  \label{figure:freeChiralBWduality}
\end{figure}

To understand the nature of the duality implied by Figure \ref{figure:BlackWhite} we analyze its impact on the local model of the singular tangle involving a single singularity.   Equality in more complicated examples follows from the locality of our constructions.  The singular tangle together with its dual choices of Seifert surface and fixed compactification data $\delta_{i}$ are shown in Figure \ref{figure:freeChiralBWduality}.
The ultraviolet field content in each case is given by the following.
\begin{itemize}
\item Figure \ref{fig:freeChiral}: There is a background $U(1)$ flavor symmetry associated to the cycle $\alpha$ and no propagating gauge fields.  Associated to the singularity there is a chiral multiplet with charge $1$ under the flavor symmetry.  This particle contributes $+\frac{1}{2}$ to the Chern-Simons level.  The scalar $x_{\alpha}$ in the background $U(1)$ multiplet is the real mass of the chiral field.
\item Figure \ref{fig:freeChiralBWflip}: There is a $U(1)$ flavor symmetry associated to the cycle $\alpha$ and a $U(1)$ gauge symmetry associated to the cycle $\beta$.  Associated to the singularity is a chiral multiplet uncharged under the flavor symmetry but with charge $-1$ under the gauge symmetry.  The level matrix, including classical contributions from the Trotter pairing as well as the fractional contributions of the particles is given by
\begin{equation}
K(\beta,\beta)=-\frac{1}{2}, \hspace{.5in} K(\alpha,\alpha)=0, \hspace{.5in} K(\alpha,\beta)=1.
\end{equation}
The off-diagonal portion of the level implies that the scalar $x_{\alpha}$ is the FI-parameter of the gauged $U(1)$.
\end{itemize}

These two field theories are indeed known to form a mirror pair \cite{KapusStrass}.  At the level of partition functions this equivalence is represented by a quantum dilogarithm identity, known as the \textit{Fourier transform identity} \cite{Faddeev:2000if}
\begin{equation}
	E_+(x_{\alpha}-c_b) = \int d x_{\beta}  \ e^{-2\pi i x_{\alpha} x_{\beta}} E_-(x_{\beta}).
\end{equation}
The fact that our geometric description of conformal field theories provides a framework where this duality is manifest is a satisfying outcome of our analysis.

To gain further insight into this duality we now study resolutions of the singularity in both theories and interpret these from the viewpoint of three-dimensional physics. These resolutions correspond to motion onto the moduli space of the conformal field theory. From the perspective of the ultraviolet Lagrangians, the various branches of the moduli space can be described as Coulomb or Higgs branches, and the effect of the mirror symmetry is to exchange the two descriptions.\footnote{Here and in the following the term \emph{Coulomb branch} will be used generally to include the expectation values of scalars in both dynamical and background vector multiplets.}  

\begin{figure}[here!]
  \centering
  \subfloat[~]{\label{fig:freeChiral_resa}\includegraphics[ width=0.3\textwidth]{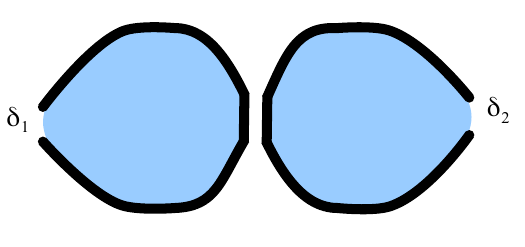}}     
  \hspace{.2in}       
  \subfloat[~]{\label{fig:freeChiral_resb}\includegraphics[ width=0.3\textwidth]{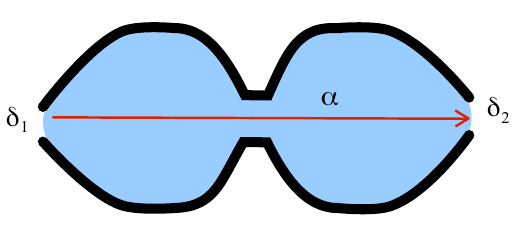}}
   \hspace{.2in}  
  \subfloat[~]{\label{fig:freeChiral_resc}\includegraphics[ width=0.3\textwidth]{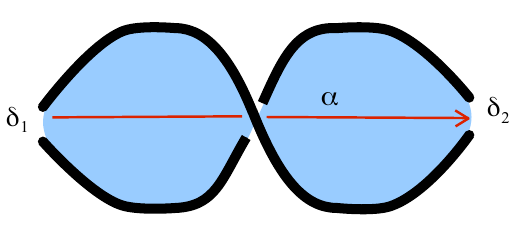}}
  \caption{The three resolutions of the free Chiral field singularity. Part (a) corresponds to the Higgs branch. Part (b) and (c) represent motion onto the two sides of the Coulomb branch.}
  \label{figure:freeChiral_res}
\end{figure}

The three different resolutions (\ref{eq:blines}) have the following effect on the geometry of branch lines, see Figure \ref{figure:freeChiral_res}.
Let us start with the case (c). One can clearly see that the self-Chern-Simons level of the field $\alpha$, as determined by the Trotter pairing, is one. This has a simple explanation from the point of view of field theory. Resolving the singularity means making the M2-brane massive with a mass $m \gg 0$. Thus the IR physics is obtained by integrating out this massive field which according to (\ref{eq:CSshift}) gives rise to a shift
\begin{equation}
	\delta k_{\alpha \alpha} = \frac{1}{2} \textrm{sign}(m) ~ q_{\alpha} q_{\alpha}  = \frac{1}{2}.
\end{equation}
Thus, as the ultraviolet Chern-Simons level was already one-half, the effective level is one exactly as the geometry of resolution (c) predicts. There is yet another way to see this. The limiting behavior of the quantum dilogarithm is as follows
\begin{equation}
	E_+(m) \stackrel{m \rightarrow \infty}{\xrightarrow{\hspace*{1.5cm}}} e^{-i \pi m^2}, 
\end{equation}
which again gives CS-level one in the effective theory as in our case $m = x_{\alpha}$. Resolution (b) corresponds to the other extreme where we take $m \ll 0$. This gives rise to 
\begin{equation}
	\delta k_{\alpha \alpha} = - \frac{1}{2},
\end{equation}
which results in an effective Chern-Simons level $k_{\alpha \alpha} = 0$. This is in complete accord with the geometry as cycle $\alpha$ has no self-linking after push-off in Figure \ref{figure:freeChiral_res}b. Equivalently, this can be again seen in the limiting behaviour of the quantum dilogarithm
\begin{equation}
	E_+(m) \stackrel{m \rightarrow -\infty}{\xrightarrow{\hspace*{1.5cm}}} 1.
\end{equation}
The two resolutions we have studied thus correspond to motion onto the Coulomb branch of the theory parameterized by the real mass $m$. 

\begin{figure}[h!]
\center\includegraphics[width=.5\textwidth]{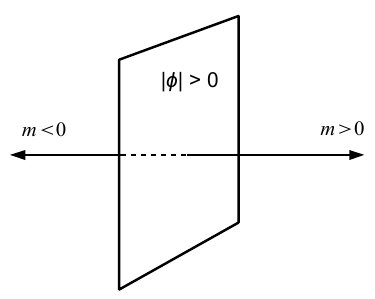}  
\caption{Moduli space of a free chiral field.}
\label{figure:CoulombHiggs}
\end{figure}

\begin{figure}[here!]
  \centering
  \subfloat[~]{\label{fig:freeChiralBWflip_resa}\includegraphics[ width=0.3\textwidth]{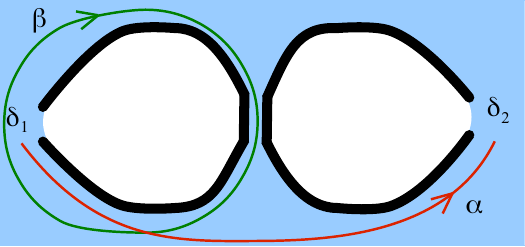}}     
  \hspace{.2in}       
  \subfloat[~]{\label{fig:freeChiralBWflip_resb}\includegraphics[ width=0.3\textwidth]{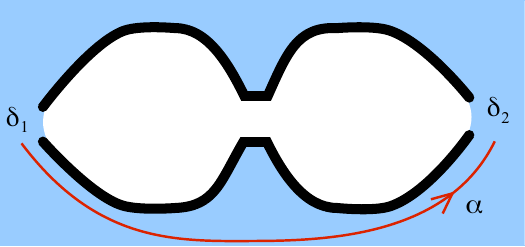}}
   \hspace{.2in}  
  \subfloat[~]{\label{fig:freeChiralBWflip_resc}\includegraphics[ width=0.3\textwidth]{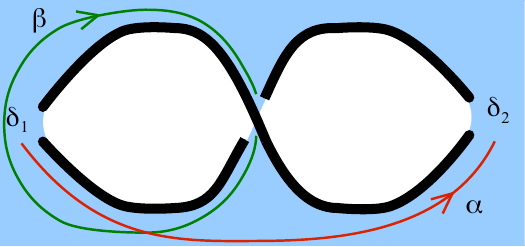}}
  \caption{Resolutions of the theory dual to a free Chiral field.}
  \label{figure:freeChiralBWflip_res}
\end{figure}

Now let us come to resolution (a) which is of a different nature. In order to understand what is happening we follow a path in the moduli space of the Joyce special Lagrangian starting from a point which corresponds to a resolution (b) or (c) to a point of resolution type (a). Along such a path the absolute value of the mass of the particle shrinks, as the volume of the M2-brane disc shrinks, until the field becomes massless at the singularity.   As long as the field is massive it is not possible to turn on a vacuum expectation value for the scalar $\phi$ of the chiral multiplet as this would lead to an infinite energy potential. However, when we sit at the CFT point and the field is massless we can deform the theory onto the Higgs branch by activating an expectation value for $\phi$.  We draw the three branches of the theory schematically in Figure \ref{figure:CoulombHiggs}.
We claim that motion onto the Higgs branch corresponds to resolution (a) on the geometry. In order to see how this comes about we flip the Seifert surface to obtain the resolutions of the dual description of the theory as shown in Figure \ref{figure:freeChiralBWflip_res}. 
In this dual theory resolution (a) arises from choosing $x_{\beta} \ll 0$ as can be seen from the limiting behavior of the negative parity quantum dilogarithm
\begin{equation}
	E_-(x_{\beta}) \stackrel{x_{\beta} \rightarrow -\infty}{\xrightarrow{\hspace*{1.5cm}}} 1.
\end{equation}
Thus in the dual channel this resolution is obtained by giving a vev to the scalar part of a vector multiplet and therefore corresponds to a point on the Coulomb branch of the dual theory. But then the D-term equation of the dual theory requires that $x_{\alpha}$ be set to zero due to the Chern-Simons coupling of the two fields. Translating back to the original theory we indeed see that $m=x_{\alpha} = 0$ and that we have a propagating massless field and are thus capturing the correct effective description of the physics on the Higgs branch. For completeness we note that the dual theory is on the Higgs branch for resolution (b) and on the Coulomb branch for resolution (c). This can be easily seen by noting the limiting behavior of the negative parity quantum dilogarithm for $x_{\beta} \gg 0$
\begin{equation}
	E_-(x_{\beta}) \stackrel{x_{\beta} \rightarrow \infty}{\xrightarrow{\hspace*{1.5cm}}} e^{i\pi x_{\beta}^2}.
\end{equation}

The fact that resolutions of singular tangles capture motion onto the moduli space of the corresponding conformal field theories is a general feature of our constructions which will be pursued in more detail in section \ref{sec:resolutions}.

\subsection{Superpotentials From Geometry}
\label{sec:potential}
There is one more ingredient in defining a three-dimensional theory with $\mathcal{N}=2$ supersymmetry that we have yet to address: the superpotential.  In this section we fill this gap.  As with previous constructions, we find that the precise form of the superpotential as an explicit expression involving fields depends on a choice of Seifert surface used to construct a Lagrangian description.

The superpotential itself has a straightforward geometric interpretation in terms of M2-brane instantons, as described in \cite{CCV}. Here we will briefly review that discussion. Consider some collection of massless chiral fields, $X_i$. Our M5-brane resides on a three-manifold $M,$ which is a double cover of $\mathbb{R}^3$ branched over a singular tangle $L.$ Meanwhile, the entire construction is embedded in an ambient Calabi-Yau $Q.$ As studied above, each of the particles $X_i$ corresponds to a singularity of the tangle $L$. 

Given this setup, a superpotential interaction for the chiral fields $X_i$ may arise from an instanton configuration of an M2-brane. This is a three-manifold $C$ in $Q$, whose boundary $\partial C$ is a two-cycle in $M$ that intersects the particle singularities $X_i$. Consider the projection of the instanton $M2$ to one sheet of the double cover, $\partial C_\pm$. This must be a polygon bounded by the tangle $L$ with vertices given by the singularities of $X_i$. A volume-minimizing configuration of this three-cycle will correspond to an interaction generated by a supersymmetric M2 instanton. This object is precisely of the correct geometric form to generate a superpotential term of the schematic form $\mathcal{W}=\prod_i X_i.$ 

\begin{figure}[here!]
  \centering
  \subfloat[Projection of BPS Instanton]{\label{fig:bpsinstproj1}\includegraphics[ width=0.4\textwidth]{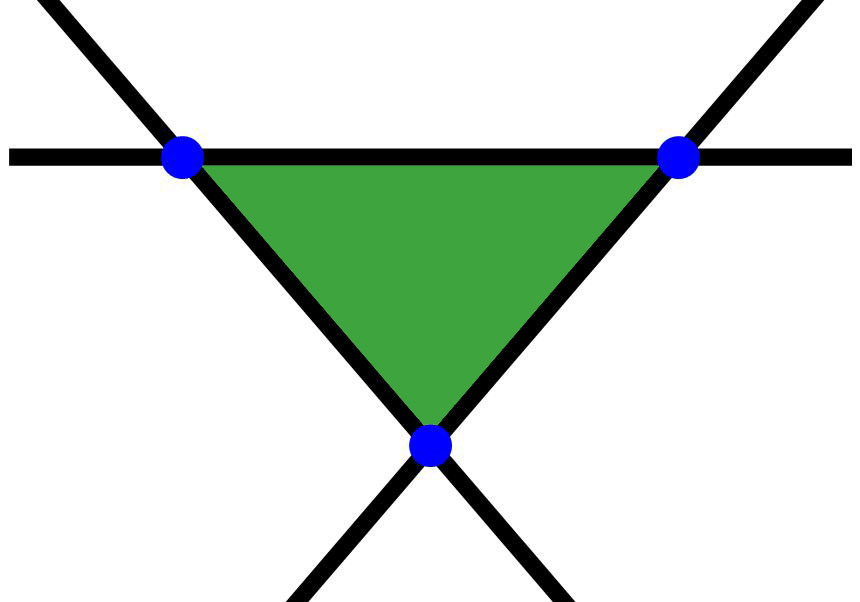}}     
  \hspace{.25in}       
  \subfloat[Lift to $Q$]{\label{fig:bpsinstproj2}\includegraphics[ width=0.4\textwidth]{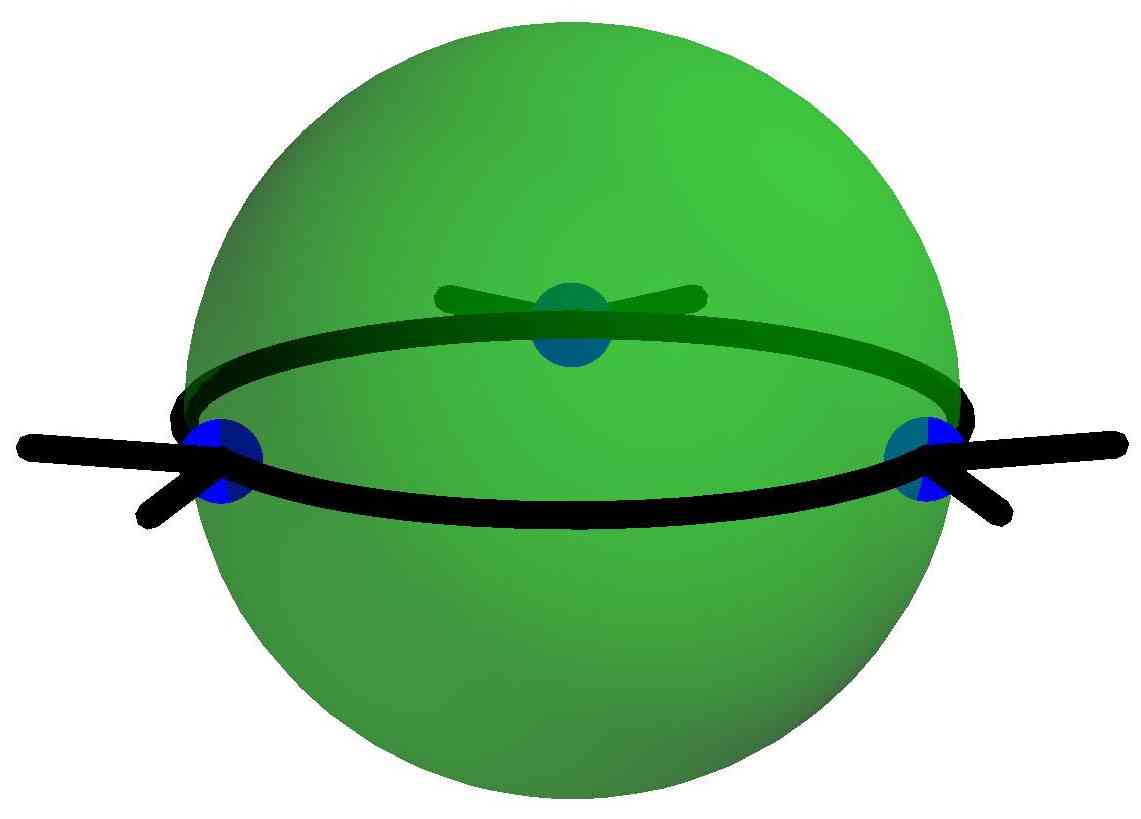}}
  \caption{Projections of BPS M2-brane instanton to the base.  A portion of the branching tangle $L$ is shown in black.  The tangle has singular self-intersections supporting massless particles shown in blue.  In (a), the interior of the polygon, shown in green, is the projection to $M$, of the boundary of an M2 instanton.  In (b), we see the lift of the M2 instanton to the ambient manifold $Q$.  Its boundary is doubled to an $S^{2}$ presented as two hemispheres glued along $L$.  In the interior, this $S^{2}$ is filled in to make a three-ball.}
  \label{fig:m2instproj}
\end{figure}
To sharpen this discussion, there are several further considerations.
\begin{itemize}
\item The coefficient of the interaction is controlled by the instanton action, which is proportional to $e^{-V},$ where $V$ is the volume of the supersymmetric three-manifold $C$. To generate a non-zero interaction, we need the three-manifold to have finite volume. Since our framework allows a non-compact manifold $M$ with $L$ going off to infinity, we must restrict our superpotential polygons on $\partial C_\pm$ to be compact.
\item The instanton action gets a contribution of $\exp{\left(i\int_{\partial C} B\right)},$ from the boundary of the M2 ending on the M5-brane. If $\partial C=0,$ that is, the boundary of the M2 is a trivial two-cycle, then this term is irrelevant. However in general, $\partial C$ is a non-trivial homology class and we find
\begin{equation}
\exp{\left(i\int_{\partial C} B\right)}=\exp{(i\gamma)}
\end{equation}
Where $\gamma$ is a scalar field dual to a photon. This indicates the presence of a monopole operator $\mathcal{M}_j=\exp{(\sigma +i\gamma)}$ in the superpotential. So in this situation, we find a superpotential $\mathcal{W}=\mathcal{M}\prod_i X_i$. Of course, more generally $\partial C$ is some integer linear combination of homology basis elements and so we might find multiple monopole operators in the superpotential.  

\item The invariance of $\mathcal{W}$ under all gauge symmetries apparent in the homology of the Seifert surface implies a compatibility condition on the discrete data living at the singularities bounding the associated polygonal region.  To analyze the charge, we make use of the fact that the exact quantum corrected charge of the monopole operator is
\begin{equation}
q_{\beta}(\mathcal{M}_{\alpha})=k_{\alpha \beta}-\frac{1}{2}\sum_{\mathrm{Chirals} \ X_{i}}|q^{i}_{\alpha}|q^{i}_{\beta}, \label{oneloopm}
\end{equation}
where $k_{\alpha \beta}$ is the Chern-Simons level including both the integral part from the Trotter form, and the fractional contribution from particle singularities.  
\end{itemize}

Given the above discussion, the next step is to analyze the explicit geometry of supersymmetric M2-brane instantons and determine which possible contributions in fact occur. This problem is important, but beyond the scope of this work.  For our purposes we simply take as an ansatz that every possible gauge invariant contribution to the superpotential present in the geometry as a polygon bounded by singularities in fact occurs.

With this hypothesis, to extract the superpotential in complete generality, we analyze a candidate contribution by expressing the boundary two-cycle $\partial C$ in a basis of two-cycles $\{\beta_{a}\}$ dictated by the Seifert surface
\begin{equation}
\partial C= \sum _{a} c_{a}\beta_{a}, \hspace{.5in} c_{a}\in \mathbb{Z}.
\end{equation}  
For example, when utilizing the planar checkerboard Seifert surfaces discussed in section \ref{checkerboards}, the sum $a$ ranges over compact un-colored regions, associated to gauge cycles, as non-compact un-colored regions associated to flavor cycles.  Then, the term in question is
\begin{equation}
\prod_{a}\mathcal{M}_{a}^{c_{a}}\prod_{i\in \partial C}X_{i}. 
\end{equation}
We include such a term in the superpotential provided it is gauge invariant as dictated by the charge formula \eqref{oneloopm}.  The full superpotential is then a sum over all gauge invariant terms associated to all polygonal regions present in the tangle diagram of $L$.

Although it may seem cumbersome to explicitly calculate which polygons yield gauge invariant contributions to $\mathcal{W}$, in practice there is a simple sufficient, but not necessary, graphical rule which ensures gauge invariance that applies to the simplest class of contributions to the superpotential namely polygons which lie entirely in the plane of a given projection of the Seifert surface.  This rule is simply that the arrows on the singularities must circulate all in one direction around the gauge cycle in question.  It may be easily derived from formula \eqref{oneloopm} as well as the charge assignments of particles dictated by Figure \ref{figure:Ep}.  Examples of this type are shown in Figure \ref{fig:m2instproj}.
\begin{figure}[here!]
  \centering
  \subfloat[Superpotential without Monopole]{\label{fig:bpsinstproj3}\includegraphics[ width=0.4\textwidth]{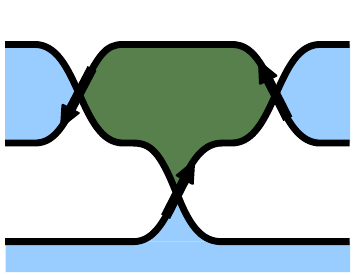}}     
  \hspace{.25in}       
  \subfloat[Superpotential with Monopole]{\label{fig:bpsinstproj4}\includegraphics[ width=0.4\textwidth]{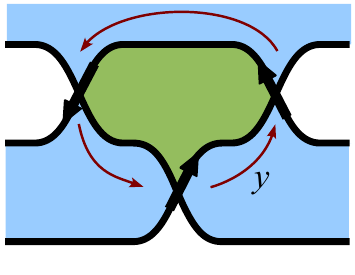}}
  \caption{Projections of BPS M2-brane instanton, with the singular tangle in black.  The particles $X_i$ are indicated by the location of the black arrows, the Seifert surface is shaded in blue, and the projection of the instanton is shown in green.  In (a), the M2 instanton projects to a trivial 2-cycle in $M,$ and therefore has no monopole contribution. We find $\mathcal{W}=X_1X_2X_3.$ In (b), the M2 projects to the non-trivial 2-cycle dual to the 1-cycle $y$ shown on the Seifert surface This contributes a monopole operator, yielding $\mathcal{W}=\mathcal{M}_y X_1X_2X_3.$ }
  \label{fig:m2instproj}
\end{figure}
We encounter more general `non-planar' superpotential terms in our analysis of examples in section \ref{sec:rflowsp}.

\subsection{Physics From Singular Tangles: A Dictionary}
\label{rules}
To conclude our discussion of singularities, we briefly summarize the algorithm for extracting an ultraviolet Lagrangian description of the physics associated to a singular tangle $L$.
\begin{itemize}
\item Pick a Seifert surface $\Sigma$.  The homology $H_{1}(\Sigma_{c},\delta,\mathbb{Z})$ specifies a basis of gauge and flavor cycles.  Boundaryless cycles are dynamical gauge variables, while cycles with boundary are background flavor fields.
\item Compute the Chern-Simons levels by computing the Trotter form on the homology  $H_{1}(\Sigma_{c},\delta,\mathbb{Z}).$  In this procedure the singularities make fractional contributions to linking numbers.  The singularities of plus type, illustrated in Figures \ref{fig:Epp} and \ref{fig:Epm}, contribute $1/2$.  The singularities of minus type, illustrated in Figures \ref{fig:Emp} and \ref{fig:Emm}, contribute $-1/2.$
\item Assign to each singularity a chiral field $X_{i}.$  The field is charged under cycles on $\Sigma$ passing through the singularity.  The charge is $+1$ (-1) if the singularity is of plus type and the cycle is oriented with (against) the arrow at the singularity. The charge is $-1$ (+1) if the singularity is of minus type and the cycle is oriented with (against) the arrow at the singularity.
\item Compute the superpotential by summing over gauge invariant contributions from closed polygonal regions in $L$.  Each monomial entering in $\mathcal{W}$ contains a product of chiral fields dictated by the vertices of the polygon, and possibly various monopole operators determined by expressing the polygon in a basis of two-cycles dual to $H_{1}(\Sigma_{c},\delta,\mathbb{Z}).$  Gauge invariance of the contribution of a given polygon is determined by application of the quantum corrected charge formula for monopole operators \eqref{oneloopm}.
\end{itemize}
The physical theory associated to $L$ is the infrared fixed point determined by this ultraviolet Lagrangian data.  Varying the choice of Seifert surface, provides mirror ultraviolet Lagrangians, but does not alter the underling infrared dynamics.

In general the resulting theory is a strongly interacting system which enjoys a $U(1)^{F}$ flavor symmetry.  The action of $Sp(2F,\mathbb{Z})$ on this conformal field theory is determined geometrically by the braid group action studied in section \ref{torelli}.  The three-sphere partition function $\mathcal{Z}$ is an invariant of the theory which is extracted from this ultraviolet Lagrangian by generalizing the quantum-mechanical framework of section \ref{sec:QM} and assigning to each singularity the quantum dilogarithm wavefunctions dictated by Figure \ref{figure:Ep}.

In the remainder of this paper we apply these rules to further analyze the geometric description of mirror symmetries, and explore applications of the framework.

\section{Dualities and Generalized Reidemeister Moves}
\label{dualities}

In the previous sections we have developed a technique for extracting conformal field theories from singular tangles.  However, there is still non-trivial redundancy in our description: as a consequence of mirror symmetry, two distinct singular tangles may give rise to equivalent quantum field theories.  In this section, we determine the equivalence relation implied on singular tangles by mirror symmetries, and explore their geometric content.   

In searching for such relationships, one may take inspiration from the case of non-singular tangles.  In that case, the basic relations are the Reidemeister moves shown below.
\begin{enumerate}
\item \vcenteredhbox{\includegraphics{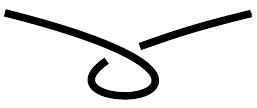}} = \vcenteredhbox{\includegraphics{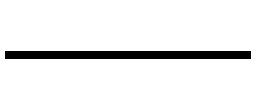}}\\
\item \vcenteredhbox{\includegraphics{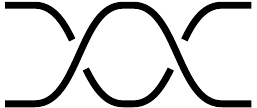}} = \vcenteredhbox{\includegraphics{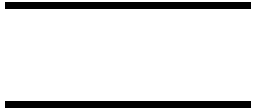}} \\ 
\item \vcenteredhbox{\includegraphics{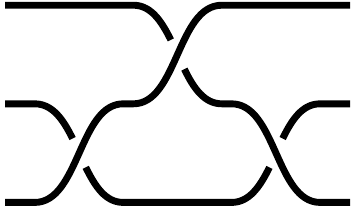}} = \vcenteredhbox{\includegraphics{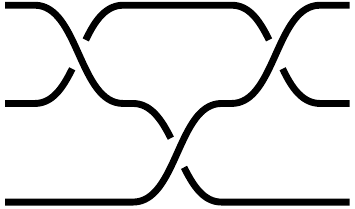}} 
\end{enumerate}
These moves are local and may be applied piecewise in any larger tangle diagram.  Further, these moves are a generating set for equivalences:  any two tangles which are isotopic may be related to one another by a sequence of Reidemeister moves.

In the case of singular tangles, we find similar structure.  Basic mirror symmetries determine relations on singular tangles which take the form of \emph{generalized} Reidemeister moves.  They are related to the moves presented above by replacing some crossings by singularities.  Further, each of these equivalences is local, and hence they may be applied piecewise in a larger singular tangle to engineer more complicated relations.  It is natural to conjecture that these generalized Reidemeister moves, together with the Torelli dualities of section \ref{torelli} provide a complete set of quantum equivalence relations on singular tangles.

In section \ref{sec:singreidemeister} we present a detailed description of the generalized Reidemeister moves as well as the associated quantum dilogarithm identities that result from application of these moves to partition functions. In section \ref{sec:resolutions} we show how deformations away from the conformal fixed point result resolve generalized Reidemeister moves into the ordinary Reidemeister moves.

\subsection{Generalized Reidemeister Moves}
\label{sec:singreidemeister}
In this section we present the list of generalized Reidemeister moves.  Each takes the form of a graphical identity involving two singular tangles.  The precise form of these equalities depends on the discrete data living at the singularities.  There are two things to note about this dependence which follow immediately from our analysis of the local model in section \ref{joyce}.
\begin{itemize}
\item If we flip all arrows by 180 degrees on both sides of an identity, it still holds.  Indeed, such a flip is equivalent to reflecting the sign of all $U(1)$ gauge and flavor groups.  Geometrically, this is equivalent to globally changing the labeling of sheets from 1 to 2 in the double cover.
\item Given any identity, if we exchange all overcross and undercross of non-singular crossings in the diagram, while at the same time exchanging all overcross vs. undercross singularities, the identity still holds.  This is true because each of our diagrams is drawn in a fixed projection with oriented normal vector $\hat{x}_{3}.$  Globally reflecting $\hat{x}_{3}\rightarrow -\hat{x}_{3}$ generates the indicated transformation on diagrams, as shown for example in Figure  \ref{figure:parityflip}.
\begin{figure}[h!]
\center\includegraphics[width=.2\textwidth]{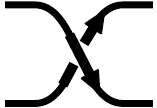}  
\caption{Reflection in the projection plane transforms an overcross singularity to an undercross singularity.}
\label{figure:parityflip}
\end{figure}
\end{itemize}
In the following, we take these two principles into account and thereby present a reduced set of generalized Reidemeister moves.  Additional dualities may be generated by changing the discrete data at the singularities as above.

\subsubsection{Rules Descending from Move 1}

Here, we consider a singular version of the first Reidemeister move. Populating the singular tangles with a Seifert surface generates partition function identities. We will look at two such choices of Seifert surface differing by black-white duality. The first choice does not contain a gauge group whereas the second choice does and is yet another version of the Fourier transform identity.
\begin{center}
\vcenteredhbox{\includegraphics[scale=1.2]{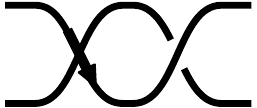}}=\vcenteredhbox{\includegraphics[scale=1.2]{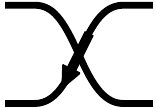}}
\end{center}
With a choice of planar Seifert surface it has the following two interpretations.

\paragraph*{$T$-transformed Singularity}

\begin{center}
\vcenteredhbox{\includegraphics[scale=1.2]{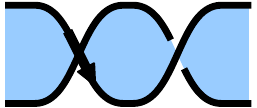}}=\vcenteredhbox{\includegraphics[scale=1.2]{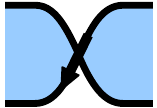}}\\
$$E_+(x-c_b(1-R))e^{i\pi x^2}=E_-(-x+c_b(1-R))$$
\end{center}

In quantum mechanics language this is equivalent to starting with a quantum dilogarithm and applying a $T$-transformation. This does not involve any integrals, as the quantum dilogarithm is an eigenstate of the $T$-operator. Hence there is also no gauge group in the 3d gauge theory interpretation. The only effect on the gauge theory is a change in the background Chern-Simons levels: they are decreased by one unit.

\paragraph*{Fourier Transform}

\begin{center}
\vcenteredhbox{\includegraphics[scale=1.2]{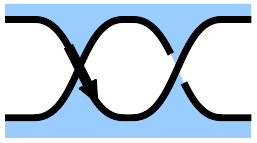}}=\vcenteredhbox{\includegraphics[scale=1.2]{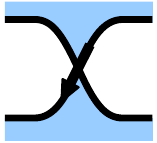}}\\
$$\int dz E_-(z-x+c_b(1-R))e^{-i\pi (z-y)^2}=E_+(y-x-c_b(1+R)/2)$$
\end{center}

This represents a duality containing a $U(1)$ gauge field on the one side but no gauge field on the other. This rule is equivalent to the Fourier transformation identity discussed in section \ref{joyce}, and is another singular-tangle representation of that duality. Here, the theory of one $U(1)$ gauge field at level one-half together with a charged chiral particle is mirror to a free chiral field. 

\subsubsection{Rules Descending from Move 2}
\label{sec:rule2}
The second Reidemeister move can be generalized to give rise to an identity between singular tangles where neighbouring singularities cancel pairwise such that on the other side of the identity there is no singularity at all. Therefore, we denote these identities with the term \textit{pairwise cancellation of singularities}. We will also examine a partition function identity inherited from the tangle identity for one choice of Seifert surface. The relevant singular tangle identities are the following.

\begin{center}
\vcenteredhbox{\includegraphics[scale=1.2]{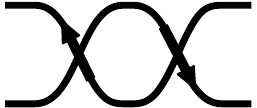}}=\vcenteredhbox{\includegraphics[scale=1.2]{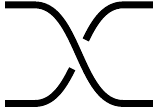}} \\
\vspace{1cm}
\vcenteredhbox{\includegraphics[scale=1.2]{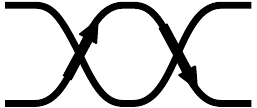}}=\vcenteredhbox{\includegraphics[scale=1.2]{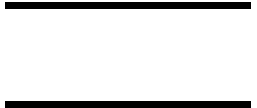}}
\end{center}

From the perspective of the 3d gauge theory these can be understood as follows. We have a closed polygonal region bounded by two singularities. As discussed in section \ref{sec:potential} this gives rise to a superpotential with the two chiral fields. Thus the particles are given mass and make no contribution to the infrared physics. The dual theory then contains no particles, but depending on the UV Chern-Simons levels it can contain background Chern-Simons levels.

Picking a Seifert surface these rules translate to the following quantum dilogarithm identities.

\begin{center}
\vcenteredhbox{\includegraphics[scale=1.2]{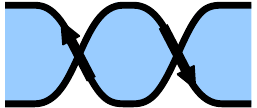}}=\vcenteredhbox{\includegraphics[scale=1.2]{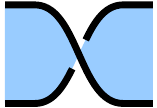}}\\
$$E_+(-x+c_b(1-R)) E_+(x-c_b(1-R))=e^{-i\pi x^2}$$
\vcenteredhbox{\includegraphics[scale=1.2]{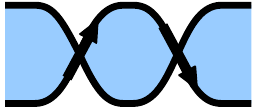}}=\vcenteredhbox{\includegraphics[scale=1.2]{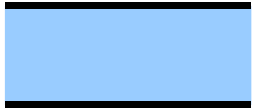}}\\
$$E_+(x-c_b(1-R))E_-(x-c_b(1-R))=1$$
\end{center}

From this perspective, the underlying identity of pairwise cancellation of singularities is equation (16) in the appendix of reference \cite{Faddeev:2000if}. 

\subsubsection{Rules Descending from Move 3}

The most important rule arises from singularization of the third Reidemeister move. This rule is called the 3-2 move and encodes a non-trivial three-dimensional mirror symmetry.  In this section we will clarify its relation to the third Reidemeister move by singularizing all crossings on one side of the identity and only two on the other side. Apart from the 3-2 move, the third Reidemeister move can be singularized by adding only one singularity on both sides.  This application follows from the previously identified Fourier transform identity and hence does not represent an independent mirror symmetry. Nevertheless, the simple application is useful when moving between Seifert surfaces in the examples of section \ref{rflow} and \ref{sec:app}.  We will turn to this simple application first and then discuss the 3-2 move. 

\paragraph*{Change of Branch sheet}
\label{branchsheetchange}
Applying the Fourier transform identity of Figure \ref{figure:BlackWhite} locally, we obtain a generalization of the third Reidmester move.  On one side of the duality we have a theory with a chiral particle charged under a $U(1)$ gauge field which in turn couples to two background gauge fields. The duality relates this theory to one with no gauge group, a chiral mulitplet and two flavor fields.   The partition function equality is again an application of Figure \ref{figure:BlackWhite}.

\begin{center}\vcenteredhbox{\includegraphics[scale=1.2]{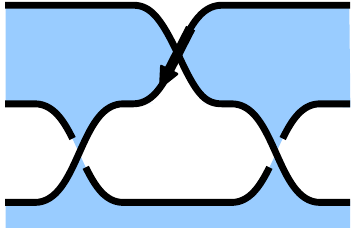}}=\vcenteredhbox{\includegraphics[scale=1.2]{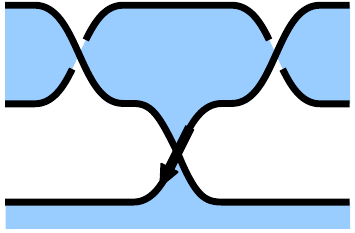}}\\
$$\int dw E_-(w-x+c_b(1-R))e^{-i\pi(w-z)^2+i\pi(w-y)^2}=E_+(y-z-c_b(1+R)/2)e^{-i\pi(z-x)^2+i\pi(x-y)^2}$$
\vcenteredhbox{\includegraphics[scale=1.2]{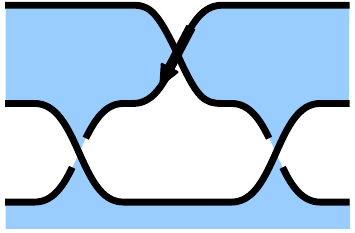}}=\vcenteredhbox{\includegraphics[scale=1.2]{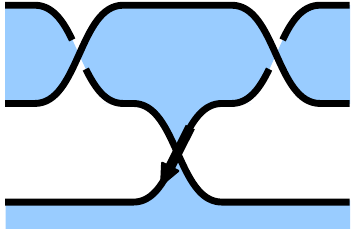}}\\
$$\int dw E_-(w-x+c_b(1-R))e^{-i\pi(w-y)^2+i\pi(w-z)^2}=E_+(z-y-c_b(1+R)/2)e^{-i\pi(y-x)^2+i\pi(x-z)^2}$$
\end{center}

\paragraph*{The 3-2 move}
\label{sec:pentagonmove}
The relevant singular tangle identity is depicted below.

\begin{center}
\vcenteredhbox{\includegraphics[scale=1.2]{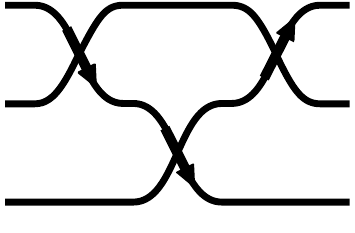}}=\vcenteredhbox{\includegraphics[scale=1.2]{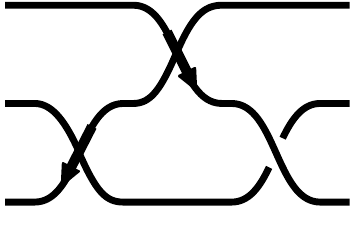}}
\end{center}

We clearly see that this identity relates a theory with three chiral fields to the one with just two chiral fields. Such theories are known to come in mirror pairs in three dimensions \cite{KapusStrass, IS, XYZ,HW}. Examining the left-hand-side we notice the presence of a closed polygonal region bounded by three singularities and hence the existence of a superpotential.  To extract the physical content we choose Seifert surfaces as shown below.
\begin{center}
\vcenteredhbox{\includegraphics[scale=1.2]{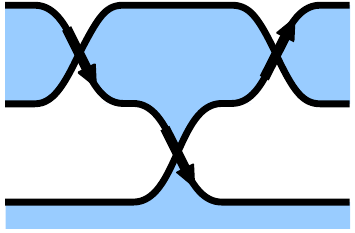}}=\vcenteredhbox{\includegraphics[scale=1.2]{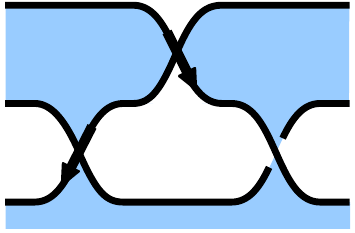}}\\
$$E_+(z-y-c_b(1-r))E_-(z-x+c_b(1-s))E_-(x-y-c_b(1-r-s))$$\\
$$=\int dw E_+(y-w-c_b)E_+(z+w-c_b(1-r))e^{i\pi (x-w)^2}$$
\end{center}
The physical theories are then read off:
\begin{itemize}
\item left-hand-side:  A theory with three chiral fields $X, Y, Z$ no gauge symmetry and a cubic superpotential $\mathcal{W}=XYZ$, known as the $XYZ$-model.
\item right-hand-side: A theory with a gauged $U(1)$ with vanishing self Chern-Simons level and two oppositely charged chiral fields $Q$ and $\widetilde{Q}$, known as $U(1)$ super-QED with $N_{f}=1$.
\end{itemize}
These theories are known to form a mirror pair \cite{XYZ}. At the level of partition functions this duality is the \textit{pentagon identity} for quantum dilogarithms \cite{Faddeev:2000if}.

\subsection{Resolutions of Dualities}
\label{sec:resolutions}
\begin{figure}[here!]
  \centering
  \subfloat[~]{\label{fig:diglogresa}\includegraphics[ width=0.4\textwidth]{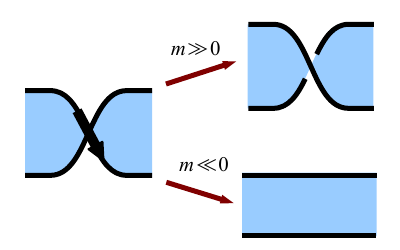}}     
  \hspace{.4in}       
  \subfloat[~]{\label{fig:dilogresb}\includegraphics[ width=0.4\textwidth]{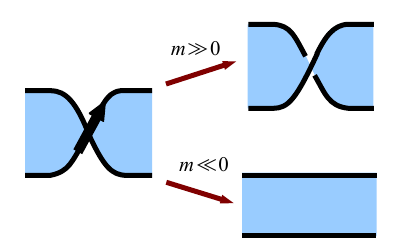}}
  \caption{Resolution of singularities by turning on Fayet-Iliopoulos or Coulomb branch parameters. In both parts, (a) and (b), $m$ is the argument of the relevant quantum dilogarithm.}
  \label{figure:dilogres}
\end{figure}

In this section we make the connection between generalized Reidemeister moves and ordinary Reidemeister moves precise.  We show that motion onto the moduli space of the conformal field theories appearing on both sides of a generalized Reidemeister move resolves them into ordinary Reidemeister moves.  To achieve this we will choose a particular Seifert surface such that all the resolutions in question are obtained as a motion onto the Coulomb branch.  In general such a deformation gives masses to all chiral fields and in the infrared they can be integrated out. Generically, this leads to a fractional shift in the Chern-Simons levels of the form \cite{XYZ}
\begin{equation}
	(K_{IJ})_{\textrm{eff}} = K_{IJ} + \frac{1}{2} \sum_{a=1}^{N_f} (q_a)_I (q_a)_J \textrm{sign}(m_a) \in \IZ, \quad I,J = 1, \cdots, G+F, 
\end{equation}
where we have noted that the effective levels are integral in order to ensure gauge invariance. These effective levels are depicted in Figure \ref{figure:dilogres} as applied to a single singularity as studied in section \ref{wavefunctions}.

 In applying this logic to study resolutions of singular tangles, one must take care to remain in a supersymmetric vacuum. In other words the $F$- and $D$-term equations have to be satisfied. This will be dealt with next.
\subsubsection{F- and D-term Equations}
\label{FDeqns}
Let us elaborate the Coulomb branch resolutions from the viewpoint of the 3d gauge theory. The singular tangle describes the CFT at the origin of the Coulomb and Higgs branches.  If we discuss only resolutions which remain at the origin of the Higgs branch then the resulting resolutions correspond to different leaves of the Coulomb branch parameterized by Fayet-Iliopoulos parameters and scalar fields in vector multiplets. 

In order to determine which resolutions are possible in a complicated singular tangle we need to solve the D- and F-term equations of the relevant 3d gauge theory. The potential $\cV$ for the theory is a sum of a D-term and an F-term contributions of the form
\begin{equation}
	\cV = \cV_D + \cV_F.
\end{equation}
In a supersymmetric vacuum this potential must vanish. As both $\cV_D$ and $\cV_F$ are non-negative, both must vanish separately. 

Let us first consider the F-term potential which reads
\begin{equation}
	\cV_F = \sum_{a=1}^{N_f} \left|\frac{\partial \mathcal{W}}{\partial \phi_a}\right|^2,
\end{equation}
where $\mathcal{W}$ is the superpotential of the theory and $\phi_a$ is the scalar component of the chiral field $X_a$. In our geometric examples, $\mathcal{W}$ arises from a sum over polygons and hence each monomial in $\mathcal{W}$ has degree larger than one.  It follows that if we remain at the origin of the Higgs branch $\phi_a = 0$
the F-term potential is trivially minimized.

Let us next turn to the D-term potential. In the following we will drop the subscript \textit{eff} from all Chern-Simons levels and assume that the IR limit has been taken. The D-term potential is then given by
\begin{eqnarray}
	\cV_D & = & \sum_{i,j} \frac{k_{ij}}{2\pi} D_i y_j + \sum_{i,\lambda} \frac{k_{\lambda i}}{2\pi} x_{\lambda} D_i \nonumber \\
	~       & ~  & + \sum_{a,i} q_{a,i} D_i |\phi_a|^2 + \sum_{a,i} |q_{a,i} y_i|^2 |\phi_a|^2,
\end{eqnarray}
where the summation is over $i,~j=1, \cdots, G$ for the gauge indices, and $\lambda =1 ,\cdots, F$ for the Fayet-Illiopoulos parameters $x_{\lambda}$.
The associated D-term equation then reads
\begin{equation} \label{Dterm}
	\frac{\partial \cV_D}{\partial D_i} =  \sum_j \frac{k_{ij}}{2\pi} y_j + \sum_{\lambda} \frac{k_{i\lambda}}{2\pi} x_{\lambda} + \sum_a q_{a,i} |\phi_a|^2 = 0.
\end{equation}
On the Coulomb branch we have that $\phi_a = 0$ which simplifies the above equation considerably. Defining
\begin{equation}
	K_{IJ} := \left(\begin{array}{cc}k_{ij} & k_{i \lambda} \\ k_{\lambda i} & k_{\lambda \mu} \end{array}\right),
	\quad \Sigma_J := \left(\begin{array}{c} y_i \\ x_{\lambda}\end{array}\right),
\end{equation}
it is possible to write equation (\ref{Dterm}) in the compact form
\begin{equation} \label{Dtermeq2}
	K_{i J} \Sigma_{J} = 0, 
\end{equation}
for $i=1, \cdots, G$.

Equation \eqref{Dtermeq2} is our desired result.  It implies that provided we are interested only in Coulomb branch deformations, we can determine which are allowed by searching for null-vectors of the effective level matrix $K$.  

\subsubsection{Resolution of Move Descending from Rule $1$}
\label{resolve}
\begin{figure}[h!]
\center\includegraphics[width=.5\textwidth]{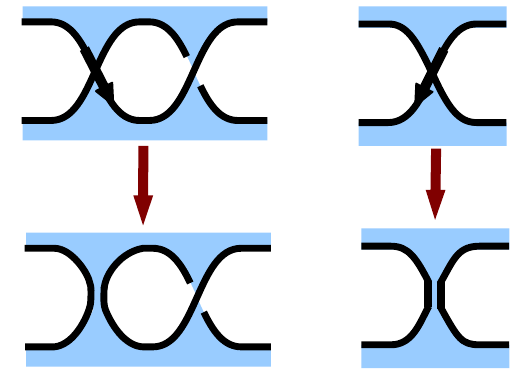}  
\caption{Resolution of the first generalized Reidemeister move.}
\label{figure:rule1res}
\end{figure}

Here, we examine how a particular resolution on the two sides of our first generalized Reidemeister move gives back the ordinary Reidemeister move of first kind. In order to proceed, we need to pick a particular Seifert surface which allows us to obtain the relevant resolution as motion onto the Coulomb branch. We will pick the second Seifert surface corresponding to the dilogarithm identity
\begin{equation}
	\int dz E_-(z-x+c_b(1-R))e^{-i\pi (z-y)^2}=E_+(y-x-c_b(1+R)/2).
\end{equation}
The limit we take is the following
\begin{equation}
	y \rightarrow - \infty, \quad x \rightarrow \infty, \quad z \rightarrow - \infty,
\end{equation}
resulting in
\begin{equation}
	E_-(y-x-c_b(1+R)/2) \rightarrow 1, \quad E_+(z-x+c_b(1-R)) \rightarrow 1.
\end{equation}
The effective Chern-Simons levels of the left-hand-side become
\begin{equation}
	k_{zz} = 1, \quad k_{zy} = -1,
\end{equation}
which in turn lead to the D-term equation (\ref{Dtermeq2})
\begin{equation}
	z - y = 0.
\end{equation}
As this is consistent with the limit taken we are indeed looking at a valid resolution satisfying the equations of motion of the gauge theory. The pictorial representation is shown in Figure \ref{figure:rule1res}.
We clearly see that the resolution reproduces the ordinary first Reidemeister move as claimed.

\subsubsection{Resolution of Moves Descending From Rule $2$}
Next, we look at resolutions of the second generalized Reidemeister move. This rule consists of two parts and we shall examine both of them. Again we have to pick a Seifert surface which we choose to be the same as in section \ref{sec:rule2}. The relevant quantum dilogarithm identity for the first subrule is
\begin{equation}
	E_+(-x+c_b(1-R)) E_+(x-c_b(1-R))=e^{-i\pi x^2}.
\end{equation}
Here we can consider the following limit
\begin{equation}
	x \rightarrow \infty: \quad E_+(-x + c_b(1-R)) E_+(x - c_b(1-R)) \rightarrow 1 \cdot e^{-i\pi x^2} = e^{-i\pi x^2}.
\end{equation}
As the limit gives the right hand side of the identity trivially there is nothing to be checked. Therefore, this resolution does not involve any Reidemeister moves. 

\begin{figure}[h!]
\center\includegraphics[width=.5\textwidth]{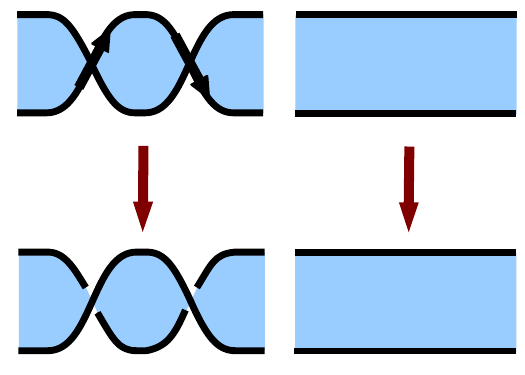}  
\caption{Resolution of the second generalized Reidemeister move.}
\label{figure:rule2res}
\end{figure}

Let us now move to the second subrule. The relevant quantum dilog identity is 
\begin{equation}
	E_+(x-c_b(1-R))E_-(x-c_b(1-R))=1.
\end{equation}
Taking the limit $x \rightarrow \infty$ the left-hand-side becomes
\begin{equation}
	E_+(x-c_b(1-R))E_-(x-c_b(1-R)) \rightarrow e^{-i\pi x^2} e^{i\pi x^2}.
\end{equation}
The pictorial representation of this resolution is the second Reidemeister rule, as shown in figure \ref{figure:rule2res}.

\subsubsection{Resolution of Move Descending From Rule $3$}
Let us now come to our last and most involved case, namely the 3-2-move. The relevant identity here is
\begin{eqnarray}
	E_+(z-y-c_b(1-r))E_-(z-x+c_b(1-s))E_-(x-y-c_b(1-r-s)) \nonumber \\
=\int dw E_+(z-w-c_b)E_+(w-y-c_b(1-r))e^{i\pi (x-w)^2}.
\end{eqnarray}

Defining
\begin{equation}
	x-y \equiv c_1, \quad z-x \equiv c_2, \quad z-y \equiv c_3,
\end{equation}
we will consider the limit
\begin{equation}
	c_i \gg 0 \textrm{~~~for~~~} i=1,2,3.
\end{equation}
As the above equation set implies the relation $c_1 = - c_2 + c_3$ we find that 
\begin{equation}
	c_3-c_2 \gg 0.
\end{equation}
Setting $w \equiv c_3$ ensures that we have the effective Chern-Simons-levels 
\begin{equation}
	k_{wx} = 1, \quad k_{wy} = -1, \quad k_{zw} = -1, \quad k_{ww}=1.
\end{equation}
The D-term equation (\ref{Dtermeq2}) thus gives
\begin{equation}
	x - y + w - z = c_3 - c_2 +c_3 + c_2 -2 c_3 = 0,
\end{equation}
and hence confirms that we are on the Coulomb-branch.

\begin{figure}[h!]
\center\includegraphics[width=.6\textwidth]{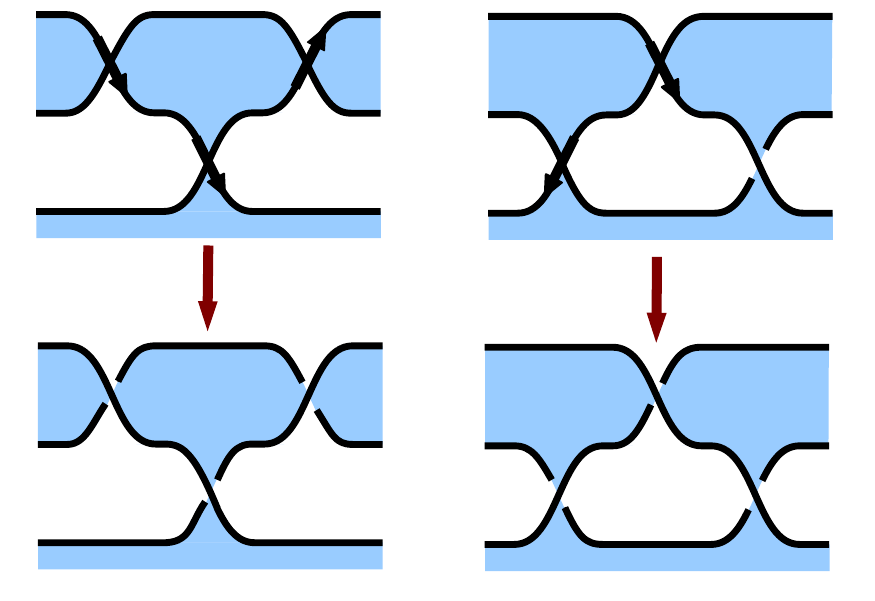}  
\caption{Resolution of the 3-2-move.}
\label{figure:rule3res}
\end{figure}

The pictorial representation of the limit discussed is the third Reidemeister move as shown in Figure \ref{figure:rule3res}.

\newpage

\section{R-flow}
\label{rflow}
We have seen how singular tangles capture the content of a 3d conformal field theory with four supercharges and that resolutions of such objects describe dynamics on the moduli space of the same theory. This is very similar to how Seiberg-Witten theory describes the Coulomb branch of 4d gauge theories with eight supercharges. In fact, the similarity goes even further.  In the Seiberg-Witten case the multi-cover of a complex curve with punctures captures all the information about the BPS states of the four-dimensional gauge theory \cite{KLMVW,DW,Witten1997,Gaiotto}. In our case a multicover (more specifically, a double cover) of $\mathbb{R}^3$ with specified boundary conditions captures the content of a three-dimensional theory. The connection of these two descriptions can be made precise by looking at specific class of examples where the three-manifolds in question arise from \textit{flows} of a Seiberg-Witten curve of a 4d theory. By this we mean that there exists a \textit{slicing} of the three-manifold along a \textit{time} direction such that each slice represents a SW-curve. It turns out, that such a flow indeed exists and is known as \textit{R-flow} \cite{CCV, CNV}. This section is devoted to the definition and properties of R-flow. It is defined on the space of central charges of certain 4d $\cN=2$ theories and describes a domain wall solution which has the interpretation of a 3d $\cN=2$  theory \cite{Gaiottowall,Janus1,Janus2}.  

\subsection{Definition of the Flow}
\label{setup}
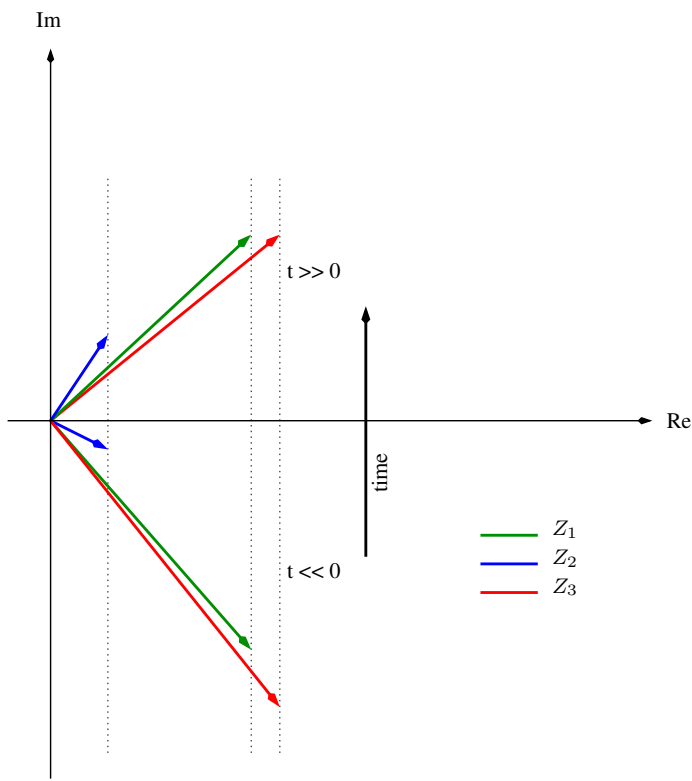
\begin{figure}[h!]
\center
\begin{picture}(0,0)%
\includegraphics{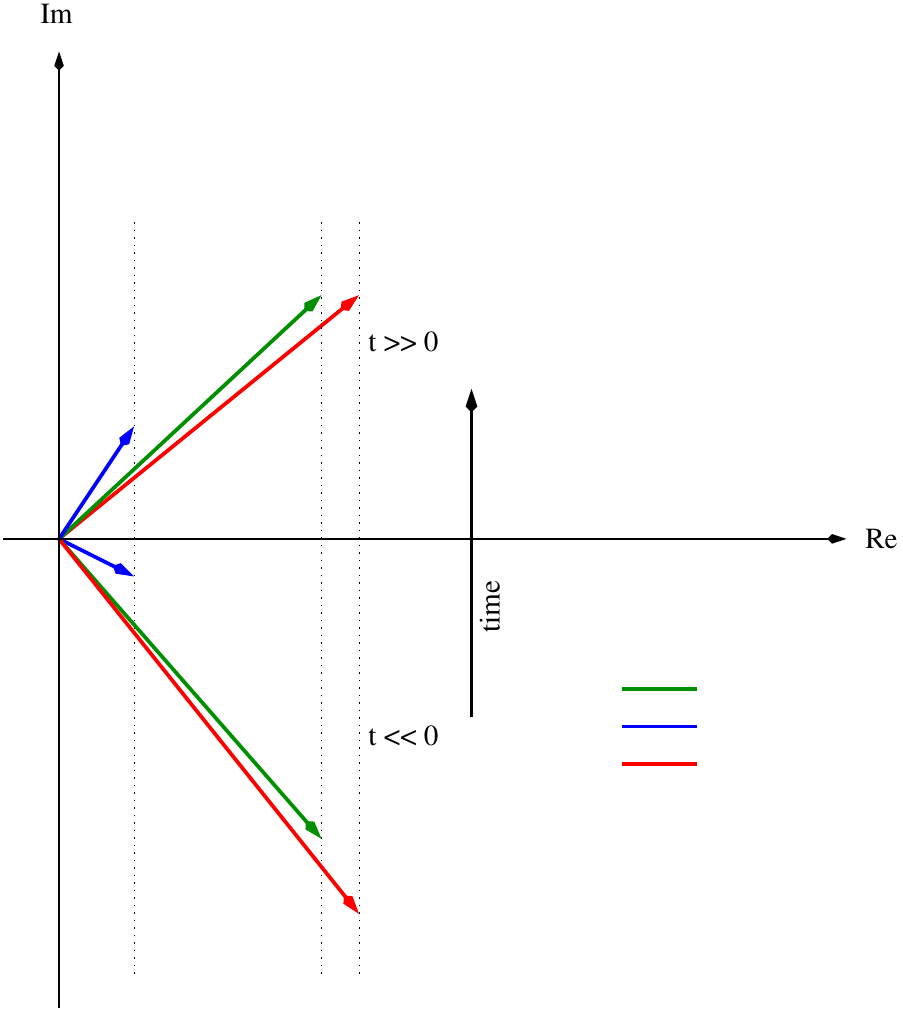}%
\end{picture}%
\setlength{\unitlength}{2368sp}%
\begingroup\makeatletter\ifx\SetFigFont\undefined%
\gdef\SetFigFont#1#2#3#4#5{%
  \reset@font\fontsize{#1}{#2pt}%
  \fontfamily{#3}\fontseries{#4}\fontshape{#5}%
  \selectfont}%
\fi\endgroup%
\begin{picture}(7213,8077)(3879,-7883)
\put(9601,-5611){\makebox(0,0)[lb]{\smash{{\SetFigFont{8}{9.6}{\rmdefault}{\mddefault}{\updefault}{\color[rgb]{0,0,0}$Z_2$}%
}}}}
\put(9601,-5911){\makebox(0,0)[lb]{\smash{{\SetFigFont{8}{9.6}{\rmdefault}{\mddefault}{\updefault}{\color[rgb]{0,0,0}$Z_3$}%
}}}}
\put(9601,-5311){\makebox(0,0)[lb]{\smash{{\SetFigFont{8}{9.6}{\rmdefault}{\mddefault}{\updefault}{\color[rgb]{0,0,0}$Z_1$}%
}}}}
\end{picture}%
\caption{R-flow for an example with three central charges.}
\label{figure:chargeflow}
\end{figure}

R-flow is a motion in the space central charges of four-dimensional theories with eight supercharges. In theories which are known to be complete \cite{CV11} deformations in the space of central charges are locally equivalent to deformations of branch points of the Seiberg-Witten curve. We define the flow to be of the following form
\begin{equation} \label{eq:Rflow}
	\frac{d}{dt} Z_i = i \textrm{Re} Z_i,
\end{equation}
where $Z_i$ is the central charge of the $i$-th charge in the $\cN=2$ 4d theory. This tells us that the central charges flow along straight lines preserving their real parts while their imaginary parts move at a rate which is proportional to their real parts. As a consequence of this flow equation, the phase ordering of central charges is preserved and hence the entire evolution takes place in a fixed BPS chamber.  In summary, we can say that \textit{phase ordering} is \textit{time ordering} and depict this in a graph shown in Figure \ref{figure:chargeflow}. This describes a three-dimensional theory as a domain-wall solution of the four-dimensional parent theory where each 4d BPS state gives rise to a 3d BPS state whose mass is given by the real part of $Z_i$.

\subsection{$A_n$ flow and the KS-operator}
\label{angeo}
In this paper we are in particular interested in flows of 4d gauge theories which arise from wrapping a M5-brane on a Riemann surface of the type $A_n$ describing Argyres-Douglas CFTs\cite{AD, Gaiotto}. These are Riemann surfaces which are double covers of the $\mathbb{C}$-plane of the form
\begin{equation}
	y^2 = (x-a_1)(x-a_2)\cdots(x-a_n) (x+a_{n+1}),
\end{equation}
where $a_{n+1} = \sum_{i=1}^n a_i$. The Seiberg-Witten differential is given by the square root of the quadratic differential 
\begin{equation}
	\phi = (x-a_1)(x-a_2)\cdots(x-a_n) (x+a_{n+1}) dx^2,
\end{equation}
i.e. $\lambda_{\textrm{SW}} = \sqrt{\phi}$. Having established the above definitions, it is straightforward to write down the central charges of the theory:
\begin{eqnarray}
	Z_1 & = & \int_{a_1}^{a_2} \sqrt{\phi}, \nonumber \\
	Z_2 & = & \int_{a_2}^{a_3} \sqrt{\phi}, \nonumber \\
	~    & \vdots & ~ \nonumber \\
	Z_n & = & \int_{a_n}^{a_{n+1}} \sqrt{\phi}. \label{centralcharges}
\end{eqnarray}

\begin{figure}[h!]
\center\includegraphics[width=.7\textwidth]{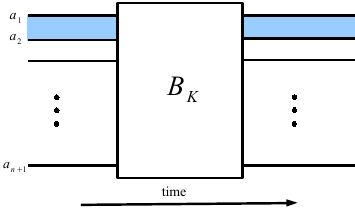}  
\caption{For each KS-operator there is an associated singular braid $B_{\IK}$.}
\label{figure:BK}
\end{figure}

Now, choosing a specific ordering of the phases of the central charges one arrives in a particular chamber of the moduli space where a specific number of BPS particles is stable. For the choice
\begin{equation}
	\textrm{arg}Z_1 < \textrm{arg} Z_{2} < \cdots < \textrm{arg} Z_n,
\end{equation}
we obtain the so called \textit{minimal chamber} with exactly $n$ stable particles. On the other hand,
the \textit{maximal chamber} is defined for the configuration
\begin{equation}
	\textrm{arg}Z_n < \textrm{arg} Z_{n-1} < \cdots < \textrm{arg} Z_1.
\end{equation}
Here the number of stable BPS particles is $\frac{1}{2} n (n+1)$ \cite{SV}. There will be also intermediate chambers with less particles and we shall refer to the number of states in a given chamber by $N$. Note that for each of these states there is a corresponding central charge which in general is a linear combination of those given in (\ref{centralcharges}). We next assign to each central charge ordering a Kontsevich-Soibelman operator of the following form \cite{KS,GMN08,GMN10}:
\begin{equation}
	\mathbb{K}(q) = \prod_i^N E_+(\hat{\gamma}_i),
\end{equation}
where $E_+$ is the non-compact quantum dilogarithm while the $\hat{\gamma}_i$ label the stable BPS states and can be interpreted as phase space variables of the quantum Hilbert space which differ by actions of $Sp(n,\mathbb{Z})$ if $n$ is even and $Sp(n-1,\mathbb{Z})$ if $n$ is odd. From the point of view of the $A_n$ curve the $\hat{\gamma}_i$ represent cycles determined by two branch points $a_k$ and $a_l$. In particular, from the point of view of the quantum mechanics description of section \ref{sec:QM}, they are linear combinations of $\hat{x}_i$ and $\hat{p}_i$ and are mapped to each other by actions of the generators
\begin{equation}
	\sigma_{2j-1} = \exp\left(-i \pi \hat{x}_j^2\right), \quad \sigma_{2j} = \exp\left(-i \pi \hat{p}_j^2\right).
\end{equation}
We can assign to each KS-operator a quantum mechanical matrix element of the form
\begin{equation}
	\cZ_{\IK} = \langle x | \IK | y \rangle,
\end{equation}
which have an interpretation as partition functions of 3d theories as discussed in section \ref{sec:doubledflavor}. These partition functions enjoy a $Sp(n,\mathbb{Z}) \times Sp(n,\mathbb{Z})$ action which has the interpretation of the braid group action on the two ends of a braid with $n+1$ strands. In our case we can thus assign a singular braid $B_{\IK}$ to the matrix element $\cZ_{\IK}$. This is depicted schematically in Figure \ref{figure:BK}. As also indicated there, the braid naturally defines a time direction which we can understand as follows. Each line of the braid describes the flow of a branch point of the $A_n$-curve along the time direction and at the singularities these branch points come close to each other and actually touch, thereby loosing their individual identities. 

Let us zoom into the braid $B_{\IK}$ to see how the strands approach each other for an isolated singularity. To this end, we rewrite the partition function as a gluing of three braids according to the formalism developed in section \ref{sec:doubledflavor}
\begin{equation}
	\cZ_{\IK} = \int dx' dy' \langle x| \cdots |y'\rangle \langle y'| E_+(\hat \gamma_{kl})|x'\rangle \langle x' | \cdots |y\rangle,
\end{equation}
where $\hat \gamma_{kl}$ represents the contribution of the 4d BPS state whose central charge is given by
\begin{equation}
	Z_{kl} = \int_{a_l}^{a_k} \sqrt{\phi}.
\end{equation}
Zooming into the braid we then have the local representation for an isolated singularity shown in Table \ref{table:localsing}.
\begin{table}
\begin{center}
\begin{tabular}{lcr}
$\langle y'|E_+(\hat \gamma_{kl})|x'\rangle$ & $=$ &
\vcenteredhbox{\includegraphics[width=.6\textwidth]{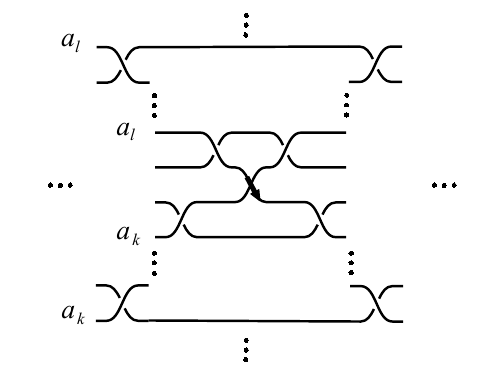} }
\end{tabular}
\caption{Braid realization of a local singularity. The relevant branch points come close to each other until they collide in the singularity and loose their individual identities. After that they depart again until they reach their original positions in the braid.}
\label{table:localsing}
\end{center}\end{table}
Resolving the singularity means turning the points at which the branch points touch to near misses. As we have seen, for each singularity there are exactly three ways to do this. R-flow, as a flow of branch points of the Seiberg-Witten curve, is equivalent to choosing the resolution of Figure \ref{figure:desingularizations} (b) for all singularities. Said differently, the singular braid $B_{\IK}$ is obtained from the flow defined by equation  (\ref{eq:Rflow}) in the limit in which all near misses are replaced by singularities.

Let us now come to the justification of this picture. The initial condition of R-flow is determined by the chamber in which the flow starts. Furthermore, as the flow continues one stays in the initial chamber due to the phase-preserving property of the flow. As central charges cross the real axis something special happens. Recall that a 4d BPS hypermultiplet has an interpretation as a geodesic on the complex plane between branch points of the Riemann surface \cite{GMN09, ACCERV1}. These geodesics obey the equation
\begin{equation} \label{gflow}
	\sqrt{\phi} = e^{i\theta_m} dt, 
\end{equation}
where $\theta_m, ~ m =1, \cdots, N$ is the phase of the $m$th BPS state, i.e.
\begin{equation}
	\theta_m = \textrm{arg} Z_m.
\end{equation}
There are two remarks in order here. First, R-flow describes a motion on the Coulomb branch (including mass parameters) of the four-dimensional gauge theory. On the other hand, the flow equation (\ref{gflow}) is a flow on the $\IC$-plane at a \textit{fixed} point in the moduli space. The Seiberg-Witten curve, being a double-branched cover of the $\IC$-plane, is not subject to change under the flow (\ref{gflow}). Therefore, in order to relate the two motions, we have to choose a fixed angle $\theta_m$ corresponding to a line in the complex plane of central charges. Secondly, the geometry of R-flow predestines exactly such a line, namely the real axis which defines a mirror axis for the flow. 
Thus we see that each time a central charge crosses the real axis there is a geodesic solution with minimal length. Thus at such points the pair of branch points corresponding to the BPS bound state whose central charge crosses the real axis are closest. 

\subsection{Examples}
\label{rflowex}
In this section present some examples of R-flow. We start with the simplest case and proceed to increasing complexity. Already in the very first example, the $A_1$ flow, we will find that R-flow gives insight into the behavior of branch lines near local singularities. 

\subsubsection*{$A_1$ flow}
\label{sec:A1flow}
\begin{figure}[h!]
\center\includegraphics[width=.5\textwidth]{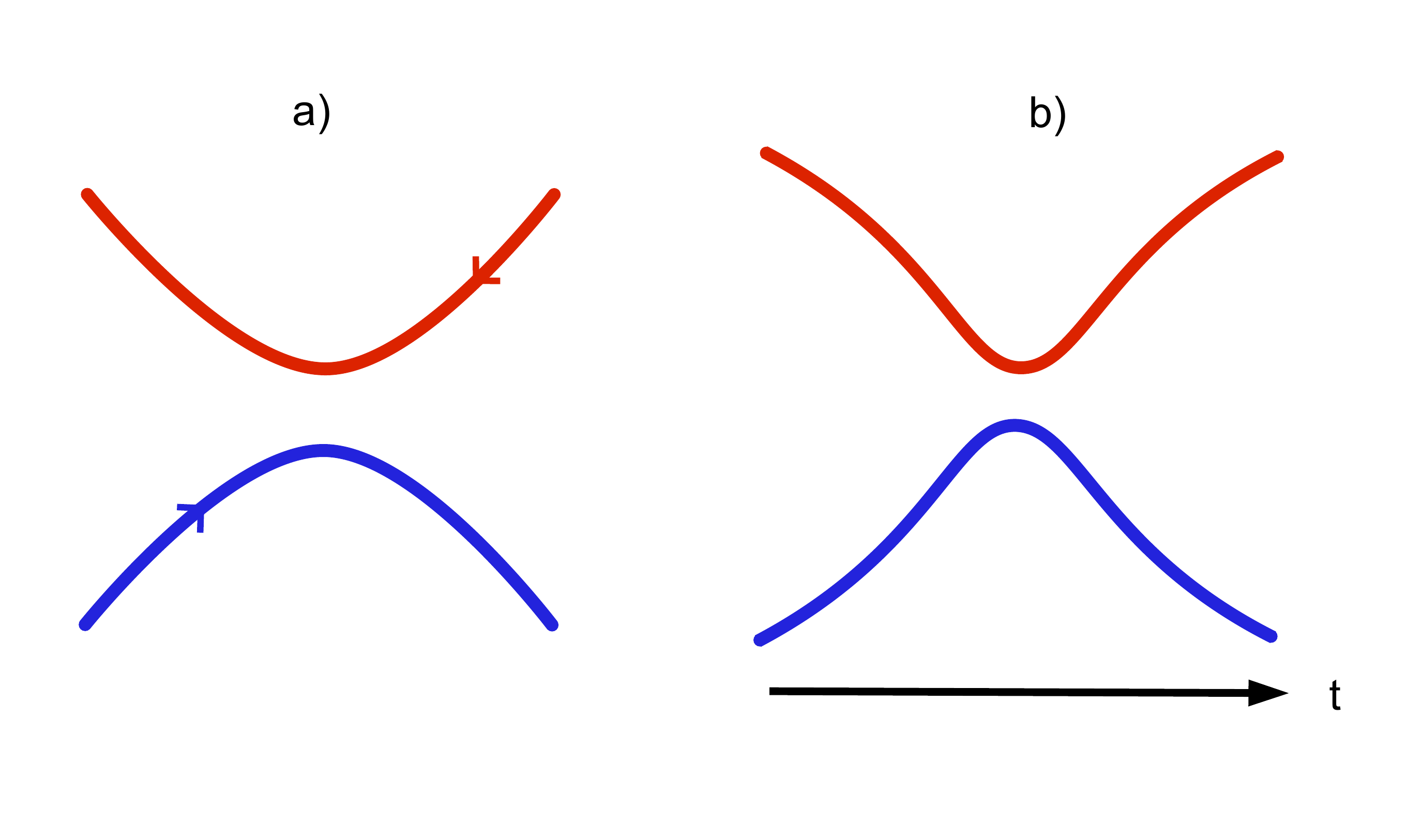}  
\caption{$m < 0$. Part a) depicts the motion of branch points of the $A_1$-curve on the $\IC$-plane while part b) describes the motion as branch lines in $\IC \times \IR$.}
\label{figure:A1flow2}
\end{figure}

As a first example we will consider the most simple case of R-flow. This is the theory corresponding to the curve
\begin{equation}
	y^2 = x^2  + \epsilon,
\end{equation}
with a single central charge, denoted by $Z_1$, given by
\begin{equation}
	Z_1 = \int_{-\sqrt{\epsilon}}^{\sqrt{\epsilon}} \sqrt{x^2 + \epsilon} dx = - \frac{\pi i}{2} \epsilon.
\end{equation}
We will find that this theory has significant importance for the resolution of arbitrary singular tangles as it predicts the possible local resolutions of an isolated singularity by turning on different values of Fayet-Iliopoulos parameters. Let us describe how this comes by. First of all, note that we can parametrize $\epsilon$ as $\epsilon = - \frac{2}{\pi} (-i m + t)$ with $m$ real and positive so that 
\begin{equation}
	Z_1 = m + i t, \quad m > 0,
\end{equation}
obeys the flow equation (\ref{eq:Rflow})\footnote{Note that the derivative with respect to $t$ is not equal to $m$. This is no problem however, as this condition was imposed initially to maintain the order of the central charges along the flow. But as the $A_1$ curve has just one central charge, we just demand that the real part stays constant.}. The motion of the branch points of the curve are then given by the law
\begin{equation} \label{eq:A1flow}
	a_1 = \alpha \sqrt{m + i t}, \quad a_2 = - \alpha \sqrt{m + i t},
\end{equation}
where $\alpha$ is a proportionality constant. We can now view this motion from two perspectives. The first is as a motion on the $\IC$-plane which forms the base of the double cover. The second perspective is obtained by looking at the motion of the two branch points as giving rise to branch lines in $\IC \times \IR$ where $\IR$ is the \textit{time}-direction parametrized by $t$. As the square root behaviour of (\ref{eq:A1flow}) is fairly simple we can depict the two perspectives easily as shown below in Figure \ref{figure:A1flow1}.
\begin{figure}[h!]
\center\includegraphics[width=.5\textwidth]{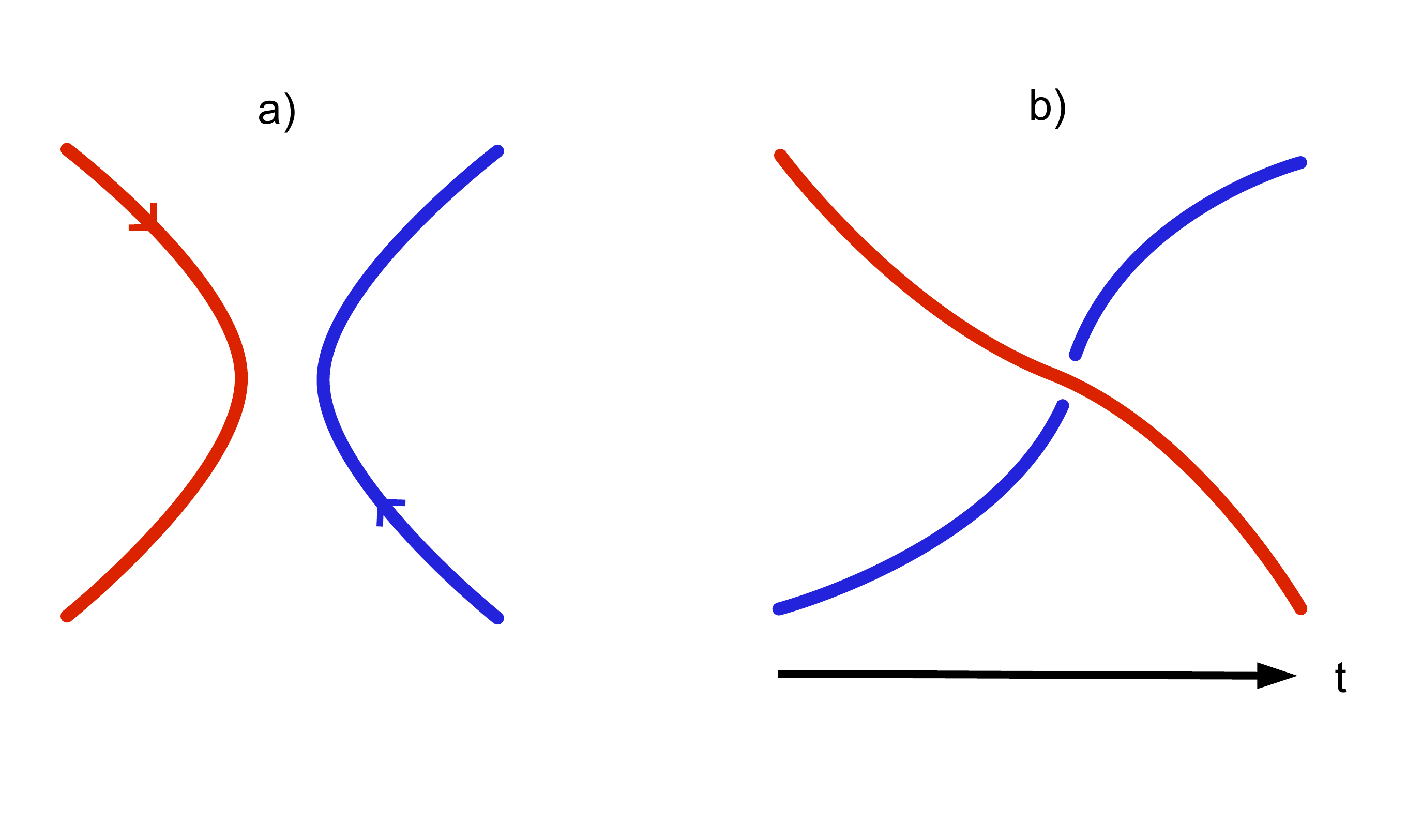}  
\caption{$m> 0$. Part a) depicts the motion of branch points of the $A_1$-curve on the $\IC$-plane while part b) describes the motion as branch lines in $\IC \times \IR$.}
\label{figure:A1flow1}
\end{figure}
A very interesting phenomenon happens when we flip the sign of the real part of the central charge, i.e. if we choose $m < 0$ instead. Fixing the projection plane, we now obtain the following picture for the branch-point flow (Figure \ref{figure:A1flow2}).

We see that this exactly mirrors two of the three possible resolutions described in section \ref{joyce}, namely resolutions \ref{figure:desingularizations} (b) and (c). Note that resolution (a) cannot be obtained in this formalism as it breaks time-flow or equivalently keeps the mass parameter $m$ at zero but deforms the theory onto the Higgs branch.

\subsubsection*{$A_2$ flow}
\label{a2flow}
We now turn to our next example, the $A_2$ curve. It is, apart from the $A_1$ case, the most important flow example as it provides insight into three-dimensional mirror symmetry in terms of flows of four-dimensional theories. In order to illustrate this we consider the two central charge orderings of this theory which provide two BPS chambers with different particle content. More precisely, we have a 2-particle chamber:
\begin{equation}
	\textrm{arg}Z_1 < \textrm{arg}Z_2 < 0, 
\end{equation}
and a three-particle chamber
\begin{equation}
	\textrm{arg}Z_2 < \textrm{arg}Z_1 < 0,
\end{equation}
where the third state is the one with charge $Z_1 + Z_2$. Looking at the Kontsevich-Soibelmann operator we see that in the first case it is given by
\begin{equation}
	E_+(\hat{p}) E_+(\hat{x}),
\end{equation}
while in the second case one has
\begin{equation}
	E_+(\hat{x}) E_+(\hat{x}+\hat{p}) E_+(\hat{p}).
\end{equation}
The crucial point here is that these two operators are actually equal if we impose the commutator 
\begin{equation}
	[\hat x, \hat p] = \frac{i}{2\pi},
\end{equation}
as was first proven in \cite{Faddeev:2000if}. This is the underlying equality leading to the 3-2-move discussed in section \ref{sec:pentagonmove}. Therefore, the 3-2-move can actually be thought of as arising from R-flow of the $A_2$ curve. However, note that the 3-2 move is obtained by looking at matrix elemts $\langle x| \IK | p\rangle$, that is position/momentum matrix elements, whereas R-flow is equivalent to matrix elements of the form $\langle x | \IK | y \rangle$, namely position/position matrix elements. Furthermore, there are many braid realizations of these matrix elements differing by the other various dualities discussed in section \ref{sec:singreidemeister}. In this section we will look at representations which are obtained from the prescription described in Table \ref{table:localsing}. That is, we will now look at the above KS-operators and their braid realizations from the perspective of branch-point flow. 

\begin{figure}[h!]
\center\includegraphics[width=.5\textwidth]{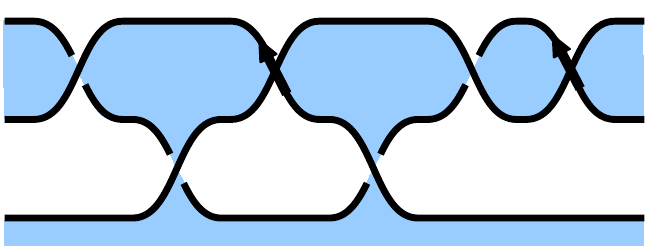}  
\caption{The singularized braid of the A2 flow in the minimal chamber.}
\label{figure:A2braidmch}
\end{figure}

\begin{figure}[here!]
  \centering
  \subfloat[Braid resolution]{\label{fig:bpsinstproj1}\vcenteredhbox{\includegraphics[ width=0.5\textwidth]{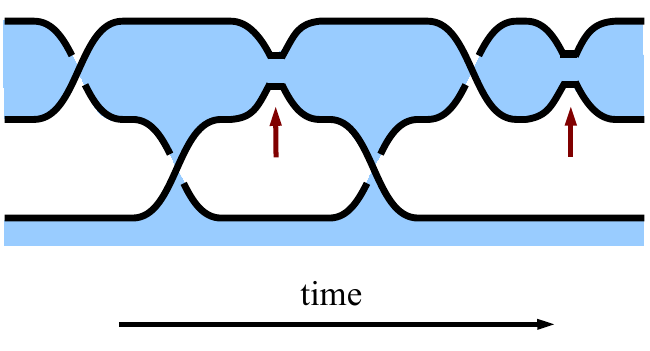}}}     
  \hspace{.25in}       
  \subfloat[Branch point flow]{\label{fig:bpsinstproj2}\vcenteredhbox{\includegraphics[ width=0.3\textwidth]{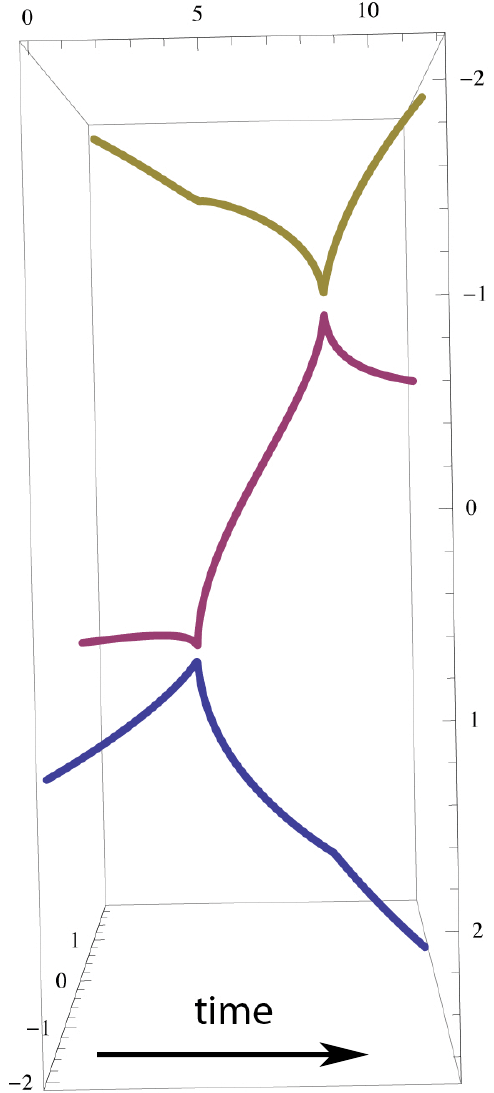}}}
  \caption{R-flow of $A_2$ in the minimal chamber. In (a) the resolved braid braid is depicted. The previous singularities appear now as cusps in the braid diagram. These are marked with red arrows. Part (b) shows the flow of branch points obtained by explicitly solving equation (\ref{eq:Rflow}).}
  \label{figure:A2braidresmch}
\end{figure}

Let us start with the minimal particle chamber. Using the identity
\begin{equation}
	E_+(\hat p) = e^{i \pi \hat x^2} e^{i \pi \hat p^2} e^{i \pi \hat x^2} E_+(\hat x) e^{-i \pi \hat x^2} e^{-i \pi \hat p^2} e^{-i \pi \hat x^2},
\end{equation}
we obtain
\begin{eqnarray}\label{ZA2mch}
	\langle x | E_+(\hat p) E_+(\hat x) | y \rangle & = & \langle x| e^{i \pi \hat x^2} e^{i \pi \hat p^2} e^{i \pi \hat x^2} \hat x e^{-i \pi \hat x^2} e^{-i \pi \hat p^2} e^{-i \pi \hat x^2} E_+(y) | y\rangle \nonumber \\
	~ & = & \int d x' \langle x| \sigma_1^{-1} \sigma_2^{-1} E_+(x') |x'\rangle \langle x'| \sigma_2 \sigma_1 E_+(y) |y \rangle.
\end{eqnarray}
This way we have rewritten the partition function in terms the $\sigma_i$ which describe actions of the braid group. The braid representation of the right-hand side of the above identity is shown in Figure \ref{figure:A2braidmch}\footnote{We have suppressed the R-charges of the singularities as these are not relevant for the present discussion.}.
The single integration variable in (\ref{ZA2mch}) corresponds to a $U(1)$ gauge group manifest as a compact white region in Figure \ref{figure:A2braidmch}. Furthermore, we have used that $\sigma_1$ and $\sigma_1^{-1}$ commute with $E_+(\hat x)$ and therefore cancel each other. Note that the theory described by the braid \ref{figure:A2braidmch} is related to $U(1)$ SQED by changing the branch sheet as discussed in section \ref{branchsheetchange}\footnote{We also need to apply an $S$-transformation to the boundary condition in order to switch from position boundary to momentum boundary.}. We will not discuss this here and rather turn our attention to a particular resolution of the singular braid.  Applying resolution rule (b) of Figure \ref{figure:desingularizations} to all singularities we obtain the Figure \ref{figure:A2braidresmch}. It is also possible to explicitly solve equation (\ref{eq:Rflow}) and compute the flow of branch points in the minimal chamber. The result is shown in the second part of Figure \ref{figure:A2braidresmch}. We see that the resolved braid and the flow of branch points are topologically equivalent and just differ by change of projection plane. That is the location of particles is represented in both pictures by cusps at which the same strands come closest.

\begin{figure}[h!]
\center\includegraphics[width=.6\textwidth]{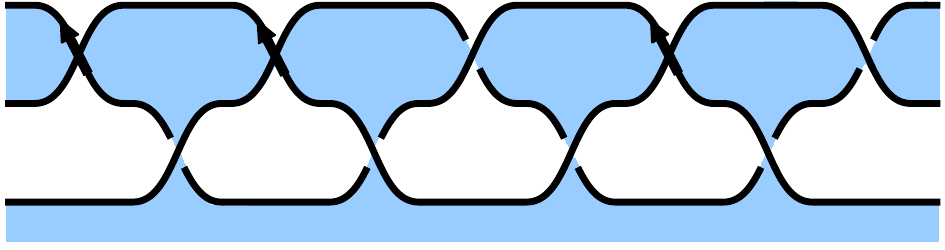}  
\caption{The singular braid of the A2 flow in the maximal chamber.}
\label{figure:A2braidmaxch}
\end{figure}

Next, we turn to the maximal chamber. Here, we need further the following identity
\begin{equation}
	E_+(\hat x \hat p) = e^{i\pi \hat p^2} E_+(\hat x) e^{-i \pi \hat p^2},
\end{equation}
which allows us to rewrite the partition function as
\begin{eqnarray}
	~ & ~ & \langle x| E_+(\hat x) E_+(\hat x + \hat p) E_+(\hat p) | y \rangle \nonumber \\
	~ & = & \langle x | E_+(\hat x) e^{i \pi \hat p^2} E_+(\hat x) e^{-i \pi \hat p^2} e^{i \pi \hat x^2} e^{i \pi \hat p^2} e^{i \pi \hat x^2} E_+(\hat x) e^{-i \pi \hat x^2} e^{- i \pi \hat p^2} e^{-i \pi \hat x^2} | y \rangle.
\end{eqnarray}
We depict the corresponding braid representation in Figure \ref{figure:A2braidmaxch}. 
One can immediately extract from this singular braid the presence of three $U(1)$ gauge fields corresponding to the three white regions in the diagram. Moreover, we see that two of the chiral fields are charged under two $U(1)$'s. Again, by a change of branch sheet (see section \ref{branchsheetchange}) and and S-duality at the boundary, we can transform this picture to the one corresponding to the $XYZ$ model discussed in section \ref{sec:pentagonmove}. We will not discuss this here but will rather analyse the connection to R-flow as branch point flow. This connection is established by looking at the particular resolution of the singular braid which corresponds to R-flow of branch points. This resolution is depicted below in Figure \ref{figure:A2braidresmaxch}. The second part of Figure \ref{figure:A2braidresmaxch} shows the flow of branch points obtained by explicitly solving equation (\ref{eq:Rflow}) in the maximal chamber. Again we see that the two Figures are topologically identical.

\begin{figure}[here!]
  \centering
  \subfloat[Braid resolution]{\label{fig:bpsinstproj1}\vcenteredhbox{\includegraphics[ width=0.6\textwidth]{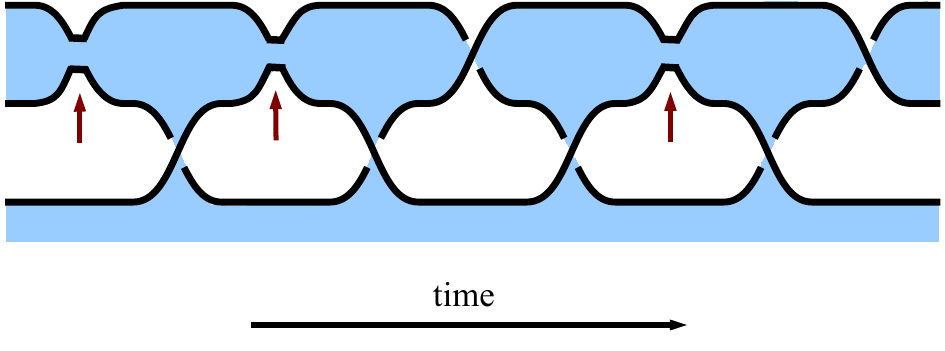}}}     
  \hspace{.25in}       
  \subfloat[Branch point flow]{\label{fig:bpsinstproj2}\vcenteredhbox{\includegraphics[ width=0.3\textwidth]{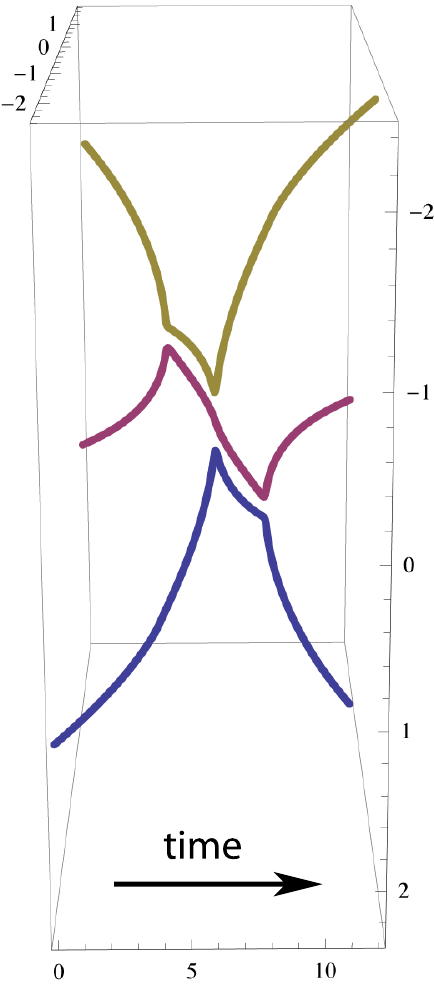}}}
  \caption{R-flow of $A_2$ in the maximal chamber. Part (a) shows the resolved braid. The locations of previous singularities are marked with red arrows. Part (b) depicts the flow of branch points as arising from a flow of central charges along straight vertical lines.}
  \label{figure:A2braidresmaxch}
\end{figure}

\subsubsection*{$A_3$ flow}
\label{a3flow}
\begin{figure}[h!]
\center\includegraphics[width=.5\textwidth]{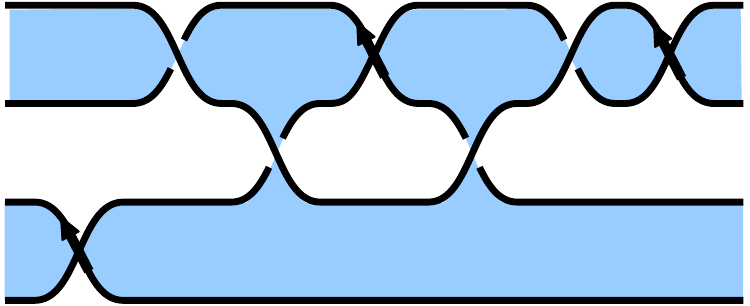}  
\caption{The singularized braid of the A3 flow in the minimal chamber.}
\label{figure:A3braid}
\end{figure}

\begin{figure}[h!]
\center\includegraphics[width=.5\textwidth]{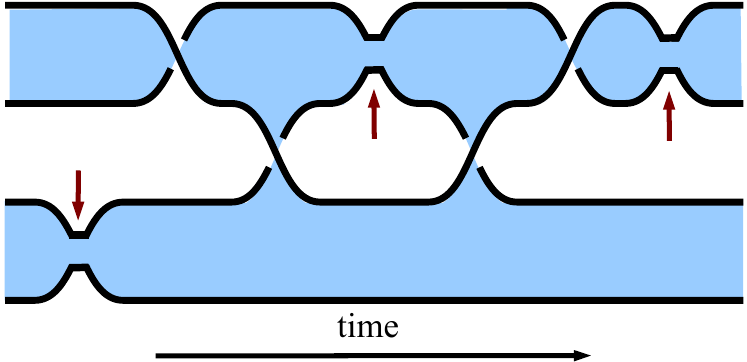}  
\caption{The desingularized braid of the A3 flow in the minimal chamber. The locations where the particles used to be are indicated by red arrows.}
\label{figure:A3braidres}
\end{figure}

\begin{wrapfigure}{r}{0.5\textwidth}
	\begin{center}
		\includegraphics[width=.3\textwidth]{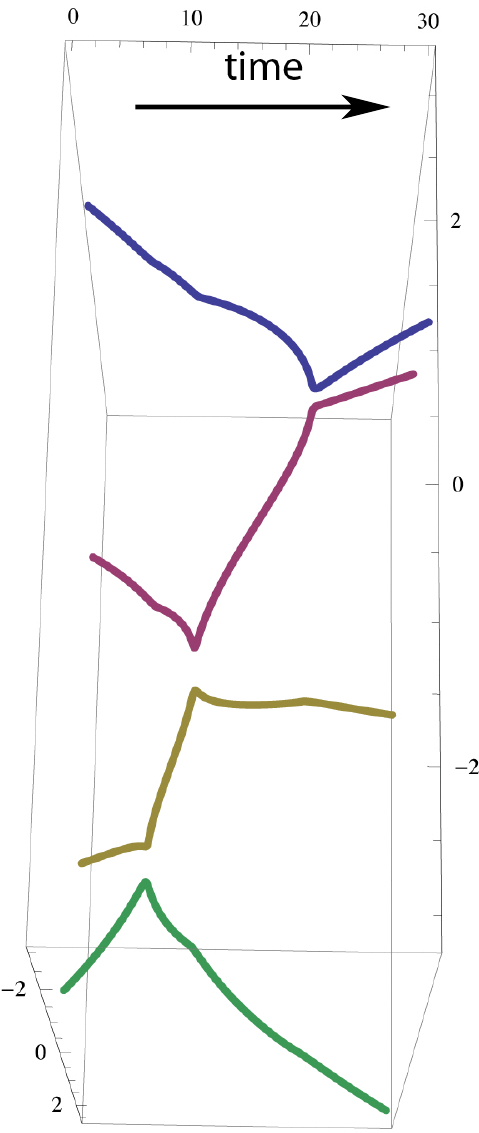}  
	\end{center}
	\caption{Flow of branch points of the minimal chamber $A_3$ by imposing a flow of central 			charges along vertical straight lines.}
\label{figure:A3flowgraph}
\end{wrapfigure}

Next in complexity is the $A_3$-flow. For clarity of presentation, we will solely concentrate on the flow in the minimal BPS chamber here. There are three cycles corresponding to the operators $\hat \gamma_1=\hat{x}$, $\hat \gamma_2=\hat{p}$, and $\hat \gamma_3=\hat x + c$, which form a central extension of the $SL(2,\IZ)$-algebra generated by $\hat{x}$ and $\hat{p}$ with commutators\footnote{We have chosen here a different commutator between $\hat  x$ and $\hat p$ compared to the $A_2$ case. This is merely a convention. We could also have worked with the former commutator.}
\begin{equation}
	[\hat{x},\hat{p}] = -\frac{i}{2\pi} , \quad [c,\hat x] = [c, \hat p] = 0.
\end{equation}
The KS-operator corresponding to the minimal particle chamber is given by
\begin{equation}
	\IK = E_+(\hat x + c) E_+(\hat p) E_+ (\hat x).
\end{equation}
A partition function can be formed from this operator by considering the wave-function
\begin{equation}
	\cZ_{\IK} = \langle x | E_+(\hat x + c) E_+(\hat p) E_+ (\hat x) | y \rangle.
\end{equation}
This partition function now represents a singular braid. In order to extract the braid, we have to rewrite it as a gluing of simple partition functions containing no gauge groups. This is done by using the identity
\begin{equation}
	E_+(\hat p) = e^{-i\pi \hat x^2} e^{-i\pi \hat p^2} e^{-i\pi \hat x^2} E_+(\hat x) e^{i\pi \hat x^2} e^{i\pi \hat p^2} e^{i\pi \hat x^2},
\end{equation}
which allows us to rewrite $\cZ_{\IK}$ in the form
\begin{equation}
	\cZ_{\IK} = \int dx' \langle x | E_+(x + c) e^{-i\pi \hat x^2} e^{-i\pi \hat p^2} e^{-i\pi \hat x^2} | x'\rangle \langle x'| E_+(x') e^{i\pi \hat x^2} e^{i\pi \hat p^2} e^{i\pi \hat x^2} E_+(y) |y \rangle.
\end{equation}
This partition function can be represented by the singularized braid shown in Figure \ref{figure:A3braid}.
We see again a $U(1)$ gauge group corresponding to the one compact white region. Furthermore, a chiral field is charged under this gauge group while the two other chiral fields are gauge neutral. Applying duality rules we can transform this picture to different ones with more or less gauge groups. Applying resolution rule (b) of Figure \ref{figure:desingularizations} to all singularities we obtain picture \ref{figure:A3braidres}.
This resolved braid can again be reproduced by letting the central charges of the $A_3$ curve R-flow as depicted in Figure \ref{figure:chargeflow}. One can carry out the flow procedure by inverting the central charges as functions of the branch points locally along the flow. The resulting flow of branch points for the minimal chamber is depicted in Figure \ref{figure:A3flowgraph}.
Comparing Figure \ref{figure:A3braidres} with Figure \ref{figure:A3flowgraph} we find that the two are topologically identical in that the strands which come closest at the location of particles are the same in both pictures, i.e. first $\gamma_3$ contracts, then $\gamma_2$ and at last $\gamma_1$. They merely differ by a change of the projection plane. 

We find that this behavior generalizes. That is, associated to the KS-operator corresponding to the $A_n$ theory in a particular chamber, there exists a resolution which arises as R-flow of the branch points. The prescription for finding the resolution corresponding to R-flow is as follows. Start with the partition function
\begin{equation}
	\cZ_{A_n} = \langle x |\mathbb{K}(q) | x' \rangle.
\end{equation}
Associate to this matrix element the particular braid-representation which contains all particles as black dots \textit{within} the Seifert-surface, where by \textit{within} we mean that the Seifert-surface goes \textit{horizontally} through the dot as depicted in table \ref{table:localsing}. Apply resolution rule of Figure \ref{figure:desingularizations} (b). Note that it is not possible to obtain other resolutions for the singular braids such as the one of figure \ref{figure:A3braid} from R-flow. The reason is that a local flip of the corresponding central charge, as described in the case of $A_1$, changes the KS-operator and will thus lead to a completely different picture.

\section{Applications}
\label{sec:app}
In this section we study some further applications of the developed rules. As a first example we examine a more complicated geometry arising from the R-flow prescription. The particular geometry contains a closed non-planar polygon, i.e. a superpotential, which is only partly shaded and thus gives rise to a monopole operator. We will establish that this monopole operator appears in the superpotential. As a second example for the application of the methods developed in this paper we will look at $U(1)$ SQED with $N_f > 1$. This example does not arise from R-flow. However, we will find that the rules presented in section \ref{sec:singreidemeister} are powerful enough to establish mirror symmetry even for these more complicated models geometrically.  

\subsection{Superpotentials from R-flow}
\label{sec:rflowsp}
\begin{figure}[h!]
\center\includegraphics[width=0.8 \textwidth]{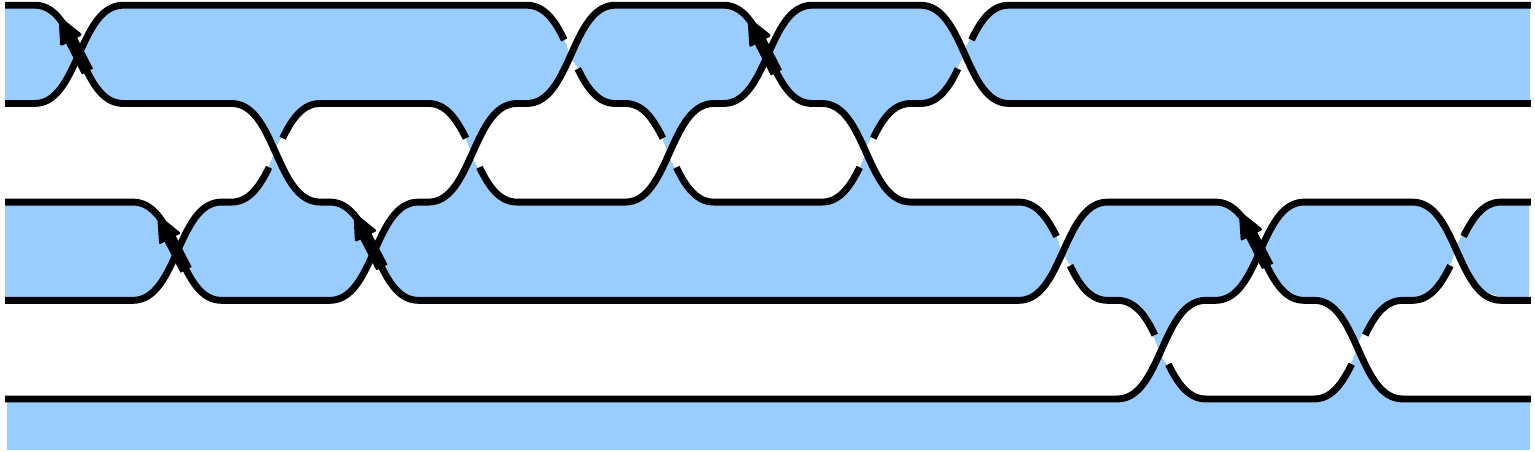}  
\caption{R-flow of $A_4$ in intermediate chamber.}
\label{figure:A4braid}
\end{figure}

\begin{figure}[here!]
  \centering
  \subfloat[$A_4$ braid]{\label{fig:bpsinstproj1}\vcenteredhbox{\includegraphics[ width=0.4\textwidth]{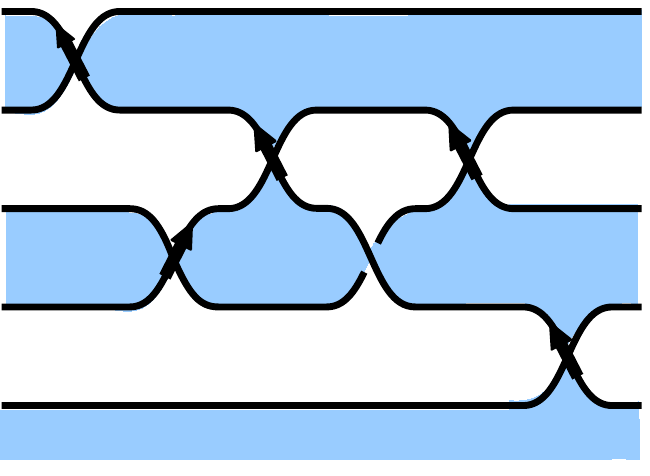}}}     
  \hspace{.25in}       
  \subfloat[$A_4$ braid with superpotential]{\label{fig:bpsinstproj2}\vcenteredhbox{\includegraphics[ width=0.4\textwidth]{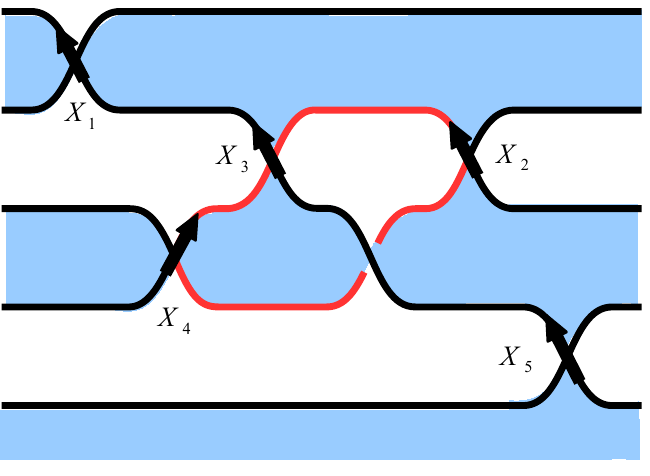}}}
  \caption{R-flow of $A_4$ in intermediate chamber, second representation. In (a) we see a dual representation of the $A_4$ braid after application of various dualities to the original R-flow braid. In (b) we see the same dual braid, now with the closed region representing the superpotential highlighted in red. Chiral Fields are indicated by $X_i$.}
  \label{figure:A4braid2}
\end{figure}

In this section we look at an example of a 3d gauge theory which arises from R-flow of an intermediate chamber of the $A_4$ theory. This example was already analyzed to some extent in \cite{CCV}. The relevant KS-operator is given by
\begin{equation}
	\IK = E_+(\hat x_1) E_+(\hat x_2) E_+(\hat p_1 + \hat x_2) E_+(\hat x_2) E_+(\hat p_2),
\end{equation}
where the phase space parameters satisfy the following commutation relations
\begin{equation}
	[\hat x_1, \hat p_1] = \frac{i}{2\pi}, \quad [\hat p_1, \hat x_2] = \frac{i}{2\pi}, \quad [\hat x_2, \hat p_2] = \frac{i}{2\pi}.
\end{equation}
The 3d partition function associated to the KS-operator is now
\begin{equation}
	\cZ_{\IK} = \langle x| E_+(\hat x_1) E_+(\hat x_2) E_+(\hat p_1 + \hat x_2) E_+(\hat x_2) E_+(\hat p_2) |x'\rangle.
\end{equation}
Its representation in terms of a singular braid is depicted in figure \ref{figure:A4braid}. 
We can clearly see 4 $U(1)$ gauge groups represented by the four white regions in the braid. Applying the Fourier transform identity twice and the $T$-transform rule of section \ref{sec:singreidemeister} we obtain the simpler braid depicted in Figure \ref{figure:A4braid2}.
This braid represents a dual description of the same quantum field theory. In this description, there is a $U(1)$ gauge group under which two chiral multiplets, denoted by $X_3$ and $X_2$, are charged oppositely. Furthermore, one can clearly see a compact polygonal region bounded by three chiral singularities. This corresponds to a superpotential in the effective 3d gauge theory to which all three chiral multiplets contribute. This theory contains a monopole operator which also participates in the superpotential term. One way to see this, is through the white region contained within the bounded polygonal region.  One can check, using the formula \eqref{oneloopm} for the charge of the monopole operator discussed in section \ref{sec:potential}, that the monopole operator $\cM$ is invariant under the $U(1)$ gauge group. This immediately tells us that we can write down a superpotential of the form
\begin{equation}
	\cW = \cM X_2 X_3 X_4,
\end{equation}
which is gauge invariant. Furthermore, this superpotential breaks exactly one $U(1)$ flavor symmetry which is consistent as there are five chiral fields but only four non-compact white regions in the geometry.

\subsection{$U(1)$ SQED with $N_f > 1$}
\label{sec:u1sqed}

Here, we will demonstrate that our rules for the singular tangles provide a convenient geometric way of encoding general mirror symmetries of 3d $N=2$ gauge theories. The example we will use to demonstrate this is the generalization of $U(1)$ SQED/$XYZ$ mirror symmetry. Start with a 3d $N=2$ gauge theory with $U(1)$ gauge group and $N_f > 1$ charged hypermultiplets. This theory has a RG fixed point with a mirror dual description as a $(U(1)^{N_f})/U(1)$ gauge theory with $N_f$ charged hypermultiplets (consisting of chiral multiplets $q_i$ and $\tilde q_i$) and $N_f$ neutral chiral multiplets $S_i$ together with a superpotential \cite{XYZ}
\begin{equation}
	W = \sum_{i=1}^{N_f} S_i q_i \tilde q_i.
\end{equation}
The charge assignments are as follows
\begin{equation} \label{table:mirrorcharges}
	\begin{array}{|c|ccccc|}
	    \hline
		~                   & U(1)_1 & U(1)_2 & U(1)_3 & \cdots & U(1)_{N_f} \\
		\hline
		q_1               & 1         & -1       & 0         & \cdots & 0 \\
		\tilde q_1      & -1       & 1        & 0         & \cdots  & 0 \\
		q_2               & 0         & 1         & -1       & \cdots & 0\\
		\tilde q_2      & 0        & -1        & 1        &  \cdots & 0\\
		\vdots           & ~        & \ddots  & ~       &   ~        & 0\\
		q_{N_f}           & -1      &  0         & 0        & \cdots & 1\\
		\tilde q_{N_f} & 1        & 0          & 0        & \cdots & -1\\
		S_i                &  0        & 0          & 0        & \cdots & 0\\
		\hline
	\end{array}
\end{equation}
The aim will now be to translate both theories into geometric tangles and transform them into each other by using ordinary as well as singularized Reidemeister moves, thereby proofing they are mirror pairs. 

\subsubsection{$U(1)$ SQED with $N_f = 2$}

\begin{figure}[h!]
\center\includegraphics[width=.4\textwidth]{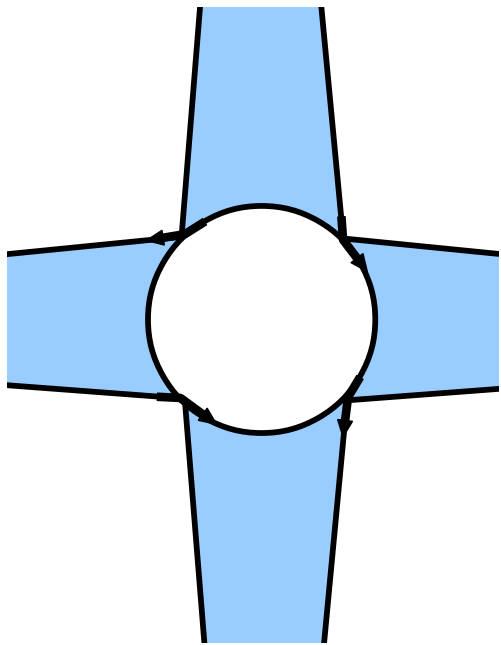}  
\caption{Diagram describing $U(1)$ SQED with $N_f = 2$.}
\label{figure:U1Nf2a}
\end{figure}

We will start with the geometry corresponding to $U(1)$ SQED and specialize to the case $N_f = 2$. The relevant diagram describing this gauge theory is depicted in Figure \ref{figure:U1Nf2a}.

\begin{figure}[h!]
\center\includegraphics[width=.4\textwidth]{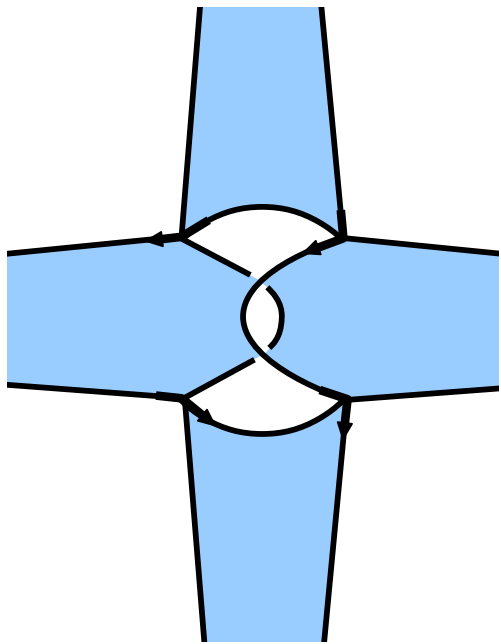}  
\caption{Application of second Reidemeister move.}
\label{figure:U1Nf2b}
\end{figure}

\begin{figure}[h!]
\center\includegraphics[width=.4\textwidth]{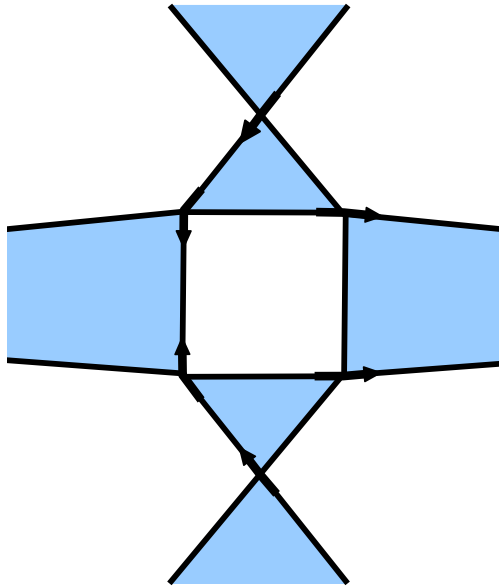}  
\caption{After applying twice the 3-2-move.}
\label{figure:U1Nf2c}
\end{figure}

The interior white region represents the $U(1)$ gauge group and each pair of singularities corresponds to a hypermultiplet whose constituents have opposite charges under the $U(1)$. Let us next apply the second Reidemeister move to this diagram. The result is depicted in Figure \ref{figure:U1Nf2b}.
Here we see that there are two extra $U(1)$'s and that two singularities are charged under the first one whereas the second pair is charged under the second. We are now in a position to apply the generalized Reidemeister move known as the 3-2 move. This move can be applied twice, once to the upper white triangle and once to the lower white triangle, resulting in Figure \ref{figure:U1Nf2c}.
This diagram simply shows a $U(1)$ gauge theory with two chiral fields charges positively under it and two fields charges negatively. Moreover, we observe two superpotential terms each combining a neutral field with two oppositely charged fields. These data exactly match those of the mirror dual which confirms the duality.

\newpage

\subsubsection{$U(1)$ SQED with $N_f = 3$}

As a second and last example we will consider the more complicated case of $U(1)$ SQED with $N_f  = 3$. The relevant diagram is

\begin{figure}[h!]
\center\includegraphics[width=.35\textwidth]{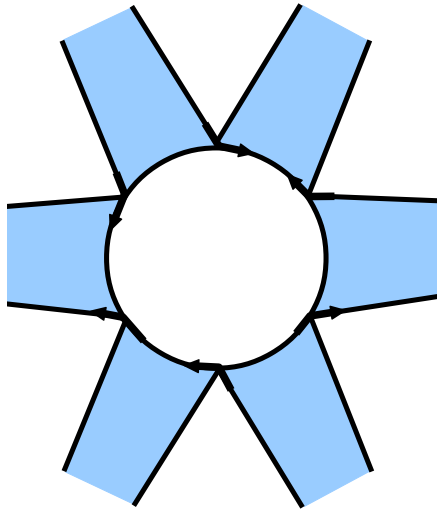}  
\caption{$U(1)$ SQED with $N_f = 3$.}
\label{figure:U1Nf3a}
\end{figure}

We can see 6 chiral multiplets charged under a $U(1)$ gauge group with the charges of the particles adding up to zero pairwise. The overcross and undercross singularities are arranged such that the net self-Chern-Simons level of the $U(1)$ is zero. 
\begin{figure}[h!]
\center\includegraphics[width=.35\textwidth]{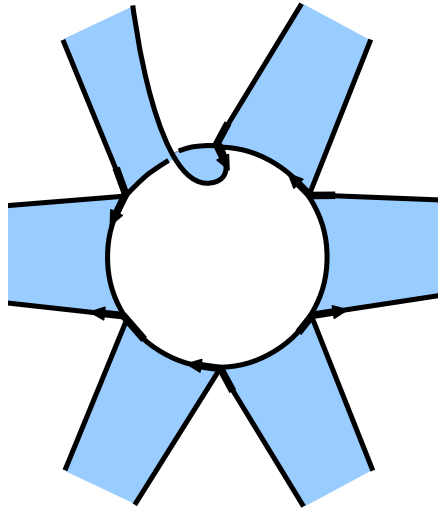}  
\caption{Adding a T-transform.}
\label{figure:U1Nf3b}
\end{figure}
We can add a T-transform to turn one type of singularity to another, as shown in Figure \ref{figure:U1Nf3b}. Next, we do a second Reidemeister move to create a white region.

\begin{figure}[h!]
\center\includegraphics[width=.4\textwidth]{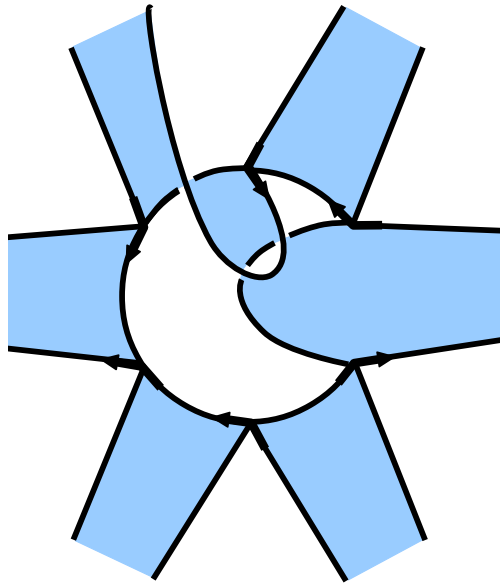}  
\caption{Applying a Reidemeister move.}
\label{figure:U1Nf3c}
\end{figure}

\begin{figure}[h!]
\center\includegraphics[width=.4\textwidth]{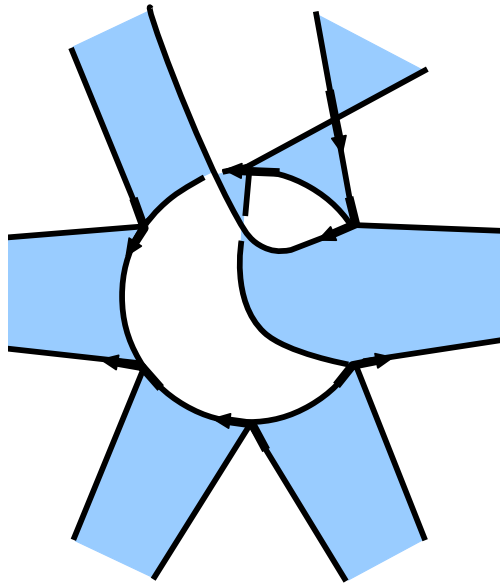}  
\caption{After first 3-2 move.}
\label{figure:U1Nf3d}
\end{figure}

Performing the 3-2 move we end up with a superpotential and an extra $U(1)$, shown in Figure \ref{figure:U1Nf3d}.

We now perform the Reidemeister move a second time to create a third white region with two charged fields.

\begin{figure}[h!]
\center\includegraphics[width=.4\textwidth]{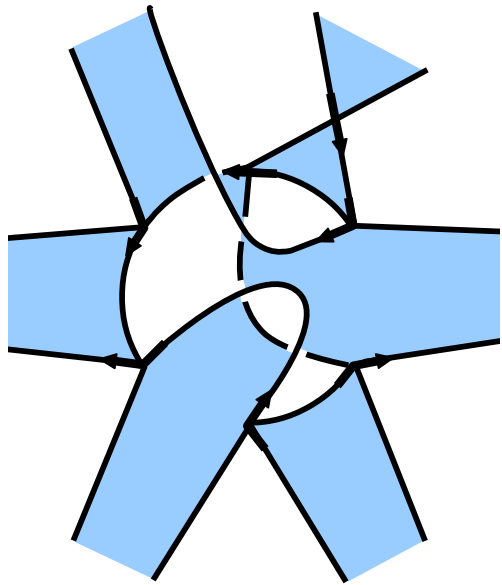}  
\caption{After second Reidemeister move.}
\label{figure:U1Nf3e}
\end{figure}

Application of the 3-2 move for a second time leads to the second superpotential term.

\begin{figure}[h!]
\center\includegraphics[width=.4\textwidth]{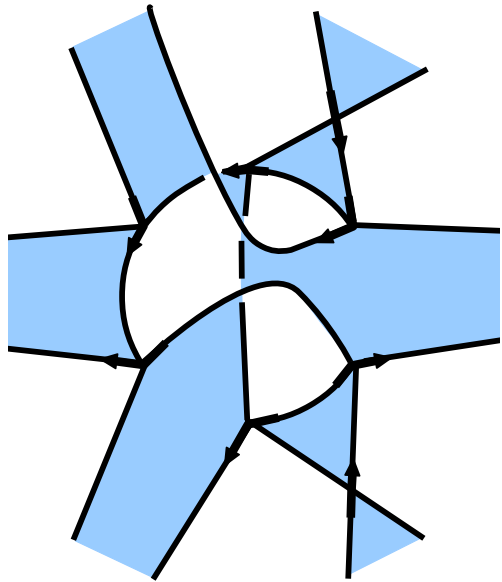}  
\caption{After second 3-2 move.}
\label{figure:U1Nf3f}
\end{figure}

\begin{figure}[h!]
\center\includegraphics[width=.4\textwidth]{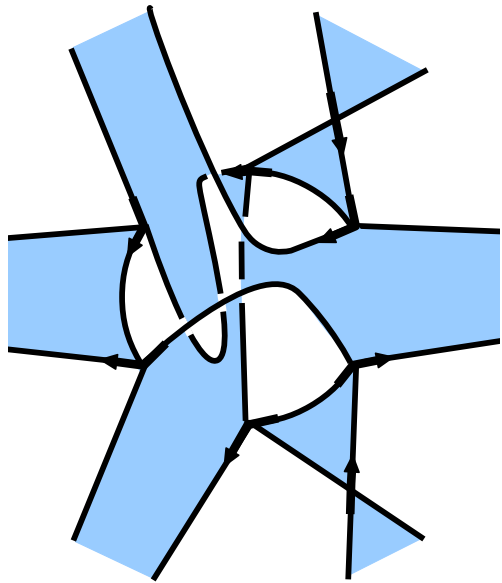}  
\caption{After third Reidemeister move.}
\label{figure:U1Nf3g}
\end{figure}

As should by now be obvious, we again perform the Reidermeister move with the result shown in Figure \ref{figure:U1Nf3g}.

\begin{figure}[here!]
\center\includegraphics[width=.4\textwidth]{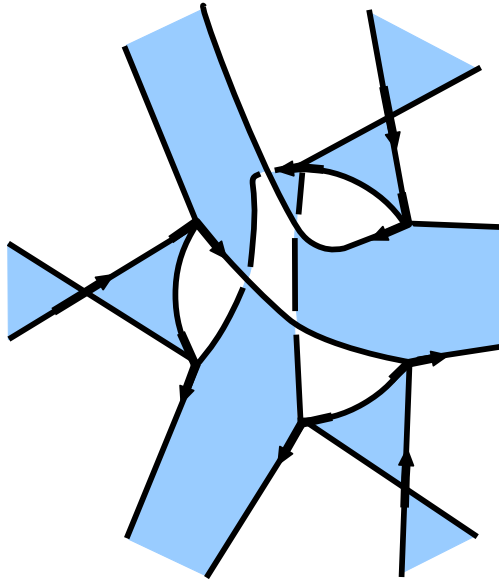}  
\caption{The mirror dual.}
\label{figure:U1Nf3h}
\end{figure}

The last step is again a 3-2 move leading to the final result depicted in Figure \ref{figure:U1Nf3h}.

As one can clearly see the above picture is the diagram describing the mirror dual of our original theory. We have three superpotentials each containing one neutral field and we have three $U(1)$'s under each of which 2 chiral fields are charged. Note that the white region in the interior, under which no particle is charged, ensures that the sum of all $U(1)$'s adds up to zero as required by the charge assignments shown in table (\ref{table:mirrorcharges}). Thus we see that the diagram captures the theory in all details. The constructions we have presented easily generalize to the case of arbitrary $N_f$.

\section*{Acknowledgements}

We thank K. Intrilligator, D. Jafferis, and A Klemm for helpful discussions. We also thank the 2012 Simons Workshop in Mathematics and Physics for hospitality during a portion of this work.    The work of C.C. is support by a Junior Fellowship at the Harvard Society of Fellows.  The work of B.H is supported by DFG fellowship HA6096/1-1.  The work C.V. is supported in part by NSF grant PHY-0244821.

\bibliography{Arxiv.bbl}{}

\providecommand{\href}[2]{#2}\begingroup\raggedright\begin{thebibliography}{10}

\bibitem{Nahm}
W.~{Nahm}, ``{Supersymmetries and their Representations},'' {\em Nucl. Phys. B}
  {\bfseries 135} (1977) 149.

\bibitem{KLMVW}
A.~Klemm, W.~Lerche, P.~Mayr, C.~Vafa, and N.~P. Warner, ``{Selfdual strings
  and N=2 supersymmetric field theory},''
  \href{http://dx.doi.org/10.1016/0550-3213(96)00353-7}{{\em Nucl.Phys.}
  {\bfseries B477} (1996) 746--766},
\href{http://arxiv.org/abs/hep-th/9604034}{{\ttfamily arXiv:hep-th/9604034
  [hep-th]}}.

\bibitem{DW}
R.~{Donagi} and E.~{Witten}, ``{Supersymmetric Yang-Mills theory and integrable
  systems},'' \href{http://arxiv.org/abs/hep-th/9510101}{{\ttfamily
  arXiv:hep-th/9510101 [hep.th]}}.

\bibitem{Witten1997}
E.~Witten, ``{Solutions of four-dimensional field theories via M theory},''
  \href{http://dx.doi.org/10.1016/S0550-3213(97)00416-1}{{\em Nucl.Phys.}
  {\bfseries B500} (1997) 3--42},
\href{http://arxiv.org/abs/hep-th/9703166}{{\ttfamily arXiv:hep-th/9703166
  [hep-th]}}.

\bibitem{Gaiotto}
D.~Gaiotto, ``{N=2 dualities},''
\href{http://arxiv.org/abs/0904.2715}{{\ttfamily arXiv:0904.2715 [hep-th]}}.

\bibitem{SW1}
N.~Seiberg and E.~Witten, ``{Electric - magnetic duality, monopole
  condensation, and confinement in N=2 supersymmetric Yang-Mills theory},''
  \href{http://dx.doi.org/10.1016/0550-3213(94)90124-4,
  10.1016/0550-3213(94)90124-4}{{\em Nucl.Phys.} {\bfseries B426} (1994)
  19--52},
\href{http://arxiv.org/abs/hep-th/9407087}{{\ttfamily arXiv:hep-th/9407087
  [hep-th]}}.

\bibitem{SW2}
N.~Seiberg and E.~Witten, ``{Monopoles, duality and chiral symmetry breaking in
  N=2 supersymmetric QCD},''
  \href{http://dx.doi.org/10.1016/0550-3213(94)90214-3}{{\em Nucl.Phys.}
  {\bfseries B431} (1994) 484--550},
\href{http://arxiv.org/abs/hep-th/9408099}{{\ttfamily arXiv:hep-th/9408099
  [hep-th]}}.

\bibitem{DGG1}
T.~{Dimofte}, D.~{Gaiotto}, and S.~{Gukov}, ``{Gauge Theories Labelled by
  Three-Manifolds},''
\href{http://arxiv.org/abs/1108.4389}{{\ttfamily arXiv:1108.4389 [hep-th]}}.

\bibitem{DGG2}
T.~{Dimofte}, D.~{Gaiotto}, and S.~{Gukov}, ``{3-Manifolds and 3d Indices},''
\href{http://arxiv.org/abs/1112.5179}{{\ttfamily arXiv:1112.5179 [hep-th]}}.

\bibitem{CCV}
S.~{Cecotti}, C.~{Cordova}, and C.~{Vafa}, ``{Braids, Walls, and Mirrors},''
  \href{http://arxiv.org/abs/1110.2115}{{\ttfamily arXiv:1110.2115 [hep-th]}}.

\bibitem{TBranes}
C.~{Cecotti}, C.~{Cordova}, J.~{Heckman}, and C.~{Vafa}, ``{T-Branes and
  Mondromy},''
\href{http://arxiv.org/abs/1010.5780}{{\ttfamily arXiv:1010.5780 [hep-th]}}.

\bibitem{WittenSL2}
E.~{Witten}, ``{SL(2,Z) Action on Three-Dimensional Conformal Field Theories
  With Abelian Symmetry},'' \href{http://arxiv.org/abs/0307.041.002}{{\ttfamily
  arXiv:0307.041.002 [hep.th]}}.

\bibitem{KapusStrass}
A.~{Kapustin} and M.~{Strassler}, ``{On Mirror Symmetry in Three Dimensional
  Abelian Gauge Theories},'' \href{http://arxiv.org/abs/9902.033}{{\ttfamily
  arXiv:9902.033 [hep.th]}}.

\bibitem{Redlich}
A.~{Redlich}, ``{Gauge Noninvariance and Parity Nonconservation of
  Three-Dimensional Fermions},'' {\em Phys. Rev. Lett.} {\bfseries 52} (1984)
  18--21.

\bibitem{IS}
K.~{Intrilligator} and N.~{Seiberg}, ``{Mirror Symmetry in Three-Dimensional
  Gauge Theories},'' \href{http://arxiv.org/abs/9607.207}{{\ttfamily
  arXiv:9607.207 [hep-th]}}.

\bibitem{XYZ}
O.~{Aharony}, A.~{Hanany}, K.~{Intrilligator}, N.~{Seiberg}, and
  M.~{Strassler}, ``{Aspects of N=2 Supersymmetric Gauge Theories in Three
  Dimensions},''
\href{http://arxiv.org/abs/9703.110}{{\ttfamily arXiv:9703.110 [hep-th]}}.

\bibitem{Jafferis}
D.~{Jafferis}, ``{The Exact Superconformal R-Symmetry Extremizes Z},'' {\em
  JHEP} {\bfseries 1205} (2010) 159,
  \href{http://arxiv.org/abs/1012.3210}{{\ttfamily arXiv:1012.3210 [hep-th]}}.

\bibitem{Kapustin}
A.~{Kapustin}, B.~{Willett}, and I.~{Yaakov}, ``{Exact Results for Wilson Loops
  in Superconformal Chern-Simons Theories with Matter},'' {\em JHEP} {\bfseries
  1003} (2010) 89, \href{http://arxiv.org/abs/0909.4559}{{\ttfamily
  arXiv:0909.4559 [hep-th]}}.

\bibitem{Faddeev:2000if}
{{Faddeev},L.D. and {Kashaev}, R.M. and {Volkov}, A.Y.}, ``{Strongly coupled
  quantum discrete Liouville theory. 1. Algebraic approach and duality},'' {\em
  Commun. Math. Phys.} {\bfseries 219} (2001) 199,
  \href{http://arxiv.org/abs/hep-th/0006156}{{\ttfamily hep-th/0006156}}.

\bibitem{MooreCS}
D.~{Belov} and G.~{Moore}, ``{Classification of Abelian Spin Chern-Simons
  Theories},''
\href{http://arxiv.org/abs/0505.235}{{\ttfamily arXiv:0505.235 [hep-th]}}.

\bibitem{contactterms}
C.~{Closset}, T.~{Dumitrescu}, G.~{Festuccia}, and N.~{Komargodski},
  Z.~{Seiberg}, ``{Contact Terms, Unitarity, and F-Maximization in
  Three-Dimensional Superconformal Theories},'' {\em JHEP} {\bfseries 1210}
  (2012) 53, \href{http://arxiv.org/abs/1205.4142}{{\ttfamily arXiv:1205.4142
  [hep.th]}}.

\bibitem{WittenFiveBrane}
E.~{Witten}, ``{Five-brane effective action in M theory},'' {\em J. Geom.
  Phys.} {\bfseries 22} (1996) 103,
  \href{http://arxiv.org/abs/hep-th/9610234}{{\ttfamily arXiv:hep-th/9610234
  [hep.th]}}.

\bibitem{Birman}
J.~{Birman}, {\em Braids, Links, and Mapping Class Groups}.
\newblock Princeton University Press, 1974.

\bibitem{Lickorish}
R.~{Lickorish}, {\em An Introduction to Knot Theory}.
\newblock Springer-Verlag, 1997.

\bibitem{AGT}
L.~{Alday}, D.~{Gaiotto}, and Y.~{Tachikawa}, ``{Liouville Correlation
  Functions from Four-dimensional Gauge Theories},'' {\em Lett. Math. Phys.}
  {\bfseries 91} (2010) 167, \href{http://arxiv.org/abs/0906.3219}{{\ttfamily
  arXiv:0906.3219 [hep.th]}}.

\bibitem{CV09}
S.~Cecotti and C.~Vafa, ``{BPS Wall Crossing and Topological Strings},''
\href{http://arxiv.org/abs/0910.2615}{{\ttfamily arXiv:0910.2615 [hep-th]}}.

\bibitem{BDP}
C.~{Beem}, T.~{Dimofte}, and S.~{Pasquetti}, ``{Holomorphic Blocks in Three
  Dimensions},'' \href{http://arxiv.org/abs/1211.1986}{{\ttfamily
  arXiv:1211.1986 [hep.th]}}.

\bibitem{WittenPhase2}
E.~Witten, ``{Phase transitions in M theory and F theory},''
\href{http://arxiv.org/abs/9603150}{{\ttfamily arXiv:9603150 [hep-th]}}.

\bibitem{AVTOP}
M.~{Aganagic} and C.~{Vafa}, ``{Mirror Symmetry, D-Branes, and Counting
  Holomorphic Discs},'' \href{http://arxiv.org/abs/0012.041}{{\ttfamily
  arXiv:0012.041 [hep.th]}}.

\bibitem{Joyce}
D.~Joyce, ``{On counting special Lagrangian homology 3-spheres},'' {\em
  Contemp. Math.} {\bfseries 314} (2002) 125--151,
\href{http://arxiv.org/abs/hep-th/9907013}{{\ttfamily arXiv:hep-th/9907013}}.

\bibitem{HHL}
{N.~Hama, K.~Hosomichi and S.~Lee}, ``{SUSY Gauge Theories on Squashed
  Three-Spheres},'' {\em JHEP} {\bfseries 1105} (2011) 014,
  \href{http://arxiv.org/abs/1102.4716}{{\ttfamily arXiv:1102.4716 [hep-th]}}.

\bibitem{Pasquetti}
S.~{Pasquetti}, ``{Factorization of N=2 Theories on the Squashed 3-Sphere},''
  {\em JHEP} {\bfseries 1204} (2012) 120,
  \href{http://arxiv.org/abs/1111.6905}{{\ttfamily arXiv:1111.6905 [hep.th]}}.

\bibitem{HW}
A.~{Hanany} and E.~{Witten}, ``{Type IIB superstrings, BPS monopoles, and
  three-dimensional gauge dynamics},''
  \href{http://arxiv.org/abs/hep-th/9611230}{{\ttfamily arXiv:hep-th/9611230
  [hep.th]}}.

\bibitem{CNV}
S.~Cecotti, A.~Neitzke, and C.~Vafa, ``{R-Twisting and 4d/2d
  Correspondences},''
\href{http://arxiv.org/abs/1006.3435}{{\ttfamily arXiv:1006.3435 [hep-th]}}.

\bibitem{Gaiottowall}
N.~{Drukker}, D.~{Gaiotto}, and J.~{Gomis}, ``{The Virtue of Defects in 4D
  Gauge Theories and 2D CFTs},''
\href{http://arxiv.org/abs/1003.1112}{{\ttfamily arXiv:1003.1112 [hep-th]}}.

\bibitem{Janus1}
D.~{Gaiotto} and E.~{Witten}, ``{Janus Configurations, Chern-Simons Couplings,
  And the Theta-Angle in N=4 Super Yang-Mills Theory},''
  \href{http://arxiv.org/abs/0804.2907}{{\ttfamily arXiv:0804.2907 [hep.th]}}.

\bibitem{Janus2}
D.~{Gaiotto} and E.~{Witten}, ``{S-Duality of Boundary Conditions In N=4 Super
  Yang-Mills Theory},'' \href{http://arxiv.org/abs/0807.3720}{{\ttfamily
  arXiv:0807.3720 [hep.th]}}.

\bibitem{CV11}
S.~Cecotti and C.~Vafa, ``{Classification of complete N=2 supersymmetric
  theories in 4 dimensions},''
\href{http://arxiv.org/abs/1103.5832}{{\ttfamily arXiv:1103.5832 [hep-th]}}.

\bibitem{AD}
P.~C. Argyres and M.~R. Douglas, ``{New phenomena in SU(3) supersymmetric gauge
  theory},'' \href{http://dx.doi.org/10.1016/0550-3213(95)00281-V}{{\em
  Nucl.Phys.} {\bfseries B448} (1995) 93--126},
\href{http://arxiv.org/abs/hep-th/9505062}{{\ttfamily arXiv:hep-th/9505062
  [hep-th]}}.

\bibitem{SV}
A.~D. Shapere and C.~Vafa, ``{BPS structure of Argyres-Douglas superconformal
  theories},''
\href{http://arxiv.org/abs/hep-th/9910182}{{\ttfamily arXiv:hep-th/9910182
  [hep-th]}}.

\bibitem{KS}
M.~{Kontsevich} and Y.~{Soibelman}, ``{Stability structures, motivic
  Donaldson-Thomas invariants and cluster transformations},'' {\em ArXiv
  e-prints} (Nov., 2008) , \href{http://arxiv.org/abs/0811.2435}{{\ttfamily
  arXiv:0811.2435 [math.AG]}}.

\bibitem{GMN08}
D.~Gaiotto, G.~W. Moore, and A.~Neitzke, ``{Four-dimensional wall-crossing via
  three-dimensional field theory},''
  \href{http://dx.doi.org/10.1007/s00220-010-1071-2}{{\em Commun.Math.Phys.}
  {\bfseries 299} (2010) 163--224},
\href{http://arxiv.org/abs/0807.4723}{{\ttfamily arXiv:0807.4723 [hep-th]}}.

\bibitem{GMN10}
D.~Gaiotto, G.~W. Moore, and A.~Neitzke, ``{Framed BPS States},''
\href{http://arxiv.org/abs/1006.0146}{{\ttfamily arXiv:1006.0146 [hep-th]}}.

\bibitem{GMN09}
D.~Gaiotto, G.~W. Moore, and A.~Neitzke, ``{Wall-crossing, Hitchin Systems, and
  the WKB Approximation},''
\href{http://arxiv.org/abs/0907.3987}{{\ttfamily arXiv:0907.3987 [hep-th]}}.

\bibitem{ACCERV1}
M.~{Alim}, S.~{Cecotti}, C.~{Cordova}, S.~{Espahbodi}, A.~{Rastogi}, and
  C.~{Vafa}, ``{BPS Quivers and Spectra of Complete N=2 Quantum Field
  Theories},''
\href{http://arxiv.org/abs/1109.4941}{{\ttfamily arXiv:1109.4941 [hep-th]}}.

\end{thebibliography}\endgroup
\bibliographystyle{utphys}

\end{document}